\documentclass[letterpaper,12pt]{article}
\usepackage[utf8]{inputenc}
\DeclareUnicodeCharacter{211D}{\ensuremath{\mathbb{R}}}
\usepackage{setspace}
\usepackage[margin=1in]{geometry}
\usepackage[parfill]{parskip}
\usepackage{authblk}
\usepackage[round]{natbib}
\usepackage{amssymb}
\usepackage{amsmath}
\usepackage{amsthm}
\usepackage{mathtools}
\usepackage{bbm}
\usepackage{gensymb}
\usepackage{url}
\usepackage[colorlinks,linkcolor=blue,citecolor=blue,urlcolor=blue,filecolor=blue]{hyperref}
\usepackage[linesnumbered,ruled]{algorithm2e}
\SetAlFnt{\small}
\usepackage{graphicx}
\usepackage{float}
\usepackage{multirow}
\usepackage[font=small]{caption}
\usepackage{appendix}
\floatstyle{plaintop}
\restylefloat{table}
\graphicspath{{./images/}}
\newcommand{\bld}[1]{\boldsymbol{\mathbf{#1}}}
\numberwithin{equation}{section}
\numberwithin{figure}{section}
\numberwithin{table}{section}
\newtheorem{prop}{Proposition}[section]

\title{Incorporating Subsampling into Bayesian Models for High-Dimensional Spatial Data}
\author[1]{Sudipto Saha}
\author[1]{Jonathan R. Bradley}
\affil[1]{Department of Statistics, Florida State University, Tallahassee, FL 32306, USA.}

\date{}

\begin{document}
\maketitle{}

\begin{abstract}
    Additive spatial statistical models with weakly stationary process assumptions have become standard in spatial statistics. However, one disadvantage of such models is the computation time, which rapidly increases with the number of data points. The goal of this article is to apply an existing subsampling strategy to standard spatial additive models and to derive the spatial statistical properties. We call this strategy the ``spatial data subset model'' (SDSM) approach, which can be applied to big datasets in a computationally feasible way. Our approach has the advantage that one does not require any additional restrictive model assumptions. That is, computational gains increase as model assumptions are removed when using our model framework. This provides one solution to the computational bottlenecks that occur when applying methods such as Kriging to ``big data''. We provide several properties of this new spatial data subset model approach in terms of moments, sill, nugget, and range under several sampling designs. An advantage of our approach is that it subsamples without throwing away data, and can be implemented using datasets of any size that can be stored. We present the results of the spatial data subset model approach on simulated datasets, and on a large dataset consists of 150,000 observations of daytime land surface temperatures measured by the MODIS instrument onboard the Terra satellite.
    
\end{abstract}

\section{Introduction} \label{sec:intro}
A usual inferential task for spatially referenced data is to predict a spatial process at observed and unobserved locations (i.e. spatial prediction). The classical approach to perform spatial prediction is known as ``Kriging''. Taking a Bayesian approach for parameter estimation and spatial prediction is a well-established and widely used approach in spatial statistics (e.g., see \citealp{more1989bayesian,omre1989bayesian,cressie1993statistics,handcock1993bayesian,banerjee2014hierarchical} for standard references). Classical spatial statistical models assume that a latent spatial process is weakly stationary with known covariogram, e.g. Mat{\'e}rn, exponential etc. \citep{banerjee2014hierarchical}. Although Bayesian prediction in this setting is possible and provides accurate spatial predictions for small datasets, the processing time rapidly increases with the increase in the size of dataset. This is a well-known problem that has motivated several methodologies \citep{vecchia1988estimation,nychka2000spatial,cressie2008fixed,rue2009approximate,xu2013efficient,nychka2015multiresolution,rue2017bayesian,vigsnes2017fast}. In this article, we provide a computationally feasible solution to analyze big data using a classical Bayesian spatial model with standard weakly stationary or non-stationary process assumptions. In particular, we develop a new ``spatial data subset model'' approach by applying the use of a recently introduced ``data subset model'' approach \citep{bradley2021approach} to spatial data. We also provide inferential properties of the data in terms of its moments, and spatial properties in terms of sill, nugget and range, under two sampling designs that are natural for spatial data.

We extend the use of data subset model \citep{bradley2021approach} and develop the spatial data subset model that allows one to implement classical Bayesian spatial models for analyzing big data quickly. To achieve this, we redefine the data model of a classical Bayesian spatial model using subsamples, without imposing any additional model assumptions. This is different than what has been done in the existing literature. For example, \cite{byers2022applied} presented a similar work where they used a Bootstrap Random Spatial Sampling (BRSS) method to estimate the full Bayesian-Kriging model over large samples. They independently generated $B$ bootstrap subsamples of size $n$, and estimated model parameters independently for each of $B$ bootstraps. In our proposed model, we redefine the full data model of the classical Bayesian spatial model to be semi-parametric using subsamples and we add a subset model to the hierarchical structure, rather than estimating model parameters independently for each bootstrap subsamples or approximating the full data model using the subsamples. This is the primary difference between our exact approach, and the other exact and approximate approaches \citep{vecchia1988estimation,stein2004approximating,gunawan2017fast,quiroz2019speeding,katzfuss2021general,byers2022applied}. Moreover, adding a subset model to the hierarchical structure itself provides the flexibility to choose a sampling method (from the survey sampling literature) inside the model.

There are many existing approaches for subsampling spatial data. One approach is to divide the spatial domain into overlapping subblocks which are obtained by moving a subsampling window over sequences of increasing index sets \citep{guan2004nonparametric}, and draw the subsamples from these subblocks. One disadvantage of this approach is that the overlapping subblocks can introduce unnecessary complexity into the model. To overcome this issue, a common approach is to divide the spatial domain into several partitions or subshapes \citep{possolo1991subsampling,sherman1994nonparametric,lahiri1999asymptotic,ekstrom2004subsampling}. One can partition a spatial domain using one of many partitioning methods. For example, equal area partition \citep{sang2011covariance}, hierarchical clustering based partition \citep{heaton2017nonstationary}, tree based partition \citep{konomi2014adaptive}, partition based on cluster centroids \citep{knorr2000bayesian,kim2005analyzing}, partition based on mixture model \citep{neelon2014multivariate} etc. However, all these spatial partitioning approaches assume independence across partitions (blockwise independence), which can be a strong assumption \citep{bradley2015comparing} in certain scenarios. To avoid any additional assumption, we choose the sampling method for our spatial data subset mode from the survey sampling literature \citep{lohr2010sampling,nassiuma2001survey}. In particular, we consider simple designs, like simple random sampling (SRS) and stratified random sampling, that are natural for spatial data. Cluster sampling is another natural choice for spatial data. However, in the survey literature it has been shown to be less efficient than stratified random sampling \citep{lohr2010sampling}, and hence, we do not consider it here.

\cite{possolo1991subsampling} is an example of a method that uses a single subsample of size $n$ and discards $(N-n)$ spatial observations, where $N$ is the size of the entire dataset. Of course this is a standard data pre-processing step for whenever a method is not scalable (e.g., a subset of the data is used). Several in the more general Gaussian process literature have given this strategy a name, i.e., the ``Subset of Data'' (SoD) method  \citep{lawrence2002fast,seeger2003bayesian,keerthi2005matching,chalupka2013framework,hayashi2020random,liu2020gaussian}. Our proposed method is different from using a single subsample, as the Data Subset Model approach iteratively re-samples the data, and if one runs the MCMC long enough, all data is used for inference. We define such method that makes use of all $N$ observations without discarding them as ``fully scalable''. To our knowledge, there is no such Bayesian spatial model to analyze large spatial data that is fully scalable (for large enough length of the Markov chain) to datasets of any size that can be stored.

Outside of subsampling there are several other strategies to make a Gaussian Process (GP) scalable in Bayesian and non-Bayesian contexts. For example, reduced rank methodologies reduce the parameter space to speed up computations \citep{cressie2008fixed}. Sparse GPs similar reduce the parameter space via sparsity (e.g., see \cite{besag1974spatial,besag1991bayesian} in the spatial context, and \cite{hensman2013gaussian} in a more general GP context). Approximate Bayesian strategies approximate the likelihood to aid with computations, for example, the Vecchia approximation \citep{katzfuss2021general} writes the likelihood as a product and truncate the product to fewer terms, and integrated nested Laplace approximations (INLA, \citealp{rue2009approximate}) use Gaussian approximations of the marginal posterior to aid with inference. Some apply spatial statistical models to spatial partitions or for a given prediction location adaptively to create a partition based on nearest neighborhoods (e.g., see \citealp{heaton2019case}, and the local GP approach of \citealp{gramacy2015local} for examples). For a broad discussion on the available strategies in spatial statistics see \cite{bradley2021approach}, \cite{heaton2019case} and in the generic GP context see \cite{liu2020gaussian}. In general, all of the fully Bayesian strategies involve additional parametric assumptions, or approximations. The data subset model, does not add assumptions, but rather removes assumptions; that is, the hold-out sample in the data subset model approach is assumed to follow its true nonparametric distribution instead of some parametric density enforced to have sparsity or dimension reduction structure. Furthermore the data subset model approach does not involve approximations used in approximate Bayesian strategies.

Our proposed spatial data subset model has a hyperparameter, subsample size $(n)$, that can be set to a value to achieve a computational goal, rather than determining an optimal subsample size. There are several ways to choose an optimal subsample size. For example, one can choose an optimal subsample size by minimizing the mean square error of variance estimation of the blocked sample variance \citep{politis1993sample}, or by minimizing the combined expansions for the bias and the variance parts of the subsample variance estimator \citep{nordman2004optimal}. A limitation of both of these approaches is that they both demand an approximation to the full likelihood. However, in our model, we \textit{redefine} the full likelihood to be semi-parametric (or more flexible) based on the subsamples instead of approximating it. So, we keep the subsample size as a hyperparameter for the users to choose based on their computational goal, which makes our model flexible to demand. This is particularly exciting because our model scales with the subsample design, and not with the size of the entire dataset. Hence, one can scale our model to a dataset of any size that can be stored by choosing the subsample size small enough.

While $n$ can be chosen to achieve a computational goal, this is not necessarily a positive. Ideally, we would choose $n$ for inference, and choosing $n$ in this manner, is to make an admission that a given model is not fully scalable. The main goal of this article is to apply the data subset approach to standard spatial models (i.e., kriging). It is well known that kriging is not fully scalable, which leads to a natural restriction on $n$, since kriging is known to break down computationally when $n\ge 10,000$ (e.g., see \citealp{rulliere2018nested}), and can be extremely slow at $n = 10,000$. In our empirical results we often see that $n\ll N$ produces similar answers compared to kriging based on $N$. This is consistent with results in \cite{bradley2021approach} and results in the SoD literature (\citealp{chalupka2013framework}, \S 5).

The choice of the sampling method affects the fundamental properties of the model. In particular, we are interested in the changes in the first and second moments, and the spatial properties of the data as we change the sampling design. Hence, we present a detailed discussion on the inferential, as well as the spatial properties of the data under two sampling strategies - SRS and stratified random sampling. That is, we investigate several spatial properties like sill, nugget and (effective) range \citep{cressie1993statistics,banerjee2014hierarchical}, including the mean, variance and covariance \citep{casella2021statistical} of the data for SRS and stratified random sampling. One exciting feature of our model is that our model retains the stationarity of the true data model, i.e. our model becomes weakly/intrinsically stationary or non-stationary when the true data model is respectively weakly/intrinsically stationary or non-stationary.

The remainder of the article is organized as follows. In Section \ref{sec:review}, we provide an extensive review of the existing data subset model in a generic GP setting. In Section \ref{sec:method}, we present the proposed spatial data subset model, along with the inferential and spatial properties under the mentioned two sampling designs and the MCMC implementation. In Section \ref{sec:simulation}, we demonstrate the performance of our model by providing an illustration through a simulated dataset, followed by a more extensive simulation study and an analysis on the scalability of the model. We demonstrate that the performance of our model gets better as we increase the number of subsamples. We also show that the proposed model is fully scalable (for large enough length of the Markov chain) to the datasets of any size that can be stored. In Section \ref{sec:application}, we apply our model to the benchmark dataset from \cite{heaton2019case}, that consists of Land Surface Temperature (LST) captured by the Terra satellite. We find that our semi-parametric version of classical Bayesian spatial models provides competitive inferential and computational results as compared to several existing strategies \citep{heaton2019case}. Finally, a concluding discussion is presented in Section \ref{sec:discussion}. We provide the proofs of all the technical results in the Appendix for convenience.

\section{Review} \label{sec:review}
In this section, we review general subsampling strategies. In particular, we review the ``data subset approach'' in the generic GP Bayesian setting from \cite{bradley2021approach}, and the SoD method \citep{chalupka2013framework}.

\subsection{Review of Data Subset Model}

Let, an $N$-dimensional data vector be $\bld{y}=(Y_1,\ldots,Y_N)'$, the latent process vector be $\bld{\nu}$, and a generic real-valued parameter vector be $\bld{\theta}$. In general, a Bayesian hierarchical model can be written as the product of the following conditional and marginal distributions:
\begin{align}
    &\text{Data Model:} \quad \prod_{i=1}^N f(Y_i |\bld{\nu},\bld{\theta}) \nonumber \\
    &\text{Process Model:} \quad f(\bld{\nu}|\bld{\theta}) \nonumber \\
    &\text{Parameter Model:} \quad f(\bld{\theta}). \label{eq:genericBHM}
\end{align}
\cite{bradley2021approach} considered $N$ to be so large that estimating $\bld{\nu}$ and $\bld{\theta}$ using \eqref{eq:genericBHM} (referred as ``full parametric model'') directly is not possible. To solve this problem, \cite{bradley2021approach} proposed to replace the data model in (\ref{eq:genericBHM}) with the data subset model (DSM) in the following way:
\begin{align}
    &\text{Data Subset Model:} \quad \left\{\prod_{i=1}^N f(Y_i |\bld{\nu},\bld{\theta},\bld{\delta})^{\delta_i}\right\} \frac{m(\bld{1}_N,\bld{y})}{m(\bld{\delta},\bld{y}_{\delta})} \nonumber \\
    &\text{Process Model:} \quad f(\bld{\nu}|\bld{\theta}) \nonumber \\
    &\text{Parameter Model:} \quad f(\bld{\theta}) \nonumber \\
    &\text{Subset Model:} \quad \text{Pr}(\bld{\delta}| n), \label{eq:datasubsetmodel}
\end{align}
where $\bld{\delta}=(\delta_1,\ldots,\delta_N)'$ be the $N$-dimensional random vector consists of ones and zeros, $\bld{1}_N$ (and $\bld{0}_N$) be the $N$-dimensional vector of ones (and zeros), and $\bld{y}_{\delta}=(Y_i:\delta_i=1)'$ be the $n$-dimensional random vector since they chose exactly $n$ number of ones in the $\bld{\delta}$ vector by setting $\sum_{i=1}^N \delta_i =n$. The subset model determines which $n$ observations to include in the likelihood through $\delta_i$ (= 0 or 1), which is treated as a parameter in the hierarchical model. The marginal distribution, $m(\bld{\delta},\bld{y}_{\delta})$, is defined as
\begin{equation} \label{eq:marginal}
    m(\bld{\delta},\bld{y}_{\delta})=\int \int \left\{\prod_{i=1}^N f(Y_i |\bld{\nu},\bld{\theta},\bld{\delta})^{\delta_i}\right\}f(\bld{\nu}|\bld{\theta})f(\bld{\theta}) d\bld{\nu} d\bld{\theta},
\end{equation}
where $m(\bld{\delta},\bld{y}_{\delta})$ is the marginal distribution of $\bld{y}_{\delta}$ from the full parametric model. $m(\bld{1}_N,\bld{y})$ is the marginal distribution of the data associated with the full parametric model. Hence, the data subset model in (\ref{eq:datasubsetmodel}) re-weights the likelihood evaluated at $n (\ll N)$ observations by the ratio of marginal densities, $m(\bld{1}_N,\bld{y})/m(\bld{\delta},\bld{y}_{\delta})$.

\cite{bradley2021approach} established several important theoretical results of the model in \eqref{eq:datasubsetmodel}. In particular, the model in \eqref{eq:datasubsetmodel} is proper given that the model in \eqref{eq:genericBHM} is proper, and the marginal distributions of data $\bld{y}$ are the same for the Bayesian models in \eqref{eq:genericBHM} and \eqref{eq:datasubsetmodel}. Thus, the Bayesian models in \eqref{eq:genericBHM} and \eqref{eq:datasubsetmodel} have the same expression of $f(\bld{\nu}|\bld{\theta})$, $f(\bld{\theta})$ and $m(\bld{1}_N,\bld{y})$. This implies that, marginally, the distributional assumptions of the data, process, and parameters are the same between \eqref{eq:datasubsetmodel} and \eqref{eq:genericBHM}. Sufficiency is a key property in Bayesian analysis, and \cite{bradley2021approach} showed that $\bld{y}_{\delta}$ is a partial sufficient statistic for $\bld{\nu}$ and $\bld{\theta}$. Another important result is the data vector $\bld{y}$ and the $\bld{\delta}$ vector are independent. This result is particularly important from a computational perspective, as it leads to an efficient sampler that we develop further in the spatial setting. In particular, to sample from the posterior $f(\bld{\nu},\bld{\theta},\bld{\delta}|\bld{y})$, one first samples from $\text{Pr}(\bld{\delta}|n)$ and then samples from the posterior using subsample $\bld{y}_{\delta}$ (i.e., $p(\bld{\nu},\bld{\theta}|\bld{y}_{\delta})$). Note that, the data are continually resampled in this sampling scheme, which is different from many existing subsampling strategies that use a single subsample. \cite{bradley2021approach} explored this sampler in truly high-dimensional settings (i.e., a dataset of size 10 Gigabytes and a simulated dataset of size 100 million).

A semi-parametric interpretation of the model in \eqref{eq:datasubsetmodel} in the generic Bayesian GP setting is also discussed in \cite{bradley2021approach}, where the following model is considered.
\begin{align}
    &\text{Data Model:} \quad w_{\delta}(\bld{y}_{-\delta}) \prod_{\{i:\delta_i=1\}} f(Y_i |\bld{\nu},\bld{\theta},\bld{\delta}) \nonumber \\
    &\text{Process Model:} \quad f(\bld{\nu}|\bld{\theta}) \nonumber \\
    &\text{Parameter Model:} \quad f(\bld{\theta}) \nonumber \\
    &\text{Subset Model:} \quad \text{Pr}(\bld{\delta}| n), \label{eq:semiparametric}
\end{align}
where $w_{\delta}(\cdot)$ is the true unknown unparameterized pdf for the $(N-n)$-dimensional random vector $\bld{y}_{-\delta}$ for a given $\bld{\delta}$. \cite{bradley2021approach} referred this model as the semi-parametric full model (SFM). When one assumes that $\bld{y}$ and $\bld{\delta}$ are independent in \eqref{eq:semiparametric}, then \cite{bradley2021approach} showed that \eqref{eq:semiparametric} has many of the same properties as \eqref{eq:datasubsetmodel}. Specifically, $\bld{y}_{\delta}$ is a partially sufficient statistic for $\bld{\nu}$ and $\bld{\theta}$. Furthermore, \cite{bradley2021approach} established that $f_{DSM}(\bld{\nu},\bld{\theta}|\bld{y},\bld{\delta},n)=f_{SFM}(\bld{\nu},\bld{\theta}|\bld{y},\bld{\delta},n)$ and $f_{DSM}(\bld{\nu},\bld{\theta}|\bld{y},n)=f_{SFM}(\bld{\nu},\bld{\theta}|\bld{y},n)$, where $f_{DSM}$ and $f_{SFM}$ be the posterior distributions under the model in \eqref{eq:datasubsetmodel} and \eqref{eq:semiparametric}, respectively. These two results are particularly interesting because together they show that the posterior distributions of DSM and SFM are equivalent when one adds the assumption that $\bld{y}$ and $\bld{\delta}$ are independent in \eqref{eq:semiparametric}.

\subsection{Review of Subset of Data Method}
It is important to understand that the data subset model does not use a single subsample defined by $\bld{\delta}$. This is because $\bld{\delta}$ is actively resampled in the Gibbs-within-composite sampler. This is different from the setting where one makes use of a single subsample. Some in the general GP literature call this approach the Subset of Data (SoD) method \citep{lawrence2002fast,seeger2003bayesian,keerthi2005matching,chalupka2013framework,hayashi2020random,liu2020gaussian}. The SoD method simply uses a single subsample of size $n (n \ll N)$ for inference. For reviews of SoD see \cite{chalupka2013framework} and \cite{liu2020gaussian}. The difference between SoD and the Data Subset Approach is important because the data subset model scales to higher dimensions in a similar way that SoD scales to higher dimension without discarding $N-n$ observations. We say a method is ``fully scalable'' if it can be implemented on $N$ observations \textit{without} discarding any observations. Thus, the SoD method is not fully scalable by definition, and the data subset model approach is fully scalable for large enough $G$, where $G$ is the length of the Markov chain. That is, as one continually samples from the Gibbs-within-composite sample, as long as each observation has a non-zero probability of being selected, all data will be used when $G$ is large enough.

Resampling in the SoD literature is sometimes based on $n$ randomly selected points, which has computational complexity of $\mathcal{O}(n)$, since one does not need to scan through all $N$ data-points to produce the subsample. Others in the SoD literature have considered clustering algorithms, which are on the order of $\mathcal{O}(nN)$ or $\mathcal{O}(N \hspace{2pt} log(n))$ \citep{chalupka2013framework}. A fully Bayesian data subset model strategy is more restrictive here, since the data cannot be used to select the $n$ points from $N$ (such a strategy would yield an empirical Bayesian approach with unchecked variability). As a result, we make use of random selected points and do not make use of any clustering algorithms.

\section{Methodology} \label{sec:method}
Suppose, we observe an $N$-dimensional spatial data vector $\bld{y}=\left(Y(\bld{s}_1),\ldots,Y(\bld{s}_N)\right)'$, where the locations are $\{\bld{s}_1,\ldots,\bld{s}_N\} \in D \subset \Re^d$. Let, the corresponding $N$-dimensional latent process of interest be $\bld{\nu}=\left(\nu(\bld{s}_1),\ldots,\nu(\bld{s}_N)\right)' \in \mathbb{R}^N$, the $N\times p$ dimensional matrix of known covariates be $\bld{X} \in \mathbb{R}^{N\times p}$, and the $p$-dimensional unknown regression parameter be $\bld{\beta} \in \mathbb{R}^p$. The additive model of the classical Bayesian spatial model is defined as,
\begin{equation}
    \bld{y}=\bld{X\beta}+\bld{\nu}+\bld{\epsilon},
\end{equation}
where $\epsilon(\bld{s}_i) \stackrel{iid}{\sim} \mathcal{N}\left(0,\tau^2\right),\forall i=1,\ldots,N$ is the white noise and $\bld{w}=\bld{X\beta}+\bld{\nu}$ is known as the latent process. Hence, the data mode of classical Bayesian spatial model refers to the conditional distribution $f(\bld{y}|\bld{\nu},\bld{\theta})=\prod_{i=1}^N f(Y(\bld{s}_i)|\bld{\nu},\bld{\theta})$, where $\bld{\theta}$ be a real-valued parameter vector. However, we consider $N$ to be so large that estimating $\bld{\nu}$ and $\bld{\theta}$ using classical Bayesian spatial model directly is not possible in real time. As a solution, we extend the data subset model \citep{bradley2021approach} by incorporating it into the classical Bayesian spatial model and the resulting model is known as spatial data subset model.

\subsection{Bayesian Hierarchical Model} \label{subsec:BHM}
We write the data subset model \citep{bradley2021approach} incorporating into the classical Bayesian spatial model, where the resulting joint spatial data subset model (SDSM) is proportional to the product of the following conditional and marginal distributions,
\begin{equation} \label{eq:BHM}
    \begin{split}
        \text{Data Subset Model: } & w_{\delta}(\bld{y}_{-\delta})\left\{\prod_{\{i:\delta(\bld{s}_i)=1\}} f\left(Y(\bld{s}_i)|\bld{\nu},\bld{\theta}\right)\right\},\\
        \text{Process Model: } \bld{\nu}|{\sigma^2,\phi} &\sim \mathcal{N}\left(\bld{0},\sigma^2\bld{H}(\phi)\right),\\
        \text{Parameter Model 1: } \bld{\beta}|\sigma_{\beta}^2 &\sim \mathcal{N}\left(\bld{0},\sigma_{\beta}^2\bld{I}_p\right),\\
        \text{Parameter Model 2: } \tau^2 &\sim \mathcal{IG}(a_{\tau},b_{\tau}),\\
        \text{Parameter Model 3: } \sigma^2 &\sim \mathcal{IG}(a_{\sigma},b_{\sigma}),\\
        \text{Parameter Model 4: } \sigma_{\beta}^2 &\sim \mathcal{IG}(a_{\beta},b_{\beta}),\\
        \text{Parameter Model 5: } \phi &\sim \pi(\phi),\\
        \text{Subset Model: } \bld{\delta} &\sim \text{Pr}(\bld{\delta}|n),
    \end{split}
\end{equation}
where $f\left(Y(\bld{s}_i)|\bld{\nu},\bld{\theta}\right)$ is the probability density function (pdf) of a $\mathcal{N}\left(\bld{x}(\bld{s}_i)'\bld{\beta}+\nu(\bld{s}_i),\tau^2\right)$ distribution, $\bld{\delta}=\left(\delta(\bld{s}_1),\ldots,\delta(\bld{s}_N)\right)'$ is an $N$-dimensional random vector consists of ones and zeros, $\bld{y}_{-\delta}=(Y(\bld{s}_i):\delta(\bld{s}_i)=0)'$ consists of all such $Y(\bld{s}_i)$ where $\delta(\bld{s}_i)=0$ for a given $\bld{\delta}$, and $w_{\delta}(\bld{y}_{-\delta})$ is a probability density that represents the true unknown unparameterized pdf for $\bld{y}_{-\delta}$. We define the real-valued parameter vector $\bld{\theta}=(\bld{\beta},\tau^2,\sigma^2,\sigma_{\beta}^2,\phi)'$. A general structure of an element of the matrix $\bld{H}(\phi)=\{h_{ij}(\phi):i,j=1,\ldots,N\} \in \mathbb{R}^{N\times N}$ is $h_{ij}(\phi)=\rho_c(\bld{s}_i-\bld{s}_j;\phi)$, where $\rho_c$ is a valid correlation function on $\Re^2$ indexed by the parameter $\phi$. $h_{ij}(\phi)$ and the density function $\pi(\phi)$ both depends on the choice of the function $\rho_c$. For example, if we choose exponential covariogram, i.e. $h_{ij}(\phi)=\exp(-\phi\|\bld{s}_i-\bld{s}_j\|)$, then a Gamma distribution might seem sensible for the choice of $\pi(\phi)$ \citep{banerjee2014hierarchical}. $\mathcal{N}(\bld{\mu},\bld{\Sigma})$ is shorthand for a multivariate Normal distribution with mean vector $\bld{\mu}$ and covariance matrix $\bld{\Sigma}$. Similarly, $\mathcal{IG}(a,b)$ is used as shorthand for inverse Gamma distribution with shape parameter $a$ and scale parameter $b$. Notice that, the hyperparameters of the $\mathcal{IG}$ distributions are different across $\tau^2$, $\sigma^2$ and $\sigma_{\beta}^2$. It is important to mention that one has the flexibility to consider other expressions of the spatial data model. For example, one can marginalize across $\bld{\nu}$. This would lead to one fewer block update of parameters in the sampler. We choose to include $\bld{\nu}$, as it leads to conjugate updates, which helps with mixing of Markov chains \citep{banerjee2014hierarchical}. Additionally, one has the flexibility to include other covariograms. We often use the exponential covariogram for illustration.

The model in \eqref{eq:BHM} shows that for a given $\bld{\delta}$, $\delta(\bld{s}_i)=1$ includes $Y(\bld{s}_i)$ in the expression of the parametric portion of the likelihood (proportionally), and $\delta(\bld{s}_i)=0$ removes $Y(\bld{s}_i)$ from the expression of the parametric portion of the likelihood (proportionally). We sample $\bld{\delta}$ from $\text{Pr}(\bld{\delta}|n)$ for a given $n$ such that $\sum_{i=1}^N \delta(\bld{s}_i)=n\ll N$, where $n$ is the ``subsample size''. Therefore, for any sample of $\bld{\delta}$, only $n$ number of $Y(\bld{s}_i)$ would be included in the expression of likelihood (proportionally). This further means that, the spatial data subset model only requires an $n$-dimensional data vector $\bld{y}_{\delta}=(Y(\bld{s}_i):\delta(\bld{s}_i)=1)'$ (at a time), instead of the larger $N$-dimensional data vector $\bld{y}$, to estimate $\bld{\nu}$ and $\bld{\theta}$. We choose $n$ in a way that the model in \eqref{eq:BHM} achieves computational goals \citep{bradley2021approach}. We specify $\text{Pr}(\bld{\delta}|n)$ in a manner that it produces simple random samples (SRS) without replacement and stratified random samples as described in Section \ref{sec:intro}.

We assume that $\bld{\delta}$ is independent of $\bld{y}$, and \cite{bradley2021approach} showed that such a model exists. This assumption leads to a very efficient composite sampling scheme for posterior inference, where one first samples from $\text{Pr}(\bld{\delta}|\bld{y},n)=\text{Pr}(\bld{\delta}|n)$, via independence, and then samples from $\text{Pr}(\bld{\nu},\bld{\theta}|\bld{y},\bld{\delta},n)$. The conditional distribution $\text{Pr}(\bld{\nu},\bld{\theta}|\bld{y},\bld{\delta},n)$ only includes $n\ll N$ observations, and hence, one can choose $n$ small enough so that sampling from $\text{Pr}(\bld{\nu},\bld{\theta}|\bld{y},\bld{\delta},n)$ is efficient. This composite sampling scheme iteratively re-samples the data, so that every observation is included provided that the composite sampling procedure runs long enough. In practice, sampling from $\text{Pr}(\bld{\nu},\bld{\theta}|\bld{y},\bld{\delta},n)$ can require MCMC, which leads to a Gibbs-within-Composite sampler. We provide more discussion on this in Section \ref{subsec:mcmc}.

\subsection{Properties of Spatial Data Subset Model when the Sampling Design is SRS Without Replacement} \label{subsec:SRS}
The properties of a classical Bayesian spatial model are well-established in numerous literature (e.g., see \citealp{cressie1993statistics,banerjee2014hierarchical}, among several others). Enforcing $\bld{\delta}$ into the model does not change the process and parameter models, and hence it does not change the inherent properties of the latent process. But it does change the properties of the data since we are subsampling a smaller number of data points from the entire dataset. Moreover, the properties vary with the choice of sampling technique for $\bld{\delta}$. This leads us to the following propositions when $\text{Pr}(\bld{\delta}|n)$ is defined by simple random samples (SRS) without replacement.

In nonparametric statistics it is often assumed that the true measure that generated $\bld{y}_{-\delta}$ is unknown and not parameterized (\citealp{hollander2013nonparametric}, pg. 2). We assume that the true (unknown) distribution is absolutely continuous so that it has density $w$, which is also unknown. Essentially, we assume that any given $n$-dimensional ``training'' dataset $\bld{y}_{\delta}$ follows a parametric model, and the corresponding holdout $\bld{y}_{-\delta}$ is distributed according to its true unknown data generating measure as one would in nonparametric statistics. We call the mean computed using $w$ the ``true mean'', the variance computed using $w$ the ``true variance'', and the covariance computed using $w$ the ``true covariance''.

\begin{prop} \label{prop:1}
When $\text{Pr}(\bld{\delta}|n)$ is defined by simple random samples without replacement, the expectation, variance and covariance of the data $Y(\bld{s}_i)$ for $i=1,\ldots,N$ given the real-valued parameter vector $\bld{\theta}=(\bld{\beta},\tau^2,\sigma^2,\sigma_{\beta}^2,\phi)'$ are given by
\begin{align*}
    \mathbb{E}[Y(\bld{s}_i)|\bld{\theta}] & =p\{\bld{x}(\bld{s}_i)'\bld{\beta}\}+(1-p)\tilde{\mu}(\bld{s}_i)=p\{\bld{x}(\bld{s}_i)'\bld{\beta}-\tilde{\mu}(\bld{s}_i)\}+\tilde{\mu}(\bld{s}_i),\\
    Var(Y(\bld{s}_i)|\bld{\theta}) &= p(\tau^2+\sigma^2)+(1-p)\tilde{\sigma}^2+p(1-p)\{\bld{x}(\bld{s}_i)'\bld{\beta}-\tilde{\mu}(\bld{s}_i)\}^2\\
    &= p(\tau^2+\sigma^2-\tilde{\sigma}^2)+p(1-p)\{\bld{x}(\bld{s}_i)'\bld{\beta}-\tilde{\mu}(\bld{s}_i)\}^2+\tilde{\sigma}^2,\\
    Cov(Y(\bld{s}_i),Y(\bld{s}_j)|\bld{\theta}) &= a\tilde{C}(\bld{s}_i,\bld{s}_j)+(b+p^2)\sigma^2h_{ij}(\phi)\\
    &+b\{\bld{x}(\bld{s}_i)'\bld{\beta}-\tilde{\mu}(\bld{s}_i)\}\{\bld{x}(\bld{s}_j)'\bld{\beta}-\tilde{\mu}(\bld{s}_j)\},
\end{align*}
where $p=\frac{n}{N},a=\left(\frac{n}{N}\frac{n-1}{N-1}-2\frac{n}{N}+1\right),b=\frac{n}{N}\left(\frac{n-1}{N-1}-\frac{n}{N}\right)$, $\tilde{\mu}(\bld{s}_i)$ is the true mean of $Y(\bld{s}_i)$, $\tilde{\sigma}^2$ is the true variance of $Y(\bld{s}_i)$, and $\tilde{C}(\bld{s}_i,\bld{s}_j)$ is the true covariance between $Y(\bld{s}_i)$ and $Y(\bld{s}_j)$.
\end{prop}

\noindent \textit{Proof:} See Appendix \ref{app:a1}.

Proposition \ref{prop:1} is useful to understand the characteristics of the spatial data subset model in \eqref{eq:BHM} from the perspective of its first and second moments. Notice that, the expectation of $Y(\bld{s}_i)|\bld{\theta}$ is a weighted sum of the parametric mean and the true mean (shown in the first expression of $\mathbb{E}[Y(\bld{s}_i)|\bld{\theta}]$). The second expression of $\mathbb{E}[Y(\bld{s}_i)|\bld{\theta}]$ shows that the true mean of $Y(\bld{s}_i)|\bld{\theta}$ is biased by the weighted difference between the parametric mean and the true mean. Furthermore, from the expression of the expectation of $Y(\bld{s}_i)|\bld{\theta}$ we obtain
\begin{equation} \label{eq:prop1exp}
    \mathbb{E}[Y(\bld{s}_i)|\bld{\theta}]=\begin{cases}
    \tilde{\mu}(\bld{s}_i) & n=0\\
    \bld{x}(\bld{s}_i)'\bld{\beta} & n=N.
    \end{cases}
\end{equation}
Equation \eqref{eq:prop1exp} shows that $Y(\bld{s}_i)|\bld{\theta}$ achieves the true mean when we do not subsample, and achieves the parametric mean when we subsample all observations. Proposition \ref{prop:1} provides two expressions for the variance. In the first expression, the variance of $Y(\bld{s}_i)|\bld{\theta}$ is a weighted sum of the parametric variance and the true variance with an additional component. The second expression of $Var(Y(\bld{s}_i)|\bld{\theta})$ shows that the true variance of $Y(\bld{s}_i)|\bld{\theta}$ is biased by two components - (a) the weighted difference between the parametric variance and the true variance, and (b) the weighted square of the difference between the parametric mean and the true mean. However, we still obtain
\begin{equation} \label{eq:prop1var}
    Var(Y(\bld{s}_i)|\bld{\theta})=\begin{cases}
    \hfil \tilde{\sigma}^2 & n=0\\
    \tau^2+\sigma^2 & n=N.
    \end{cases}
\end{equation}
Again we see from Equation \eqref{eq:prop1var} that the variance of $Y(\bld{s}_i)|\bld{\theta}$ attains the true variance when we do not subsample, and attains the parametric variance when we subsample all observations. Lastly, Proposition \ref{prop:1} also provides the covariance between $Y(\bld{s}_i)$ and $Y(\bld{s}_j)$ given $\bld{\theta}$ is a weighted sum of the true covariance, the parametric covariance, and the cross-product between the difference between the parametric and true means. Furthermore, we have
\begin{equation} \label{eq:prop1cov}
    Cov(Y(\bld{s}_i),Y(\bld{s}_j)|\bld{\theta}) = \begin{cases}
    \hfil \tilde{C}(\bld{s}_i,\bld{s}_j) & n=0\\
    \sigma^2h_{ij}(\phi) & n=N.
    \end{cases}
\end{equation}
Equation \ref{eq:prop1cov} shows that the covariance between $Y(\bld{s}_i)$ and $Y(\bld{s}_j)$ given $\bld{\theta}$ also attains the true covariance when we do not subsample, and attains the parametric covariance when we subsample all observations. Notice that, the joint spatial data subset model from \eqref{eq:BHM} becomes the true unparameterized model, $w_{\delta}(\bld{y})$, when $n=0$, and becomes the full parametric classical Bayesian spatial model, $\prod_{i=1}^N f\left(Y(\bld{s}_i),\bld{\nu},\bld{\theta}\right)$, when $n=N$. Hence, Equations \eqref{eq:prop1exp}, \eqref{eq:prop1var} and \eqref{eq:prop1cov} exactly show the characteristics that we expect from the joint spatial data subset model derived from \eqref{eq:BHM} under the conditions of no subsampling and subsampling all observations.

Typically, the spatial properties of a data are considered for the ``de-trended'' data process that has a zero mean. Hence, we define the de-trended version of $Y(\bld{s}_i), \forall i=1,\ldots,N$ given $\bld{\theta}$ and $\bld{\delta}$ as
\begin{equation} \label{eq:detrended}
    Z(\bld{s}_i)|\bld{\theta},\bld{\delta} = \begin{cases}
        Y(\bld{s}_i)-\bld{x}(\bld{s}_i)'\bld{\beta} & \delta(\bld{s}_i)=1\\
        Y(\bld{s}_i)-\tilde{\mu}(\bld{s}_i) & \delta(\bld{s}_i)=0.
    \end{cases}
\end{equation}
In a similar manner, we have that $\tilde{C}$ is the covariance of the true de-trended data process $Y(\bld{s}_i)-\tilde{\mu}(\bld{s}_i), \forall i=1,\ldots,N$. This leads to Proposition \ref{prop:2}.

\begin{prop}[Spatial Property] \label{prop:2}
Under SRS without replacement, if the true de-trended data process model is weakly and/or intrinsically stationary, then the de-trended data model defined in \eqref{eq:detrended} is respectively weakly and/or intrinsically stationary. If the true de-trended data process model is non-stationary, then the de-trended data model defined in \eqref{eq:detrended} is also non-stationary.
\end{prop}

\noindent \textit{Proof:} See Appendix \ref{app:a1}.

Proposition \ref{prop:2} implies that the de-trended data model defined in \eqref{eq:detrended} retains the spatial characteristics of the true de-trended data process model, since the covariance between $Z(\bld{s}_i)$ and $Z(\bld{s}_j)$ given $\bld{\theta}$ depends on the true covariance between $Y(\bld{s}_i)-\tilde{\mu}(\bld{s}_i)$ and $Y(\bld{s}_j)-\tilde{\mu}(\bld{s}_j)$ in the following way,
\begin{equation} \label{eq:prop2cov}
    Cov(Z(\bld{s}_i),Z(\bld{s}_j)|\bld{\theta}) = a\tilde{C}(\bld{s}_i,\bld{s}_j)+(b+p^2)\sigma^2h_{ij}(\phi),
\end{equation}
where $\tilde{C}(\bld{s}_i,\bld{s}_j)$ is the covariance of the true de-trended data process model, and $p,a,b$ are defined in Proposition \ref{prop:1} (see Appendix \ref{app:a1} for the proof of Equation \eqref{eq:prop2cov}). Notice that, the covariace between $Z(\bld{s}_i)$ and $Z(\bld{s}_j)$ given $\bld{\theta}$ is also a weighted sum of the true and parametric covariances of the de-trended data. In the case where the true de-trended data process model is intrinsically stationary, we get the variogram as
\begin{equation}
    2\gamma(\bld{s}_i-\bld{s}_j)=2\{p(\tau^2+\sigma^2)+(1-p)\tilde{\sigma}^2\}-2\left\{a\tilde{C}(\bld{s}_i,\bld{s}_j)+(b+p^2)\sigma^2h_{ij}(\phi)\right\},
\end{equation}
where $p,a,b,\tilde{\sigma}^2$ are defined in Proposition \ref{prop:1}, $\bld{s}_i-\bld{s}_j$ is the spatial lag between $\bld{s}_i$ and $\bld{s}_j$, and $2\gamma(\bld{s}_i-\bld{s}_j)$ is called the variogram (see Appendix \ref{app:a1} for the proof of variogram). This helps us to visualize the spatial properties of the model in \eqref{eq:BHM}. If we assume that the true de-trended data process model is weakly stationary, i.e. if the true covariance $\tilde{C}(\bld{s}_i,\bld{s}_j)$ depends on the lag between $\bld{s}_i$ and $\bld{s}_j$, then we have the usual definitions of the sill, nugget and range,
\begin{equation} \label{eq:spatprop}
    \begin{split}
        &\text{Sill}=\lim_{\lVert \bld{d}\rVert\to \infty}\gamma(\bld{d})=p(\tau^2+\sigma^2)+(1-p)\tilde{\sigma}^2,\\
        &\text{Nugget}=\lim_{\lVert \bld{d}\rVert\to 0^+}\gamma(\bld{d})=\left[p\tau^2+(p-b-p^2)\sigma^2\right]+\left[(1-p-a)\tilde{\sigma}^2+a\tilde{\tau}^2\right],\\
        &\text{Range}=\inf\left\{\bld{d}:a\tilde{C}(\bld{d})+(b+p^2)\sigma^2\rho_c(\bld{d};\phi)=0\right\},
    \end{split}
\end{equation}
where $\bld{d}=\bld{s}_i-\bld{s}_j$ is the spatial lag and $\tilde{\tau}^2$ is the nugget of the true de-trended data process. Notice that, the nugget is dependent on both parametric nugget ($\tau^2$) and the true nugget ($\tilde{\tau}^2$). Refer to Appendix \ref{app:a1} for the proof of sill, nugget and range. Equation \eqref{eq:spatprop} reproduces the sill, nugget, and range of the (true) parametric model when ($n=0$) $n=N$. Thus, we again achieve a type of balance between the true model and parametric model.

\subsection{Properties of Spatial Data Subset Model when the Sampling Design is Stratified Random Sampling} \label{subsec:Strat}
In the case of stratified random sampling, the properties of the spatial data subset model discussed in Section \ref{subsec:SRS} change, which is important to explore. We start with the moment properties when $\text{Pr}(\bld{\delta}|n)$ is defined by stratified random samples.

\begin{prop} \label{prop:3}
Let, $D_1,\ldots,D_R$ represent $R$ disjoint strata such that the spatial domain $D=\cup_{r=1}^R D_r$. Let, $\text{Pr}(\bld{\delta}|n)$ be defined by stratified random samples with $R$ strata. Then the expectation, variance and covariance of the data $Y(\bld{s}_i)$ for $i=1,\ldots,N$ given the real-valued parameter vector $\bld{\theta}=(\bld{\beta},\tau^2,\sigma^2,\sigma_{\beta}^2,\phi)'$ are given by

\noindent When $\bld{s}_i \in D_r$ for $r=1,\ldots,R$,
\begin{align*}
    \mathbb{E}[Y(\bld{s}_i)|\bld{\theta}] &= p_r\{\bld{x}(\bld{s}_i)'\bld{\beta}\}+(1-p_r)\tilde{\mu}(\bld{s}_i)=p_r\{\bld{x}(\bld{s}_i)'\bld{\beta}-\tilde{\mu}(\bld{s}_i)\}+\tilde{\mu}(\bld{s}_i),\\
    Var(Y(\bld{s}_i)|\bld{\theta}) &= p_r(\tau^2+\sigma^2)+(1-p_r)\tilde{\sigma}^2+p_r(1-p_r)\{\bld{x}(\bld{s}_i)'\bld{\beta}-\tilde{\mu}(\bld{s}_i)\}^2\\
    &=p_r(\tau^2+\sigma^2-\tilde{\sigma}^2)+\tilde{\sigma}^2+p_r(1-p_r)\{\bld{x}(\bld{s}_i)'\bld{\beta}-\tilde{\mu}(\bld{s}_i)\}^2;
\end{align*}
When $\bld{s}_i,\bld{s}_j \in D_r$ for $r=1,\ldots,R$,
\begin{align*}
    Cov(Y(\bld{s}_i),Y(\bld{s}_j)|\bld{\theta}) &= a_r\tilde{C}(\bld{s}_i,\bld{s}_j)+(b_r+p_r^2)\sigma^2h_{ij}(\phi)\\
    & +b_r\{\bld{x}(\bld{s}_i)'\bld{\beta}-\tilde{\mu}(\bld{s}_i)\}\{\bld{x}(\bld{s}_j)'\bld{\beta}-\tilde{\mu}(\bld{s}_j)\};
\end{align*}
When $\bld{s}_i \in D_r,\bld{s}_j \in D_t$ and $r\neq t$ for $r,t \in \{1,\ldots,R\}$,
\begin{equation*}
    Cov(Y(\bld{s}_i),Y(\bld{s}_j)|\bld{\theta})=(p_rp_t-p_r-p_t+1)\tilde{C}(\bld{s}_i,\bld{s}_j)+p_rp_t\sigma^2h_{ij}(\phi),
\end{equation*}
where $p_r=\frac{n_r}{N_r},p_t=\frac{n_t}{N_t},a_r=\left(\frac{n_r}{N_r}\frac{n_r-1}{N_r-1}-2\frac{n_r}{N_r}+1\right),b_r=\frac{n_r}{N_r}\left(\frac{n_r-1}{N_r-1}-\frac{n_r}{N_r}\right)$, $n_r$ is the number of subsamples drawn from stratum $r$ with size $N_r$, $n_t$ is the number of subsamples drawn from stratum $t$ with size $N_t$, and $\tilde{\mu}(\bld{s}_i),\tilde{\sigma}^2,\tilde{C}(\bld{s}_i,\bld{s}_j)$ are defined in Proposition \ref{prop:1}.
\end{prop}

\noindent \textit{Proof:} See Appendix \ref{app:a2}.

Notice that, the expectation of $Y(\bld{s}_i)|\bld{\theta}$ is a weighted sum of the parametric mean and the true mean (shown in the first expression of $\mathbb{E}[Y(\bld{s}_i)|\bld{\theta}]$) like it was for SRS without replacement. Also, the variance of $Y(\bld{s}_i)|\bld{\theta}$ is a weighted sum of the parametric variance and the true variance with an additional component as shown in the first expression of $Var(Y(\bld{s}_i)|\bld{\theta})$ (similar to SRS without replacement). On the other hand, the covariance between $Y(\bld{s}_i)$ and $Y(\bld{s}_j)$ depends on whether $\bld{s}_i$ and $\bld{s}_j$ are in the same stratum or not. In both the cases, the covariance is a weighted sum of the parametric covariance and the true covariance (plus a weighted cross-product between the difference between the parametric and true means, when $\bld{s}_i$ and $\bld{s}_j$ are in same stratum). Furthermore, we arrive at the same equations as \eqref{eq:prop1exp}, \eqref{eq:prop1var} and \eqref{eq:prop1cov} from Proposition \ref{prop:3} under the conditions $n_r=n_t=0$, $n_r=N_r$ and $n_t=N_t$. This implies that even if the sampling design is different, the properties of the model in \eqref{eq:BHM} are fully non-parametric when we do not subsample and fully parametric when we subsample all observations.

It is also noticeable that $p_r$ (and $p_t$) is dependent on the number of subsamples drawn from stratum $r$ (and $t$) and size of the corresponding stratum. If the sizes of all strata are same, and equal number of subsamples are drawn from each stratum, then the probability of sampling one data point using SRS without replacement is same as the probability of sampling one data point using stratified random sampling, since $p=n/N=\frac{n/R}{N/R}=n_r/N_r=p_r$ for $r=1,\ldots,R$ (same relation can be found for $p_t$). So, $p_r$ (or $p_t$) is different than $p$ when the sizes of strata are different or unequal number of subsamples are drawn from each stratum. Thus, the expectation and variance of $Y(\bld{s}_i)|\bld{\theta}$ in Propositions \ref{prop:1} and \ref{prop:3} are the same when equal number of subsamples are drawn from equal sized strata. However, the covariances between $Y(\bld{s}_i)$ and $Y(\bld{s}_j)$ given $\bld{\theta}$ under two sampling techniques are unequal.

\begin{prop}[Spatial Property] \label{prop:4}
Under stratified random sampling, if the true de-trended data process model is weakly and/or intrinsically stationary, then the de-trended data model defined in \eqref{eq:detrended} is respectively weakly and/or intrinsically stationary. Similarly, if the true de-trended data process model is non-stationary, then the de-trended data model defined in \eqref{eq:detrended} is also non-stationary.
\end{prop}

\noindent \textit{Proof:} See Appendix \ref{app:a2}.

Proposition \ref{prop:4} also implies that the de-trended data model defined in \eqref{eq:detrended} always retains the stationary properties of the true data process model. Under stratified random sampling, the covariance between $Z(\bld{s}_i)$ and $Z(\bld{s}_j)$ given $\bld{\theta}$ becomes
\begin{equation} \label{eq:prop4cov}
    Cov(Z(\bld{s}_i),Z(\bld{s}_j)|\bld{\theta}) = \begin{cases}
        a_r\tilde{C}(\bld{s}_i,\bld{s}_j) + (b_r+p_r^2)\sigma^2h_{ij}(\phi) & \text{when } \bld{s}_i,\bld{s}_j \in D_r \\
        a_{rt}\tilde{C}(\bld{s}_i,\bld{s}_j) + p_rp_t\sigma^2h_{ij}(\phi) & \text{when } \bld{s}_i \in D_r, \bld{s}_j \in D_t,r \neq t,
    \end{cases}
\end{equation}
where $a_{rt}=p_rp_t-p_r-p_t+1$, $\tilde{C}(\bld{s}_i,\bld{s}_j)$ is the covariance of the true de-trended data process model, and $p_r,a_r,b_r,p_t$ are defined in Proposition \ref{prop:3}. Equation \eqref{eq:prop4cov} (see Appendix \ref{app:a2} for the proof) shows that under stratified random sampling, the covariance between $Z(\bld{s}_i)$ and $Z(\bld{s}_j)$ given $\bld{\theta}$ is again a weighted sum of the true and parametric covariances of the de-trended data, but the weights change based on $\bld{s}_i$ and $\bld{s}_j$ are in same stratum or not. If the true data process model is assumed to be intrinsically stationary, then we get two different variograms depending on whether $\bld{s}_i$ and $\bld{s}_j$ are in the same stratum or not. When $\bld{s}_i,\bld{s}_j \in D_r$ for $r=1,\ldots,R$, we obtain the variogram as
\begin{equation} \label{eq:variostrat1}
    2\gamma(\bld{s}_i-\bld{s}_j)=2\{p_r(\tau^2+\sigma^2)+(1-p_r)\tilde{\sigma}^2\}-2\left\{a_r\tilde{C}(\bld{s}_i,\bld{s}_j)+(b_r+p_r^2)\sigma^2h_{ij}(\phi)\right\},
\end{equation}
where $p_r,a_r$ and $b_r$ are defined in Proposition \ref{prop:3}. When $\bld{s}_i \in D_r,\bld{s}_j \in D_t$ and $r \neq t$ for $r,t \in \{1,\ldots,R\}$, the variogram becomes
\begin{equation} \label{eq:variostrat2}
    2\gamma(\bld{s}_i-\bld{s}_j)=(p_r+p_t)(\tau^2+\sigma^2)+(2-p_r-p_t)\tilde{\sigma}^2-2\left\{a_{rt}\tilde{C}(\bld{s}_i,\bld{s}_j)+p_rp_t\sigma^2h_{ij}(\phi)\right\},
\end{equation}
where $a_{rt}=p_rp_t-p_r-p_t+1$. The proof of Equations \eqref{eq:variostrat1} and \eqref{eq:variostrat2} are shown in Appendix \ref{app:a2}. If the true de-trended data process model is assumed to be weakly stationary, i.e. if the true covariance $\tilde{C}(\bld{s}_i,\bld{s}_j)$ depends on the lag between $\bld{s}_i$ and $\bld{s}_j$, then we have the usual definitions of the sill, nugget and range,
\begin{equation} \label{eq:spatprop1}
    \begin{split}
        &\text{Sill}=\begin{cases}
        p_r(\tau^2+\sigma^2)+(1-p_r)\tilde{\sigma}^2 & \bld{s}_i,\bld{s}_j \in D_r\\
        \frac{p_r+p_t}{2}(\tau^2+\sigma^2)+\left(1-\frac{p_r+p_t}{2}\right)\tilde{\sigma}^2 & \bld{s}_i \in D_r,\bld{s}_j \in D_t,r\neq t,
        \end{cases}\\
        &\text{Nugget}=\begin{cases}
        \left[p_r\tau^2+(p_r-b_r-p_r^2)\sigma^2\right]+\left[(1-p_r-a_r)\tilde{\sigma}^2+a_r\tilde{\tau}^2\right] & \bld{s}_i,\bld{s}_j \in D_r\\
        \left[\frac{p_r+p_t}{2}\tau^2+\frac{p_r+p_t-2p_rp_t}{2}\sigma^2\right]+\left[\frac{p_r+p_t-2p_rp_t}{2}\tilde{\sigma}^2+a_{rt}\tilde{\tau}^2\right] & \bld{s}_i \in D_r,\bld{s}_j \in D_t,\\
        & \hfill r \neq t,
        \end{cases}\\
        &\text{Range}=\begin{cases}
            \inf\left\{\bld{d}:a_r\tilde{C}(\bld{d})+(b_r+p_r^2)\sigma^2\rho_c(\bld{d};\phi)=0\right\} & \bld{s}_i,\bld{s}_j \in D_r\\
            \inf\left\{\bld{d}:a_{rt}\tilde{C}(\bld{d})+p_rp_t\sigma^2\rho_c(\bld{d};\phi)=0\right\} & \bld{s}_i \in D_r,\bld{s}_j \in D_t,r\neq t,
        \end{cases}
    \end{split}
\end{equation}
where $\bld{d}=\bld{s}_i-\bld{s}_j$ is the spatial lag and $\tilde{\tau}^2$ is the nugget of the true de-trended data process. This is to emphasize that the spatial characteristics change based on whether $\bld{s}_i$ and $\bld{s}_j$ belong to same stratum or not (see Appendix \ref{app:a2} for the proof of sill, nugget and range). Equation \eqref{eq:spatprop1} reproduces the sill, nugget, and range of the (true) parametric model when ($n_r=0$ and $n_t=0$) $n_r=N_r$ and $n_t=N_t$. Thus, we again achieve a type of balance between the true model and parametric model.

Propositions \ref{prop:2} and \ref{prop:4} show that the SDSM assumes the data is stationary when $w$ implies a stationary process, and the SDSM assumes the data is nonstationary when $w$ implies a nonstationary process. This consequence is particularly important for motivating the use of the SDSM. There are times when one would just prefer to implement a standard model. This is because the properties of standard models (e.g., weak stationarity, sill, nugget, and range) are well-known and easier to communicate to the general public and collaborators in different fields. However, it is difficult to apply these models in high-dimensional settings because of computational problems and the assumptions on weak stationarity may be suspect in high-dimensions. Our proposed method allows one to fit a traditional Kriging model (to each subsample) in a way that respects the stationarity properties of the entire spatial dataset (via our propositions), without throwing away data, and in a way that is fully scalable.

One important point to mention is that we can not use these propositions in practice. This is because $w$ is the true nonparametric density associated with the measure that generated the realization $\{Y(\bld{s}_i)\}$ and is generally unknown (a common assumption in nonparametric statistics, see, \citealp{hollander2013nonparametric}). The main purpose of these propositions is that it aids with interpretation (i.e., interpretation of model assumptions), which subsequently, motivates the use of our SDSM model. Specifically, we are assuming the subsample (i.e., all $Y(\bld{s}_i)$ such that $\delta(\bld{s}_i) = 1$) follows a parametric model, and the holdout data (i.e., all $Y(\bld{s}_i)$ such that $\delta(\bld{s}_i) = 0$) is assumed to be distributed according to the actual data generating mechanism that produced it (assuming it has a density), which is unknown, and this is true for every split of the data into training and holdout. Moreover, we can strike a balance between these models through our choice of $n$ (i.e., when $n = N$ we obtain a parametric model, and with $n = 0$ we obtain a nonparametric model).

One can plot the sill, nugget, and range associated with the covariance function given by $cov(Y(\bld{s}_i),Y(\bld{s}_j)\vert \bld{\theta},\delta(\bld{s}_i) = 1, \delta(\bld{s}_j) = 1) = \sigma^{2}h_{ij}(\phi)$, where $h_{ij}(\phi)$ is the covariogram evaluated at $\bld{s}_{i} - \bld{s}_{j}$. Note that $cov(Y(\bld{s}_i),Y(\bld{s}_j)\vert \bld{\theta},\delta(\bld{s}_i) = 1, \delta(\bld{s}_j) = 1)$ is different than $cov(Y(\bld{s}_i),Y(\bld{s}_j)\vert \bld{\theta})$. One cannot plot the sill, nugget, and range associated with the covariance function $cov(Y(\bld{s}_i),Y(\bld{s}_j)\vert \bld{\theta})$, which contains the unknown $w$ in its expression. The fact that $cov(Y(\bld{s}_i),Y(\bld{s}_j)\vert \bld{\theta},\delta(\bld{s}_i) = 1, \delta(\bld{s}_j) = 1)$ is different from $cov(Y(\bld{s}_i),Y(\bld{s}_j)\vert \bld{\theta})$, restricts our interpretation of the sill, nugget, and range. Traditionally, these parameters describe a covariance function at any two locations in the spatial domain. For example, the traditional interpretation of the sill and nugget describes the covariance function of the response at any two locations in the spatial domain as the spatial lag goes to infinity and zero, respectively. In our model, these interpretations hold only for the locations in the training set (i.e., $\delta(\bld{s}_i) = \delta(\bld{s}_j) = 1$). That is, the sill and nugget in our model describe the covariance function of the response at any two locations with $\delta(\bld{s}_i)=\delta(\bld{s}_j) = 1$ as the spatial lag goes to infinity and zero, respectively.

\begin{algorithm}
Define $n \ll N$, $G$, $g_0$ (burn-in), and a set of prediction locations $A \subset D$.\\
Initialize $\bld{\theta}^{[0]}=\left(\bld{\beta}^{[0]},{\tau^2}^{[0]},{\sigma^2}^{[0]},{\sigma_{\beta}^2}^{[0]},{\phi}^{[0]}\right)'$.\\
\For{$g=1:G$}{
Sample $\bld{\delta}^{[g]} \sim \text{Pr}(\bld{\delta}|n)$.\\
Set the $n$-dimensional vector $\bld{y}_{\delta}^{[g]}=\left(Y(\bld{s}_i):\delta(\bld{s}_i)^{[g]}=1\right)'$.\\
Set the $n\times p$ matrix $\bld{X}_{\delta}^{[g]}=\left(\bld{x}(\bld{s}_i):\delta(\bld{s}_i)^{[g]}=1\right)'$.\\
Calculate the $n\times n$ matrix $\bld{H}_{\delta}\left(\phi^{[g-1]}\right)=\left(h_{jl}\left({\phi}^{[g-1]}\right):j,l\in \{i:\delta(\bld{s}_i)^{[g]}=1\}\right)$.\\
Sample ${\bld{\nu}_{\delta}}^{[g]}=\left(\nu(\bld{s}_i)^{[g]}:\delta(\bld{s}_i)^{[g]}=1\right)'$ from
\begin{equation*}
    \begin{split}
        &\mathcal{N}\Biggl\{\left(\bld{I}_n+\frac{{\tau^2}^{[g-1]}}{{\sigma^2}^{[g-1]}}\bld{H}_{\delta}^{-1}\left(\phi^{[g-1]}\right)\right)^{-1}\left(\bld{y}_{\delta}^{[g]}-\bld{X}_{\delta}^{[g]}\bld{\beta}^{[g-1]}\right),\\
        &\qquad \qquad \qquad \qquad \left(\frac{1}{{\tau^2}^{[g-1]}}\bld{I}_n+\frac{1}{{\sigma^2}^{[g-1]}}\bld{H}_{\delta}^{-1}\left(\phi^{[g-1]}\right)\right)^{-1}\Biggr\}.
    \end{split}
\end{equation*}
\\
Sample $\bld{\beta}^{[g]}$ from
\begin{equation*}
    \begin{split}
        &\mathcal{N}\Biggl\{\left({\bld{X}_{\delta}^{[g]}}'\bld{X}_{\delta}^{[g]}+\frac{{\tau^2}^{[g-1]}}{{\sigma_{\beta}^2}^{[g-1]}}\bld{I}_p\right)^{-1}{\bld{X}_{\delta}^{[g]}}'\left(\bld{y}_{\delta}^{[g]}-{\bld{\nu}_{\delta}}^{[g]}\right),\\
        &\qquad \qquad \qquad \qquad  \left(\frac{1}{{\tau^2}^{[g-1]}}{\bld{X}_{\delta}^{[g]}}'\bld{X}_{\delta}^{[g]}+\frac{1}{{\sigma_{\beta}^2}^{[g-1]}}\bld{I}_p\right)^{-1}\Biggr\}.
    \end{split}
\end{equation*}
\\
Sample ${\tau^2}^{[g]}$ from $\mathcal{IG}\left\{a_{\tau}+\frac{n}{2},b_{\tau}+\frac{1}{2}\left(\bld{y}_{\delta}^{[g]}-\bld{X}_{\delta}^{[g]}{\bld{\beta}}^{[g]}-{\bld{\nu}_{\delta}}^{[g]}\right)'\left(\bld{y}_{\delta}^{[g]}-\bld{X}_{\delta}^{[g]}{\bld{\beta}}^{[g]}-{\bld{\nu}_{\delta}}^{[g]}\right)\right\}$.\\
Sample ${\sigma^2}^{[g]}$ from $\mathcal{IG}\left\{a_{\sigma}+\frac{n}{2},b_{\sigma}+\frac{1}{2}{{\bld{\nu}_{\delta}}^{[g]}}'\bld{H}_{\delta}^{-1}\left(\phi^{[g-1]}\right){\bld{\nu}_{\delta}}^{[g]}\right\}$.\\
Sample ${\sigma_{\beta}^2}^{[g]}$ from $\mathcal{IG}\left\{a_{\beta}+\frac{p}{2},b_{\beta}+\frac{1}{2}{\bld{\beta}^{[g]}}'\bld{\beta}^{[g]}\right\}$.\\
Sample $\phi^{[g]}$ from its full conditional distribution, where $\pi(\phi)$ is discrete uniform.
}
\SetKwProg{Fn}{Prediction Steps}{:}{end}
\Fn{}{
\For{$g=g_{0}:G$}{
Calculate $\bld{H}_m\left(\phi^{[g-1]}\right)=\left(h_{jl}\left({\phi}^{[g-1]}\right):j\in A,l\in \left\{i:\delta(\bld{s}_i)^{[g]}=1\right\}\right)$.\\
Calculate $\bld{H}_A\left(\phi^{[g-1]}\right)=\left(h_{jl}\left({\phi}^{[g-1]}\right):j,l\in A\right)$.\\
\SetKwProg{Gn}{Parallel computation at each $i \in A$}{:}{end}
\Gn{}{
Set $\bld{h}_{mi\cdot}'$ = $i$-th row of $\bld{H}_m\left(\phi^{[g-1]}\right)$.\\
Set $h_{Aii}$ = $i$-th diagonal element of $\bld{H}_A\left(\phi^{[g-1]}\right)$.\\
Sample $\nu(\bld{s}_i)^{[g]}$ from $\mathcal{N}\left\{\bld{h}_{mi\cdot}'\bld{H}_{\delta}^{-1}\left(\phi^{[g-1]}\right){\bld{\nu}_{\delta}}^{[g]},{\sigma^2}^{[g-1]}\left[h_{Aii}-\bld{h}_{mi\cdot}'\bld{H}_{\delta}^{-1}\left(\phi^{[g-1]}\right)\left(\bld{h}_{mi\cdot}'\right)'\right]\right\}$.\\
Compute $W(\bld{s}_i)^{[g]}=\bld{x}(\bld{s}_i)'{\bld{\beta}}^{[g]}+\nu(\bld{s}_i)^{[g]}$.
}
}
}

\caption{Implementation of the model in \eqref{eq:BHM} for a specific $n$.}
\label{algo:gibbs}
\end{algorithm}

\subsection{MCMC Implementation} \label{subsec:mcmc}
The full conditional distributions of $\bld{\nu}_{\delta},\bld{\nu}_A,\bld{\beta},\tau^2,\sigma^2,\sigma_{\beta}^2$, and $\phi$ are derived (see Appendix \ref{app:b}) using a standard proportionality argument, and we reiterate that for any given $\bld{\delta}$ the full conditional distribution is only implemented using a single dataset of size $n\ll N$. The prior distribution for all the variance parameters ($\tau^2,\sigma^2,\sigma_{\beta}^2$) are chosen to be a non-informative inverse Gamma distribution, $\mathcal{IG}(a,b)$, among different choices \citep{gelman1995bayesian,daniels1999prior,gelman2006prior,banerjee2014hierarchical}. A standard choice for the prior of $\bld{\beta}$ is independent Normal distribution with mean 0 and variance $\sigma_{\beta}^2$ \citep{banerjee2014hierarchical}. Learning the range parameter, $\phi$, has been a challenge in spatial statistics over decades \citep{berger2001objective,paulo2005default,pilz2008we,kazianka2012objective}. \cite{berger2001objective} investigated that common choices of prior distribution for $\phi$, like a constant prior $\pi(\phi)=1$ or independent Jeffreys prior typically result in an improper posterior distribution for $\bld{\theta}=(\bld{\beta},\tau^2,\sigma^2,\sigma_{\beta}^2,\phi)'$. As a solution, they suggested that a Jeffrey rule prior or a reference prior for $\phi$ yields a proper posterior distribution. It is also often suggested to choose an informative prior for $\phi$ \citep{banerjee2014hierarchical} such as, a uniform distribution over a specified small interval, or a discrete uniform distribution over a specific finite set of points. We specify $\pi(\phi)$ as a discrete uniform distribution with a support of a finite set of points.

To sample from the joint distribution of random variables say $X$ and $Y$, one can first sample $Y$ from its density $f(Y)$ and then sample $X$ from the conditional density $f(X|Y)$. This type of sampler is well-known (e.g., see \citealp{dunn2022exploring,van2008partially}), and we call this a ``composite sampler'' borrowing terminology from the composite likelihood literature (e.g., see \citealp{varin2011overview}). In our case, we first sample from $p(\bld{\delta}|\bld{y},n) = p(\bld{\delta}|n)$ and then sample from $p(\bld{\nu},\bld{\theta}|\bld{y},\bld{\delta},n)$ using a Gibbs sampler. We call this a ``Gibbs-within-composite sampler'', since a Gibbs sampling step is used within our composite sampler. This sampling approach is needed to avoid specifying the value of the true density $w$. The density $w$ does not contain $\bld{\theta}$ and $\bld{\nu}$, and hence is a proportionality constant when deriving the full-conditionals for $\bld{\theta}$ and $\bld{\nu}$. As a result, those full-conditionals do not contain $w$ in its expression. We also assume $\bld{\delta}$ is independent of the data $\bld{y}$, where note, \cite{bradley2021approach} showed that there exists Bayesian hierarchical models that allow $\bld{\delta}$ to be independent of $\bld{y}$. Consequently $p(\bld{\delta}|\bld{y},n) = p(\bld{\delta}|n)$, which does not contain $w$ in its expression. This allows for one to adopt a Gibbs-within-composite sampler strategy to avoid specifying the true density for the holdout $\bld{y}_{-\delta}$.

\cite{bradley2021approach} suggested to use a Gibbs sampler, which is reasonable to do when the full conditional distribution for $\bld{\delta}$ has the following property, $p(\bld{\delta}|\bld{\nu},\bld{\theta},\bld{y},n)=p(\bld{\delta}|n)$; that is, if $\bld{\delta}$ is independent of $\bld{\nu}$, $\bld{\theta}$ and $\bld{y}$. When $p(\bld{\delta}|\bld{\nu},\bld{\theta},\bld{y},n)=p(\bld{\delta}|n)$, then the Gibbs-within-composite sampler becomes a Gibbs sampler. This assumption is reasonable when $n$ is sufficiently large, so that the distribution $p(\bld{\nu},\bld{\theta}|\bld{y},\bld{\delta},n)$ is roughly constant over different values of $\bld{\delta}$, which occurs when the spatial statistical model is posterior consistent. This can be seen through the following heuristic,
\begin{equation} \label{eq:heuristic}
    \begin{split}
        p(\bld{\delta}|\bld{\nu},\bld{\theta},\bld{y},n) &= \frac{p(\bld{\delta},\bld{\nu},\bld{\theta}|\bld{y},n)}{p(\bld{\nu},\bld{\theta}|\bld{y},n)} = \frac{p(\bld{\delta}|\bld{y},n)p(\bld{\nu},\bld{\theta}|\bld{\delta},\bld{y},n)}{p(\bld{\nu},\bld{\theta}|\bld{y},n)}\\
        &=\frac{p(\bld{\delta}|\bld{y},n)p(\bld{\nu},\bld{\theta}|\bld{\delta},\bld{y},n)}{\sum_{\bld{\delta^*}}p(\bld{\delta}^*|\bld{y},n)p(\bld{\nu},\bld{\theta}|\bld{\delta}^*,\bld{y},n)}\\
        &=\frac{p(\bld{\delta}|\bld{y},n)}{\sum_{\bld{\delta^*}}p(\bld{\delta^*}|\bld{y},n)\left\{p(\bld{\nu},\bld{\theta}|\bld{\delta}^*,\bld{y},n)/p(\bld{\nu},\bld{\theta}|\bld{\delta},\bld{y},n)\right\}}\\
        &\approx \frac{p(\bld{\delta}|\bld{y},n)}{\sum_{\bld{\delta^*}}p(\bld{\delta^*}|\bld{y},n)}=p(\bld{\delta}|\bld{y},n)=p(\bld{\delta}|n),
    \end{split}
\end{equation}
when $\lim_{n \to \infty} \frac{p(\bld{\nu},\bld{\theta}|\bld{\delta}^*,\bld{y},n)}{p(\bld{\nu},\bld{\theta}|\bld{\delta},\bld{y},n)}=1$ if the spatial statistical model is posterior consistent. In practice, one needs to check that $p(\bld{\nu},\bld{\theta}|\bld{\delta}^*,\bld{y},n)$ approaches $p(\bld{\nu},\bld{\theta}|\bld{\delta},\bld{y},n)$ if $n$ is sufficiently large, which can be checked empirically by comparing the replicates from the Gibbs sampler shown in Algorithm \ref{algo:gibbs} to the replicates from the Gibbs-within-composite sampler. We provide examples of this empirical investigation for both our simulations and application in Appendix \ref{app:c}. In both studies we find it reasonable to use a Gibbs sampler, but emphasize that this assumption should be checked empirically (or theoretically) in practice.

\section{Simulations} \label{sec:simulation}
In this section, we use simulation to demonstrate the predictive and computational performance of the proposed spatial data subset model in \eqref{eq:BHM} under different model specifications. The simulation setup is provided in Section \ref{subsec:simsetup}. We specify $\text{Pr}(\bld{\delta}|n)$ in a way that it generates (a) SRS without replacement, and (b) stratified random samples. We illustrate the results generated by the spatial data subset model under both sampling techniques in Section \ref{subsec:illus}. We also compare the spatial data subset model to the full model by demonstrating a simulation study on a smaller dataset in Section \ref{subsec:simstudy}. Finally, we give a detailed analysis on the scalability of our proposed model in Section \ref{subsec:simscale}.

\begin{figure}[t]
    \centering
    \includegraphics[scale=0.5]{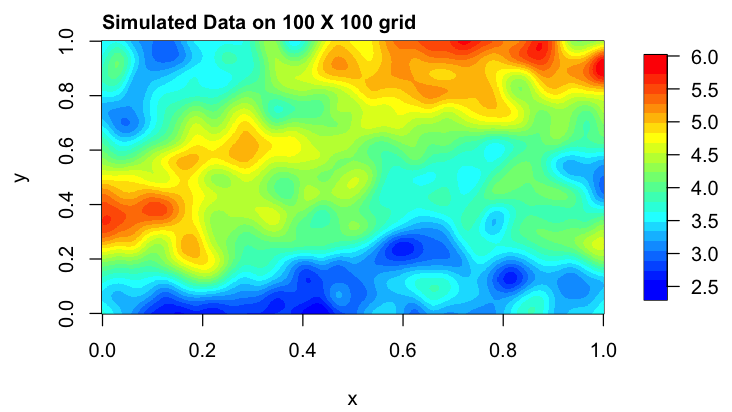}
    \caption{Image plot of the simulated spatial data on a $100\times 100$ grid over $[0,1]\times [0,1]$. This data is simulated based on the simulation setup mentioned in Section \ref{subsec:simsetup}.}
    \label{fig:data}
\end{figure}

\begin{figure}[t]
    \centering
    \includegraphics[scale=0.4]{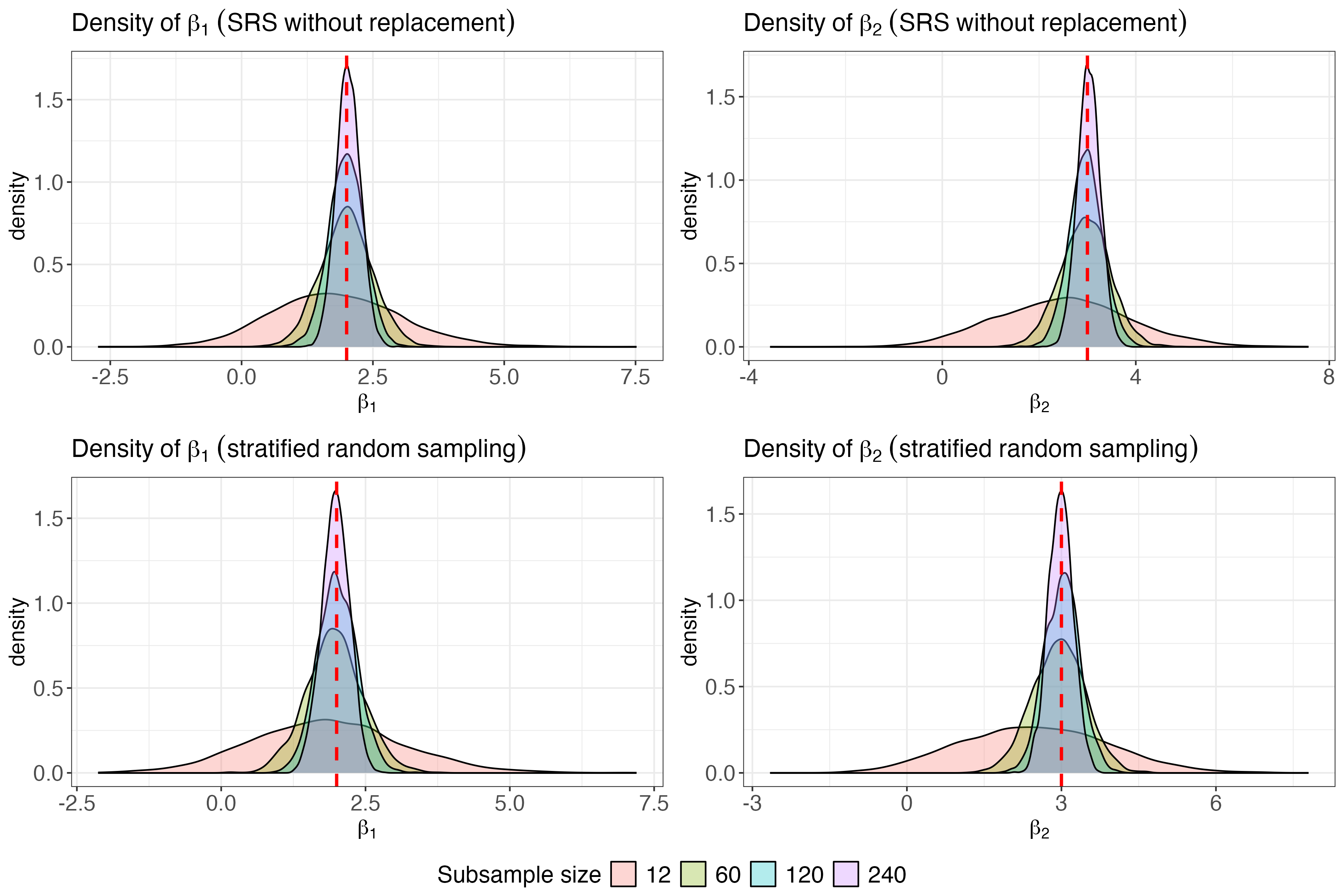}
    \caption{We plot the density of MCMC replicates for $\beta_1$ (left column) and $\beta_2$ (right column) for different subsample sizes when $\text{Pr}(\bld{\delta}|n)$ is defined by SRS without replacement (top row) and by stratified random sampling (bottom row). The red dashed lines represent the true values $\beta_1=2$ (left column) and $\beta_2=3$ (right column).}
    \label{fig:betadist}
\end{figure}

\subsection{Simulation Setup} \label{subsec:simsetup}
We fix the spatial domain $D \subset \Re^2$, where $D$ is a $100\times 100$ grid over $[0,1]\times [0,1]$. Thus, there are 10,000 spatial locations in total, all equally spaced. First, we simulate $\bld{X}$, which is a $10,000 \times 2$ dimensional matrix, from an independent $\mathcal{U}(0,1)$ distribution. Then, we simulate the process model $\bld{\nu}$, a vector of size 10,000, from a non-stationary model using \texttt{NSconvo\_sim} function from \texttt{convoSPAT} library in \texttt{R}, with true $\bld{\beta}=(2,3)'$ and exponential covariance function, and calculate $\bld{w}=\bld{X\beta}+\bld{\nu}$. Finally, we simulate the data vector $\bld{y}$, a vector of size 10,000, from the latent process and choose $\tau^2$ so that we have a signal-to-noise-ratio of 3 ($\text{SNR}=\text{Var}(\bld{w})/\text{Var}(\bld{\epsilon})$). Figure \ref{fig:data} shows the simulated spatial data $\bld{y}$ based on this simulation setup. We also enforce 20\% missing values at random, and 10\% are enforced to be missing on an arbitrary $25 \times 40$ grid (total number of missing locations = 3,000). In practice, missing data may occur over large regions similar to our $25 \times 40$ grid specification. We consider two specifications of $\text{Pr}(\bld{\delta}|n)$: (a) SRS without replacement, and (b) stratified random sampling. For the latter part, we define four strata: $\{\bld{s}_i:x\leq 0.5,y\leq 0.5\}$, $\{\bld{s}_i:x\leq 0.5,y>0.5\}$, $\{\bld{s}_i:x>0.5,y\leq 0.5\}$ and $\{\bld{s}_i:x>0.5,y>0.5\}$. The same missing data structure is used for both sampling designs.

\begin{figure}[t]
    \centering
    \includegraphics[scale=0.4]{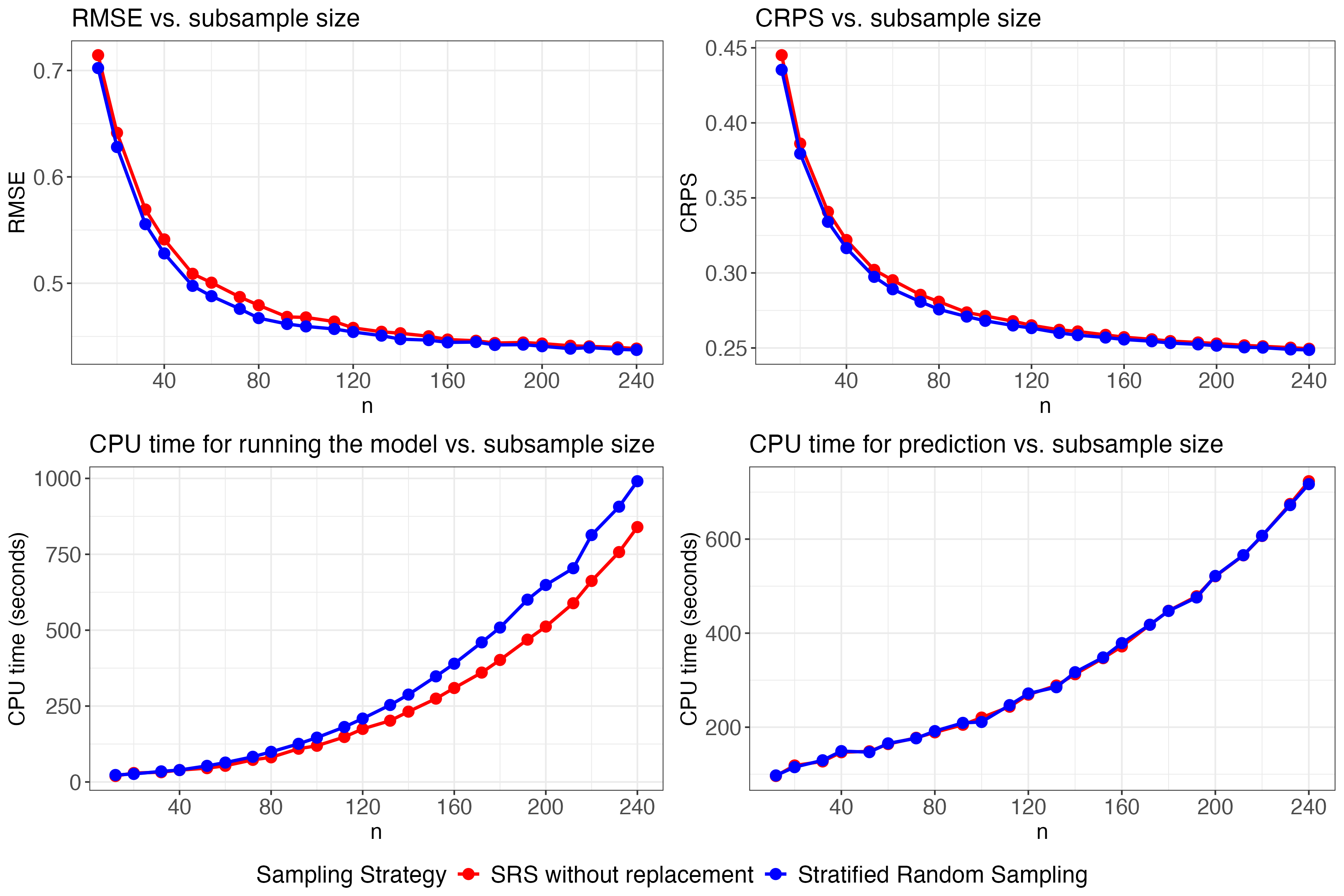}
    \caption{We plot the RMSE (top left panel), CRPS (top right panel), CPU time for running the model (bottom left panel), and CPU time for prediction (bottom right panel) at 3,000 missing locations over different subsample size, for both SRS without replacement (red) and stratified random sampling (blue).}
    \label{fig:performance}
\end{figure}

\begin{figure}[b!]
    \centering
    \includegraphics[scale=0.4]{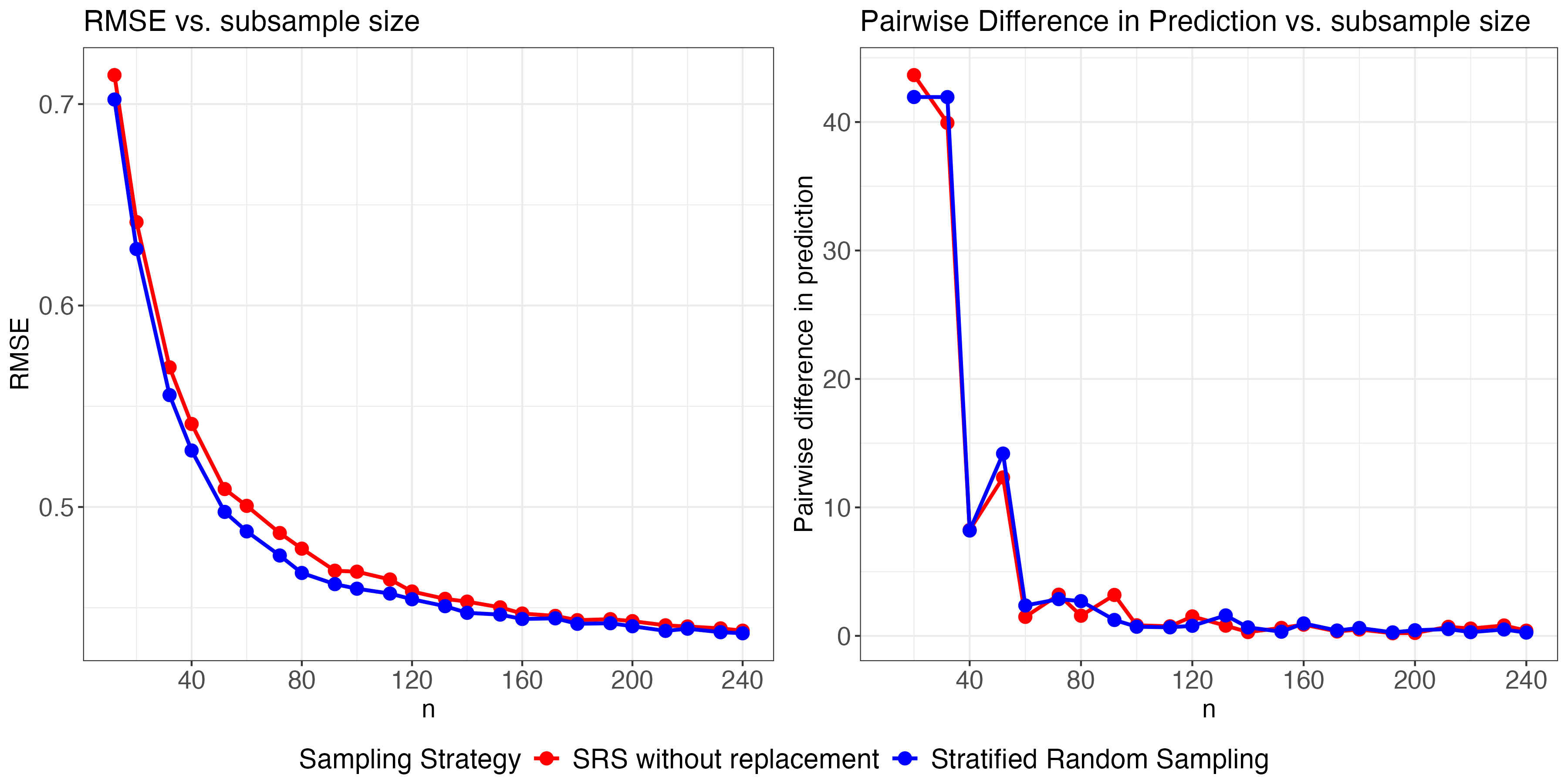}
    \caption{We plot the RMSE (left column) and pairwise difference in predictions (right column) over different subsample size under SRS without replacement (top row) and stratified random sampling (bottom row). The elbow of pairwise difference in predictions is much sharper in both sampling designs.}
    \label{fig:pairwise}
\end{figure}

\subsection{An Illustration} \label{subsec:illus}
We implement the spatial data subset model in \eqref{eq:BHM} on the simulated data with missing values defined in Section \ref{subsec:simsetup}. We denote the locations of these missing values as $A\subset D$, and we aim to make prediction at these locations. The covariogram for the model in \eqref{eq:BHM} is chosen to be exponential, i.e. $h_{ij}(\phi)=\exp(-\phi\|\bld{s}_i-\bld{s}_j\|)$. We use standard conjugate prior specifications (see Appendix \ref{app:b} for more details). The model in \eqref{eq:BHM} is implemented in the Gibbs sampler using Algorithm \ref{algo:gibbs} (since stationarity reaches irrespective of using a Gibbs-within-composite sampler; see Appendix \ref{app:c} for a detailed description) with $G=10,000$ iterations and a burn-in of $g_0=2,000$. Finally, we predict the response at each location in $A$ using Algorithm \ref{algo:gibbs}. We run the same algorithm (Algorithm \ref{algo:gibbs}) when $\text{Pr}(\bld{\delta}|n)$ is defined by (a) SRS without replacement, and (b) stratified random samples. We use the author's laptop computer with the following specification to run the algorithm: 11 Gen Intel(R) Core(TM) i7-1165G7 CPU with 2.80GHz and 16 GB RAM.

To investigate the effect of subsample size on the estimates of $\bld{\beta}$, we plot the density of MCMC replicates for $\beta_1$ and $\beta_2$ for different values of $n$ under both SRS without replacement and stratified random sampling in Figure \ref{fig:betadist}. We see that the densities of MCMC replicates for both $\beta_1$ and $\beta_2$ have high variance, non-zero skewness and platykurtic peak when $n$ is small. As we increase $n$, the densities become sharper with less variability, and become more symmetric about the corresponding true values of $\beta_1$ and $\beta_2$. The densities achieve least variability when $n=240$. These properties show that the posterior samples of $\bld{\beta}$ tend to have less variability with its mean approaching to the true value as $n$ grows, which is consistent with the results in \cite{bradley2021approach}. We see similar properties within the posterior samples of $\bld{\beta}$ across the sampling strategies.

To evaluate the inferential performance of our model, we compute the root mean square error (RMSE) between the process $W(\bld{s}_i)$ and the prediction $\widehat{W}(\bld{s}_i)$ for each $i \in A$, using the following equation,
\begin{equation} \label{eq:rmse}
    \text{RMSE} = \left\{\sum_{i\in A} \left[W(\bld{s}_i)-\widehat{W}(\bld{s}_i)\right]^2\right\}^{1/2},
\end{equation}
where $W(\bld{s}_i)$ represents the true process and $\widehat{W}(\bld{s}_i)$ represents the prediction at location $\bld{s}_i$. Since, we know the true process in case of simulated data, it is possible to use \eqref{eq:rmse} to calculate the RMSE. To evaluate the computational performance of our model, we calculate the CPU time to run the model in \eqref{eq:BHM} and perform predictions. We calculate the CPU time for predictions separately because the CPU time increases linearly with the number of missing locations for same subsample size. We compare RMSE and CPU times across the sampling strategies for $\bld{\delta}$.

\begin{figure}[t]
    \centering
    \includegraphics[scale=0.35]{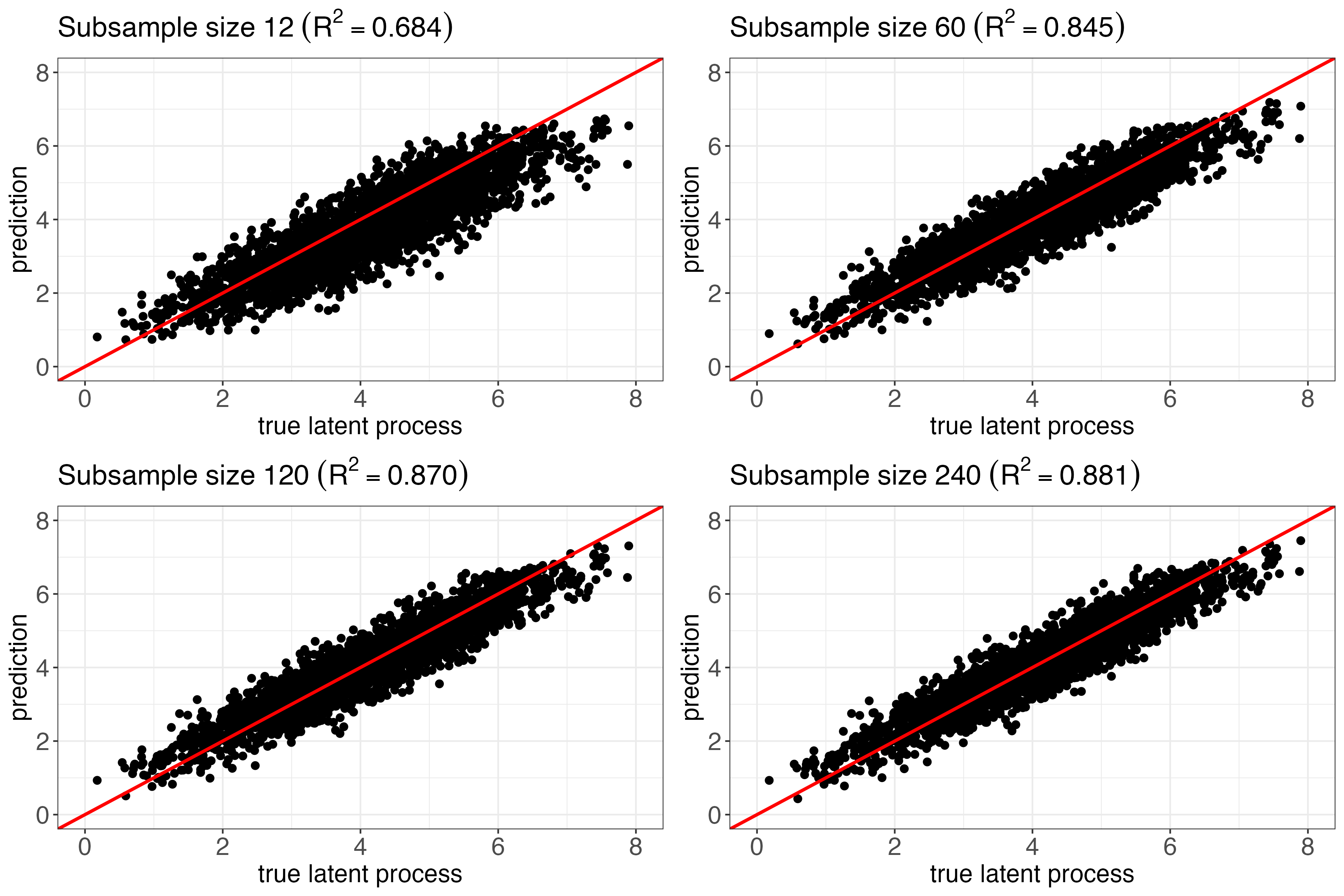}
    \caption{We plot the predictions $\{\widehat{W}(\bld{s}_i):i\in A\}$ vs. the true process $\left\{W(\bld{s}_i):i\in A\right\}$ for different choices of $n$, when $\text{Pr}(\bld{\delta}|n)$ is defined by SRS without replacement. The straight line represents the 45{\degree} reference line. The corresponding $R^2$ value is shown in the title of each plot.}
    \label{fig:pred_srs}
\end{figure}

\begin{figure}[t]
    \centering
    \includegraphics[scale=0.35]{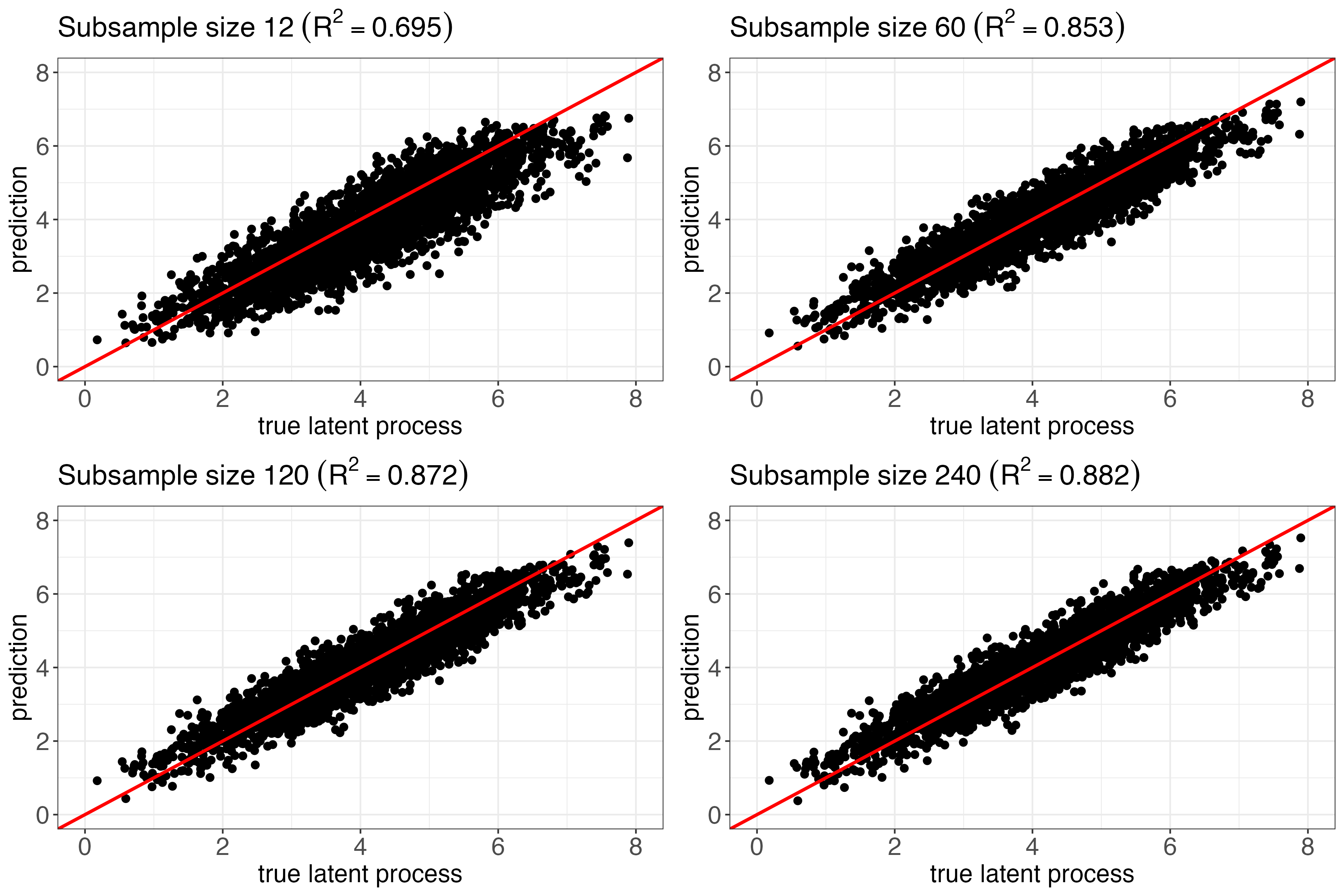}
    \caption{We plot the predictions $\{\widehat{W}(\bld{s}_i):i\in A\}$ vs. the true process $\left\{W(\bld{s}_i):i\in A\right\}$ for different choices of $n$, when $\text{Pr}(\bld{\delta}|n)$ is defined by stratified random samples. The straight line represents the 45{\degree} reference line. The corresponding $R^2$ value is shown in the title of each plot.}
    \label{fig:pred_strat}
\end{figure}

In Figure \ref{fig:performance}, we plot the RMSE, Continuous Rank Probability Score (CRPS; see \citealp{gneiting2005calibrated}, to incorporate uncertainty quantification in the metric), CPU time to run the model, and CPU time for predictions over different subsample size, when $\text{Pr}(\bld{\delta}|n)$ is defined by (a) SRS without replacement, and (b) stratified random samples. We choose $n$ to be $\{12,20,32,\ldots,220,232,240\}$ which are multiples of 4, since there are four strata in this simulation setting, and we draw equal number of samples from each stratum (in case of stratified random sampling). We keep $n$ same for SRS without replacement to make a fair performance comparison between two sampling designs. For both the sampling strategies, we see that the RMSE and CRPS decrease, and the CPU times (for both running the model and prediction) increase as the subsample size increases, and the RMSE (and CRPS) is at its smallest value when $n=240$. Hence, it can be inferred that if we increase $n$ even further, the RMSE and CRPS are expected to keep on decreasing, but at a very small rate. But at the same time, the CPU times increase with $n$. It takes roughly 13 minutes for running the model and roughly 1.5 minutes for prediction when $n=240$ for SRS without replacement. In case of stratified random sampling, it takes roughly 14 minutes for running the model and roughly 2 minutes for prediction when $n=240$. So, we see a trade-off between the model accuracy and the CPU time as $n$ changes. However, both RMSE and CRPS show an ``elbow'' pattern (in the first and second column of Figure \ref{fig:performance}), i.e. the decreasing rate is very less as subsample size increases after the elbow. Therefore, to get the best prediction, one should choose $n$ to be a value after the elbow seen in the RMSE (or in the CRPS). However, it is suggested by \cite{bradley2021approach} to simultaneously look at the pairwise difference between $\widehat{\bld{w}}_n$ and $\widehat{\bld{w}}_{n+1}$ ($\widehat{\bld{w}}_n$ is the vector of predictions for subsample size $n$) for $n=n_1,\ldots,n_B$. In Figure \ref{fig:pairwise}, we plot the pairwise difference between $\widehat{\bld{w}}_n$ and $\widehat{\bld{w}}_{n+1}$, or $(\widehat{\bld{w}}_n-\widehat{\bld{w}}_{n+1})'(\widehat{\bld{w}}_n-\widehat{\bld{w}}_{n+1})$ (in the right column) along with the RMSE (in the left column), under both sampling strategies. Notice that, the elbow of pairwise difference for stratified random sampling is much sharper than it is for SRS without replacement. Considering the plots, when $\text{Pr}(\bld{\delta}|n)$ is defined by SRS without replacement (or stratified random samples) one choice of $n$ can be around 100, for which it takes roughly 2 minutes (or 2.5 minutes) to run the model, and roughly 3.6 minutes (or 3.5 minutes) for prediction.

In Figures \ref{fig:pred_srs} and \ref{fig:pred_strat}, we plot the predictions $(\{\widehat{W}(\bld{s}_i):i\in A\})$ vs. the true process $\left(\left\{W(\bld{s}_i):i\in A\right\}\right)$ for different choices of $n$, when $\text{Pr}(\bld{\delta}|n)$ is defined by SRS without replacement and by stratified random sampling respectively. In both figures, we see that the scatter plots are mostly aligned with the 45{\degree} reference line even for the smaller values of $n$. This suggests that the model is able to predict accurately even with small subsample size. We also see an increase in $R^2$ value as $n$ increases, which does indicate that the predictions become more precise as we increase $n$. Thus in terms of prediction, we see similar properties for both sampling designs in this simulation study.

\begin{figure}[t]
    \centering
    \includegraphics[scale=0.37]{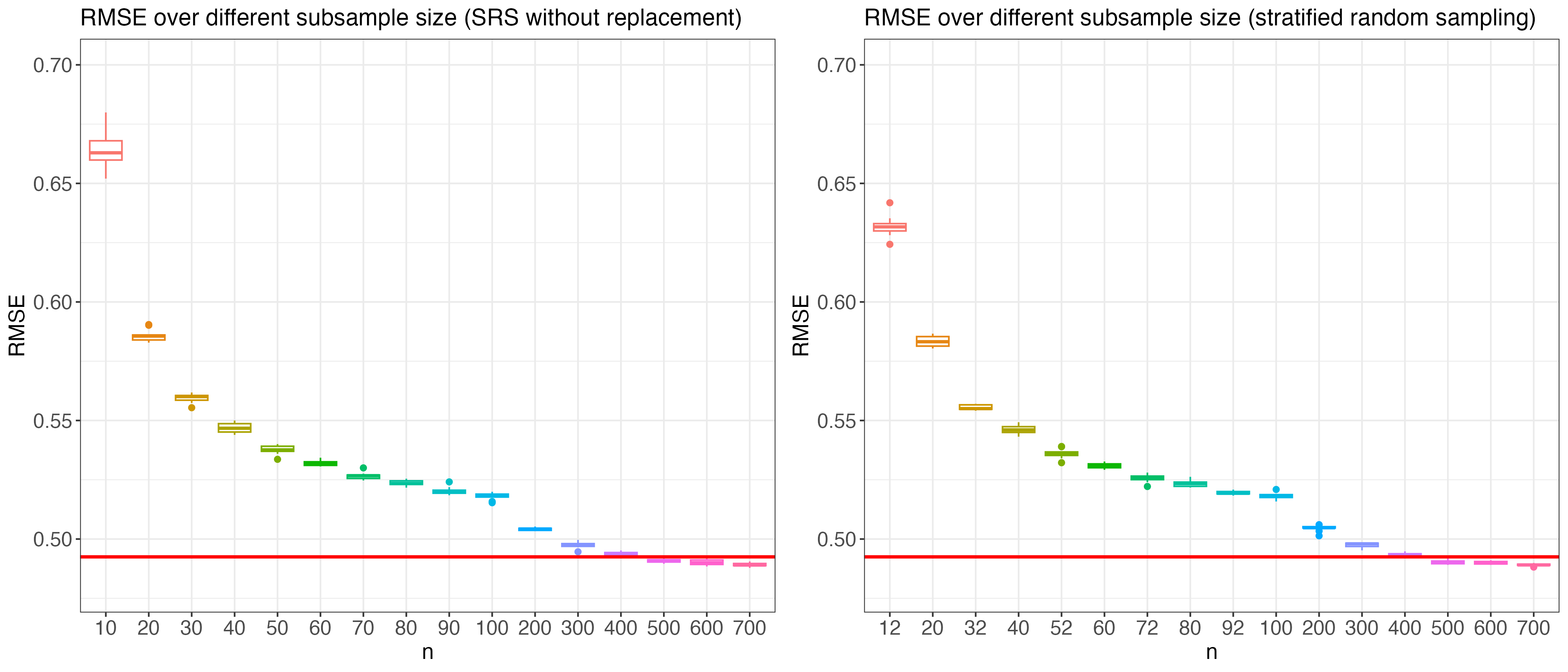}
    \caption{We plot the Boxplots of 10 replicates of RMSE that we get from the spatial data subset model for each subsample size, and the average RMSE (the red horizontal line) of 10 replicates that we get from the full model. The left and right panels show the result under SRS without replacement and under stratified random sampling respectively.}
    \label{fig:simstudy}
\end{figure}

\subsection{Simulation Study} \label{subsec:simstudy}
We now assess the performance of our proposed model over multiple simulated replicates. Since the spatial data subset model is a semi-parametric model, it is important to compare it with the full model using simulation study. By ``full model'' we mean $n=N$, which is the weakly stationary spatial model. This comparison is particularly interesting, since the full model is misspecified and our spatial data subset model approach has marginal distributions at any given subset (for the holdout) that are correctly specified. Thus, we simulate a smaller dimensional dataset of size 900 on a $30 \times 30$ grid over $[0,1] \times [0,1]$ the same way as we described in Section \ref{subsec:simsetup} so that we can implement the full model. Here, we enforce 20\% (or, 180) missing values at random locations over the grid. Hence, we generate random subsamples of sizes $10,20,\ldots,100,200,\ldots,700$ from the remaining 720 data points and predicted the response at the 180 missing locations using Algorithm \ref{algo:gibbs} with $G=5,000$ and $g_0=1,000$ under (a) SRS without replacement and (b) stratified random sampling. We also run the full model using all 720 data points and we predict the response at the missing locations using the full model. We compare both models with respect to the RMSE between the true process and the predictions at the missing locations. We generate 10 replicates of each of the mentioned results.

In Figure \ref{fig:simstudy}, we plot the Boxplots of the 10 replicates of RMSE that we get from our proposed spatial data subset model using Algorithm \ref{algo:gibbs}. The results under SRS without replacement and stratified random sampling are shown on the left and right panel respectively. Also, the red horizontal line on both panels represents the average RMSE of the 10 replicates that we get from the full model. As expected, the range of the Boxplots decreases with increase in $n$. An interesting observation from Figure \ref{fig:simstudy} is that the spatial data subset model for $n=500$, 600 and 700 performs even better than the full model which was trained with all 720 data points. This may be because the data is simulated from a non-stationary model whereas the full model is a stationary model, but our proposed model becomes non-stationary because of its semi-parametric nature.

\begin{table}[t]
    \renewcommand{\arraystretch}{1.2}
    \centering
    \begin{tabular}{@{\hspace{0.5in}}c@{\hspace{1in}}c@{\hspace{0.5in}}}
        \hline
        \hline
        \textbf{Spatial Method} & \textbf{Crash Point}\\
        \hline
        FRK & 550 K\\
        Gapfill & 8 M\\
        Lattice Kriging & 110 K\\
        LAGP & 20 M\\
        Metakriging & 100 K\\
        MRA & 1 M\\
        NNGP & 5 M\\
        Partition & 600 K\\
        Predictive Process & 3 M\\
        SPDE & 150 K\\
        Tapering & 100 K\\
        \hline
        \hline
    \end{tabular}
    \caption{Crash points (value of $N$ at which a method crash) for different spatial methods.}
    \label{table:modelcrash}
\end{table}

\subsection{Scalability of the Spatial Data Subset Model} \label{subsec:simscale}
In principle, our proposed approach can be applied to a spatial dataset of \textit{any size} that can be stored. A common target for complexity time for spatial models is $\mathcal{O}(N)$, where $N$ is the size of the dataset (see \citealp{cooley1965algorithm,mallat1989theory,cressie2008fixed} for standard references). In general, this goal of $\mathcal{O}(N)$ still can lead to computational failures, since in the era of big data the size of $N$ is continuously increasing. But, our proposed spatial data subset model (which is also a fully Bayesian model) uses $n \ll N$ data points at a time (and all $N$ for large enough $G$), which makes it a $\mathcal{O}(n^{3}G)$ spatial model. Thus, no matter what the value of $N$ is, we can always choose an $n$ scalable in the presence of any data size $N$. To investigate this, we simulate large datasets with different sizes and we fit the spatial methods for large dataset from a benchmark study in \cite{heaton2019case}. Our goal is to find the $N$ such that the methods from \cite{heaton2019case} crash the first author's laptop or stop with an error message regarding size allocation. We refer to this $N$ as the ``crash point''. For this study, we use the author's laptop computer with the same specifications mentioned in Section \ref{subsec:illus} (11 Gen Intel(R) Core(TM) i7-1165G7 CPU with 2.80GHz and 16 GB RAM).

\begin{figure}[t]
    \centering
    \includegraphics[scale=0.4]{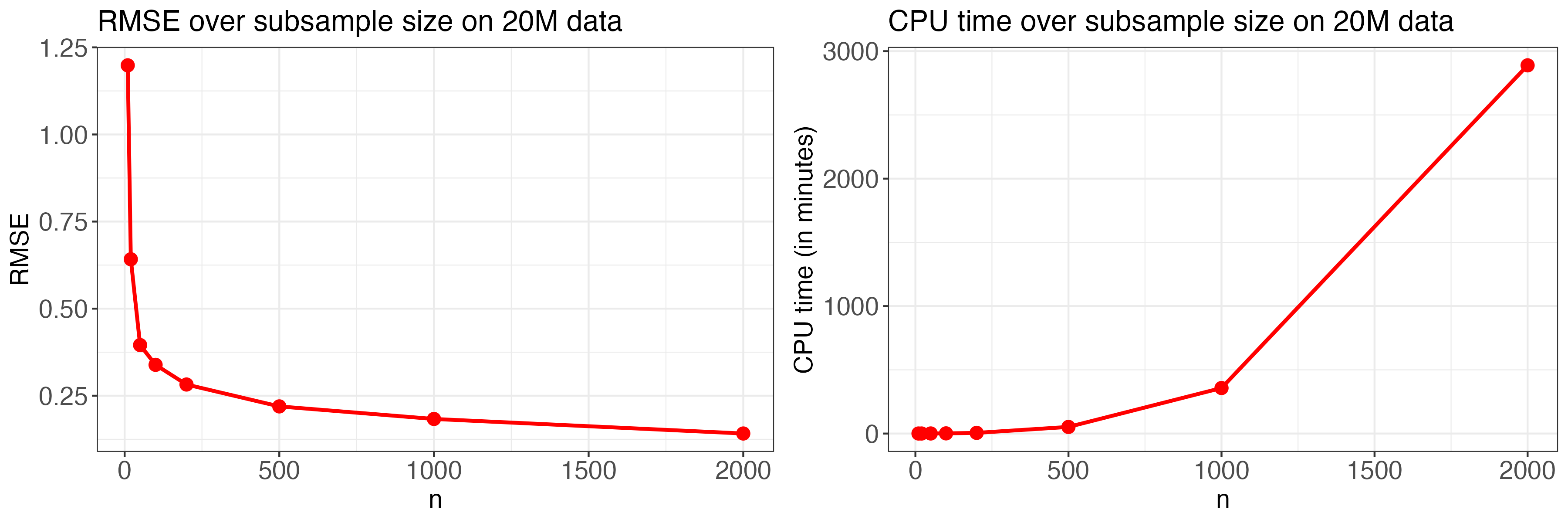}
    \caption{We plot the RMSE of predictions at 1,000 missing locations on the left panel, and the total CPU time (in minutes) on the right panel for different subsample sizes on a simulated dataset of size 20M.}
    \label{fig:performance_20M}
\end{figure}

We simulate large datasets of size $N$ using an efficient spectral simulation method (see \citealp{mejia1974synthesis,cressie1993statistics}, pg. 204; \citealp{yang2021bayesian} for more details) with the exponential covariogram with range equals to 3 and spatial domain as the unit square. An independent error component is included with an SNR equals to 3. We also enforce 1,000 missing values to perform predictions. We keep this number small to focus on the computation time on training the model. We fit the spatial methods mentioned in \cite{heaton2019case} one by one to datasets with different sizes until we reach the crash point. In Table \ref{table:modelcrash}, we list the crash points of each method. We find that the methods crash at different data sizes, and LAGP scales to the largest $N$ at 20M. However, the proposed spatial data subset model is still able to process that much data for different values of $n$. We provide the \texttt{R} code that generates the 20M dimensional spatial dataset and fits the SDSM method to it on GitHub at \url{https://github.com/sudiptosaha/Spatial-Data-Subset-Model}. The code to implement the competing methods to the 20M observations can be found at \url{https://github.com/finnlindgren/heatoncomparison} \citep{heaton2019case}. In Figure \ref{fig:performance_20M}, we plot the performance of the spatial data subset model on the same 20M data. On the left column, we plot the RMSE of the predictions as 1,000 missing locations for different subsample sizes. On the right column, we plot the total CPU time (CPU time to run the model and to make predictions) in minutes for different subsample sizes. We see that, not only our proposed model is able to fit a data of size 20M, it also takes less CPU time (at lower values of $n$) to run and make predictions, and shows compelling performance, whereas all the other methods crash for the same dataset.

We emphasize that we do not throw away any of 20M observations, and simply only consider $n \ll N$ at each step of our sampler. The crash point can be interpreted as the value of $N$ such that the corresponding model can be fully scalable, where recall, we say a method is fully scalable if it makes use of all $N$ observations without discarding them. For $N = 20\text{M}$, we have checked that 99.998\% observations are used after $G=106\text{K}$ when implementing in Algorithm \ref{algo:gibbs}. What we have observed is that our sampler for posterior replicates of $(\bld{\theta},\bld{\nu})$ reaches its stationary distribution well before $G=106\text{K}$. That is, it appears that one does not need to be fully scalable in order to appropriately search through parameter space. In practice, we suggest choosing $G$ based on standard MCMC diagnostic criteria (e.g., trace plots and Gelman-Rubin diagnostics) instead of choosing $G$ to obtain full scalability.

We reiterate that the crash points depend on the system that is used. Thus, if we run the methods on a different system, the crash points will change. But, our proposed method does not crash for any size of dataset for an $n \ll N$ provided $N$ observations can be stored. We have also considered $N = 30\text{M}$ and found that our method can still produce posterior inferences in this case (i.e., the algorithm does not crash). To our knowledge, there is no spatial model other than our proposed spatial data subset model that is fully scalable (for large enough $G$) to a spatial dataset of any size that can be stored.

\section{Application} \label{sec:application}
In this section, we present an analysis of a real dataset that consists of land surface temperatures collected by MODIS, a remote sensing instrument onboard the Terra satellite. We discuss the dataset in detail in Section \ref{subsec:dataset}, and we provide a detailed analysis of the dataset in Section \ref{subsec:analysis}.

\subsection{The Dataset} \label{subsec:dataset}
On December 18, 1999 the National Aeronautics and Space Administration (NASA) launched the Terra satellite, which is the Earth Observing System (EOS) flagship satellite orbiting at an altitude of 705 kilometers (\url{https://terra.nasa.gov/}; see \citealp{ranson2003nasa}). The Moderate Resolution Imaging Spectroradiometer (MODIS) is an important remote sensing instrument onboard the Terra satellite. It primarily captures Earth's atmospheric conditions which helps scientists in monitoring the changes in the biosphere. MODIS measures several different variables including land and sea surface temperatures. In particular, MODIS retrieves the Land Surface Temperature (LST) data daily at 1 km pixels by using the generalized split-window algorithm \citep{wan1996generalized}, and at 6 km grids by using the day/night algorithm \citep{wan1997physics}. The pair of daytime and nighttime LSTs are recorded by the day/night algorithm in seven Thermal Infrared (TIR) bands. We observe the Level-3 \citep{wan2006modis,hulley2016moderate} daytime LSTs as measured by MODIS on August 4, 2016. \cite{heaton2019case} considered this dataset as large spatial data and provided a comparative study among several spatial methods applied on this dataset. The data were originally downloaded from the MODIS reprojection tool web interface (MRTweb), and later made available at a public repository at \url{https://github.com/finnlindgren/heatoncomparison} \citep{heaton2019case}.

\begin{figure}[t]
    \centering
    \includegraphics[scale=0.43]{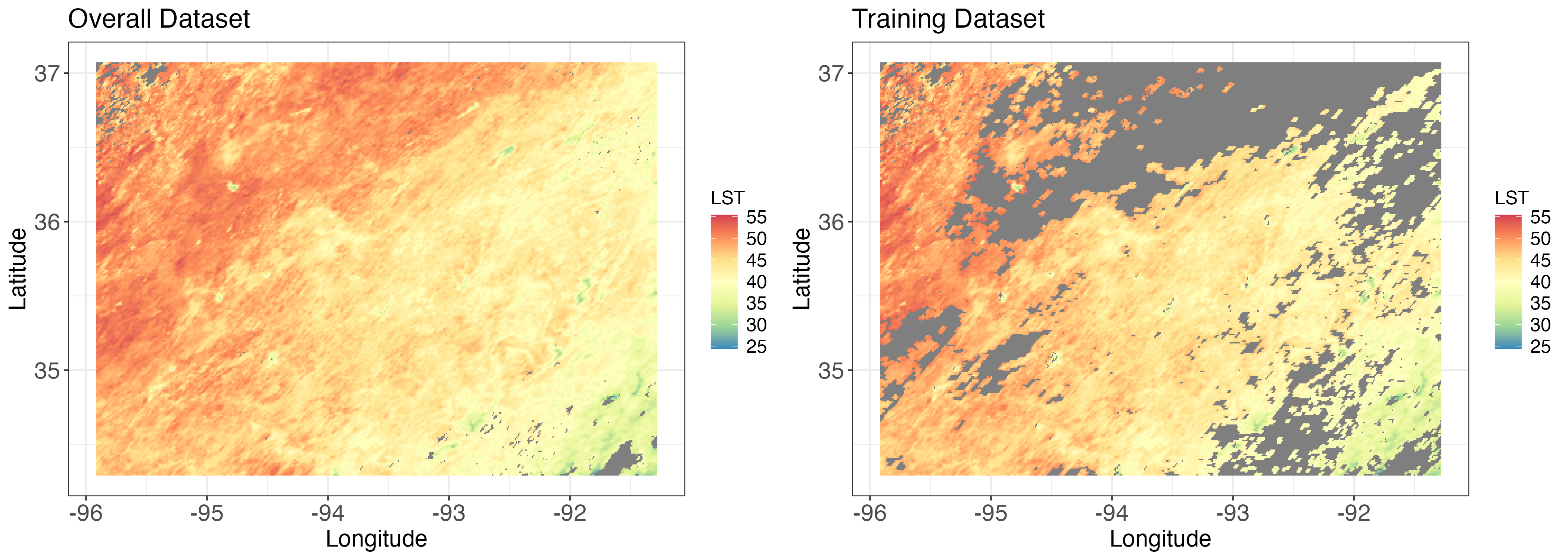}
    \caption{We plot the overall LST data on the left panel, and the training LST data on the right panel. The grey region on the left panel represents the missing data. The grey region on the right panel represents the missing data and the test data together.}
    \label{fig:satdata}
\end{figure}

This benchmark dataset is particularly useful because it allows us to compare our modification (i.e., using a data subset model) of a standard spatial model (i.e., based on covariograms) to a large list of current spatial methods including FRK \citep{cressie2008fixed}, Gapfill \citep{gerber2018predicting}, Lattice Kriging \citep{nychka2015multiresolution}, Local Approximate Gaussian Process (LAGP; \citealp{gramacy2015local}), Metakriging \citep{guhaniyogi2018meta}, Multiresolution Approximations (MRA; \citealp{jurek2021multi}), Nearest Neighbor Process (NNGP; \citealp{datta2016hierarchical}), Spatial Partitioning \citep{sang2011covariance}, Predictive Process \citep{banerjee2008gaussian}, SPDE \citep{lindgren2011explicit}, Covariance Tapering \citep{furrer2006covariance}, Periodic Embedding \citep{guinness2019spectral}, and Multi-scale Vecchia (MSV) approximation \citep{zhang2022multi}. We also compare our proposed model's performance with SoD \citep{chalupka2013framework} for subsample sizes 2,000 and 5,000. The author's laptop computer crashed when we ran SoD with $n=10,000$, which is not surprising because of kriging's natural restriction on $n$ \citep{rulliere2018nested}. Our goal is to show that we can take a familiar standard spatial model for low-dimensional stationary spatial data, and modify it with a data subset model to make it produce similar predictions as the state-of-the-art methods. This is particularly exciting, as the spatial data subset model can be scaled to effectively any dataset that can be stored, and hence, similar predictive performance on a well-known benchmark dataset would suggest that our approach would be reasonable to consider it in extremely high-dimensional settings.

The dataset consists of LST observations on a $500\times 300$ grid (150,000 observations in total) ranging longitude values from -95.91153 to -91.28381, and latitude values from 34.29519 to 37.06811. The ranges of longitude and latitude, and also the date on which the LST is measured are chosen because of the sparse cloud cover over the region on that day. We have 1.1\% missing data due to cloud cover, leaving 148,309 out of 150,000 observed values to be used. We use the same train and test dataset split as it is given in \cite{heaton2019case}. We keep 105,569 observations in the training set, and 42,740 observations in the out-of-sample test set. In Figure \ref{fig:satdata}, we plot the entire dataset on the left panel where the 1.1\% missing observations are represented by the grey region. We also plot the training dataset on the right panel where the grey region represents the missing data and the test data.

\begin{table}[t]
    \renewcommand{\arraystretch}{1.2}
    \centering
    \begin{tabular}{cccccccc}
        \hline
        \hline
        $\boldsymbol{n}$ & \textbf{MAE} & \textbf{RMSE} & \textbf{CRPS} & \textbf{INT} & \textbf{CVG} &
        \multicolumn{1}{p{2.5cm}}{\centering \textbf{CPU Time \\ to run \\ the model \\ (min.)}} & \multicolumn{1}{p{2.5cm}}{\centering \textbf{CPU Time \\ for prediction \\ (min.)}}\\
        \hline
        96 & 2.35 & 2.71 & 1.63 & 11.37 & 0.80 & 0.42 & 31.18\\
        208 & 2.25 & 2.59 & 1.62 & 14.73 & 0.72 & 1.85 & 115.75\\
        512 & 2.13 & 2.47 & 1.60 & 19.48 & 0.59 & 16.16 & 457.56\\
        1024 & 2.00 & 2.38 & 1.38 & 9.75 & 0.88 & 98.14 & 2304.19\\
        2000 & 1.93 & 2.32 & 1.35 & 10.41 & 0.83 & 468.26 & 4183.77\\
        \hline
        \hline
    \end{tabular}
    \caption{Model performance and CPU time on LST dataset over different subsample size.}
    \label{table:LST}
\end{table}

\subsection{Analysis of daytime LST data captured by MODIS} \label{subsec:analysis}
We implement Algorithm \ref{algo:gibbs} (since stationarity reaches irrespective of using a Gibbs-within-composite sampler; see Appendix \ref{app:c} for a detailed description) on the training dataset, where we set $G=2,000$, $g_0=800$, discrete uniform distribution with support $\{0.001,0.0002,\ldots,0.1\}$ for the prior of $\phi$ and $\text{Pr}(\bld{\delta}|n)$ is defined by stratified random samples with 16 equal sized strata. Notice that we have 42,740 locations in the test set (and an additional 1,691 locations for the missing data) and the CPU time increases with the large number of prediction locations. The algorithm takes comparatively longer time for prediction, which also increases as the subsample size $n$ increases. We run Algorithm \ref{algo:gibbs} for different values of $n$. For each $n$, we generate equal number of random samples ($n/16$) from each of the 16 strata, and we predict the LST values on the grey region shown on the right panel of Figure \ref{fig:satdata}. We also calculate mean absolute error (MAE), root mean squared error (RMSE), continuous rank probability score (CRPS; see \citealp{gneiting2005calibrated}), interval score (INT; see \citealp{gneiting2007strictly}) and prediction interval coverage (CVG) at $\alpha=0.05$ on the out-of-sample test dataset. We keep track of the CPU times to run/fit the model and to perform prediction separately for each $n$. In Table \ref{table:LST}, we show these metrics for different values of $n$. The values of $n$ are chosen in a way that they are multiples of 16 (number of strata) so that we can draw equal number of samples from each stratum. We see gradual decrease in MAE, RMSE and CRPS as $n$ increases, but simultaneously the CPU times for both running the model and prediction increase. It takes very less time to run/fit the model than to predict because of the large test set. We also check trace plots and credible intervals informally. We do not find any sign of lack of convergence, and the credible intervals decrease as $n$ increases.

\begin{figure}[t]
    \centering
    \includegraphics[scale=0.43]{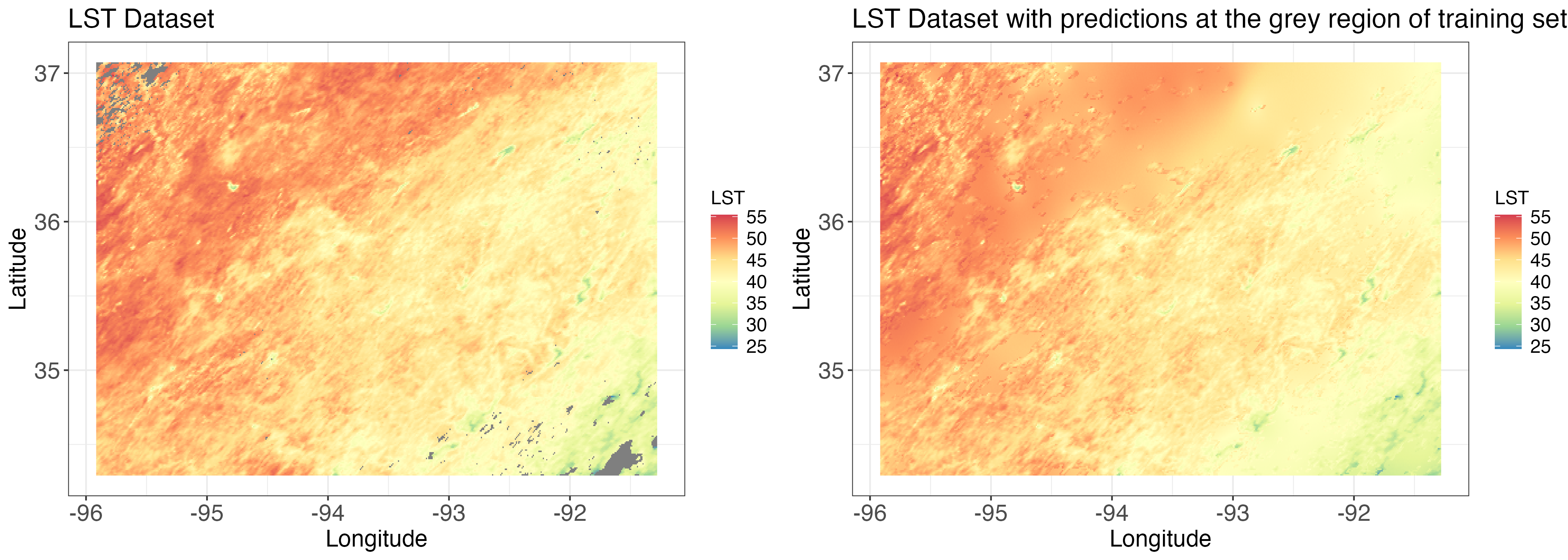}
    \caption{We plot the actual LST dataset on the left panel. We filled in the grey region of training set with the predictions, from the spatial data subset model, on those locations with $n=2,000$ and plot that on the right panel.}
    \label{fig:pred_satdata}
\end{figure}

In Figure \ref{fig:pred_satdata}, we plot the actual LST dataset on the left panel. Note that, the 1.1\% missing data are represented by the grey region in the plot. On the right panel, we filled in the grey region of the ``training dataset'' plot of Figure \ref{fig:satdata} (represents the test data and missing data together) with the predicted LSTs, from the spatial data subset model, on those locations using $n=2,000$. For a given computational goal, one can choose a different value of $n$ and that will produce a different set of predictions. We see that the model correctly identifies the spatial property of the data, i.e. the nearby locations have similar value than the locations at far.

\begin{table}[t]
    \renewcommand{\arraystretch}{1.3}
    \footnotesize
    \centering
    \begin{tabular}{p{0.1\textwidth}cccccccc}
        \hline
        \hline
        \textbf{Method} & \textbf{Reference} & \textbf{MAE} & \textbf{RMSE} & \textbf{CRPS} & \textbf{INT} & \textbf{CVG} &
        \multicolumn{1}{p{1cm}}{\centering \textbf{Run time\\(min)}} & \multicolumn{1}{p{1cm}}{\centering \textbf{Cores\\used}}\\
        \hline
        SDSM & - & 1.93 & 2.32 & 1.35 & 10.41 & 0.83 & 4652.03 & 1\\
        FRK & \tiny{\cite{cressie2008fixed}} & 1.96 & 2.44 & 1.44 & 14.08 & 0.79 & 2.32 & 1\\
        Gapfill & \tiny{\cite{gerber2018predicting}} & 1.33 & 1.86 & 1.17 & 34.78 & 0.36 & 1.39 & 40\\
        LatticeKrig & \tiny{\cite{nychka2015multiresolution}} & 1.22 & 1.68 & 0.87 & 7.55 & 0.96 & 27.92 & 1\\
        LAGP & \tiny{\cite{gramacy2015local}} & 1.65 & 2.08 & 1.17 & 10.81 & 0.83 & 2.27 & 40\\
        Metakriging & \tiny{\cite{guhaniyogi2018meta}} & 2.08 & 2.50 & 1.44 & 10.77 & 0.89 & 2888.52 & 30\\
        MRA & \tiny{\cite{jurek2021multi}} & 1.33 & 1.85 & 0.94 & 8.00 & 0.92 & 15.61 & 1\\
        NNGP & \tiny{\cite{datta2016hierarchical}} & 1.21 & 1.64 & 0.85 & 7.57 & 0.95 & 2.06 & 10\\
        Partition & \tiny{\cite{sang2011covariance}} & 1.41 & 1.80 & 1.02 & 10.49 & 0.86 & 79.98 & 55\\
        Pred. Proc. & \tiny{\cite{banerjee2008gaussian}} & 2.15 & 2.64 & 1.55 & 15.51 & 0.83 & 160.24 & 10\\
        SPDE & \tiny{\cite{lindgren2011explicit}} & 1.10 & 1.53 & 0.83 & 8.85 & 0.97 & 120.33 & 2\\
        Tapering & \tiny{\cite{furrer2006covariance}} & 1.87 & 2.45 & 1.32 & 10.31 & 0.93 & 133.26 & 1\\
        Period. Emb. & \tiny{\cite{guinness2019spectral}} & 1.29 & 1.79 & 0.91 & 7.44 & 0.93 & 9.81 & 1\\
        MSV & \tiny{\cite{zhang2022multi}} & 1.11 & 1.42 & - & 7.32 & 0.88 & - & -\\
        SoD with $n=2000$ & \tiny{\cite{chalupka2013framework}} & 50.48 & 75.37 & 32.64 & 145.8 & 0.97 & 796.95 & 1\\
        SoD with $n=5000$ & \tiny{\cite{chalupka2013framework}} & 35.36 & 57.52 & 22.90 & 105.86 & 0.88 & 4822.23 & 1\\
        \hline
        \hline
    \end{tabular}
    \caption{Comparison of spatial data subset model (when $n=2,000$) with the other methods.}
    \label{table:LSTcompare}
\end{table}

In Table \ref{table:LSTcompare}, we show a comparison of the performance of our proposed spatial data subset model (SDSM) when $n=2,000$ with all the methods listed in Section \ref{subsec:dataset} (methods described in \citealp{heaton2019case,zhang2022multi} and \citealp{chalupka2013framework}). While our method does not consistently dominate all other methods in all criteria, it is competitive. We find that in terms of RMSE, our proposed model does perform better than FRK, Metakriging, Predictive Process, Tapering and SoD with just $n=2,000$. In terms of both MAE and CRPS, our proposed model outperforms FRK, Metakriging, Predictive Process and SoD. It is also noticeable that in terms of INT, our proposed model outpeforms FRK, Gapfill, LAGP, Metakriging, Partition, Predictive Process and SoD. However, in terms of CVG, it outperforms only FRK and Gapfill, and it matches the performance of LAGP and Predictive Process. These results are particularly exciting for spatial statisticians. Our model allows one to implement a traditional spatial model (on subsamples) so that it can be implemented on high-dimensional data, respecting the spatial properties of the entire dataset, and is competitive with several state-of-the art methods used in high-dimensional setting.

The total run time for our proposed model with $n=2,000$ is 4,652 minutes. But, the run time depends on number of cores, RAM size, CPU speed and number of jobs running in the system. The author's laptop computer has different specification (mentioned in Section \ref{subsec:illus}) than what it was used for the methods in \cite{heaton2019case} and \cite{zhang2022multi}, which is why the run time is not comparable. Moreover, in Section \ref{subsec:simscale}, we show that our proposed method is fully scalable (for large enough $G$) to spatial datasets of any size that can be stored, and the CPU time stays fixed (for an $n$) as $N$ increases. There is no other method that is scalable at this extent. Depending on the system used, other methods will crash for large enough $N$, but the spatial data subset model will not. Although our proposed method does not outperform all the methods in terms of performance, it is competitive and it has a big advantage in terms of scalability.

Each method in Table \ref{table:LSTcompare} used different computational resources. Thus, one should be careful not to use the reported CPU times to rank value the methods computationally, since each method could be made faster by capitalizing on improved computational resources. For example, in Table \ref{table:LSTcompare} we provide the number of cores used, and not all methods were implemented on more than one core. Moreover, SDSM uses one core and was implemented on a different computer than the other methods. In Table \ref{table:LST}, metrics are provided for implementations of SDSM in our application as we vary the choice of $n$. In Table \ref{table:LST}, we see that SDSM is considerably faster when $n = 512$ (CPU time of roughly 32 minutes) than when $n = 2,000$ (CPU time of 4652 minutes). Previous applications of the data subset model approach in \citet{bradley2021approach} suggest that the rate of increase of SDSM's CPU time, as a function of $n$, would be smaller if more computational resources (e.g., cores) were available. Thus, the use of additional computational resources (e.g., cores) would allow one to choose larger values of $n$ while taking on a smaller computational cost.

The results from Section \ref{subsec:simscale} are particularly important for evaluating SDSM relative to the state-of-the-art methods. In particular, when $N$ is so large that it is beyond the largest crash-point of the state-of-the-art methods, SDSM is the \textit{only} approach that is possible to compute that does not throw away data. This is because SDSM is fully scalable in the limit. Moreover, the method that throws away data, i.e. SoD, performed considerably worse than SDSM in terms of inferential metrics in Table \ref{table:LSTcompare}.

There are times when one would just prefer to implement a standard spatial statistical model. This is because the properties of standard models (e.g., weak stationarity, sill, nugget, and range) are well-known and easy to communicate to the general public and collaborators in different fields. Bayesian Kriging is the traditional spatial model that we focus on in this article. However, in the big $N$ case, it is unclear if the standard model assumptions are appropriate for such high-dimensional spatial data with spatial domains with complicated terrains, and would ideally like to use all $N$ observations. Propositions \ref{prop:1} -- \ref{prop:4}, show that when the data is generated from a model with non-stationary covariances the SDSM is non-stationary, and when the data is generated from a model with stationary covariances the SDSM assumes a stationary covariance. Thus, the assumptions on the spatial properties of this SDSM model are well aligned with the data, we obtain inferences on parameters that arise in traditional spatial models, and use all $N$ observations in a fully Bayesian context. However, it is unclear how well this SDSM performs in practice. In this section, we show that the SDSM performs similar (i.e., does not produce the worst value of a metric, but not the best) to several of the state-of-the art methods. Thus, one is able to obtain reasonable inferential results (i.e., the metrics considered are not always the best, but are not always the worst) when assume a traditional model for subsamples of the data (scaling it to large $N$), without necessarily throwing out data, and while still respecting the true spatial properties of the data.

\section{Discussion} \label{sec:discussion}
In this article, we propose a new spatial data subset model (SDSM) that allows one to implement the classical Bayesian spatial model on big data in a computationally feasible way. We do this by redefining the full likelihood of the data model using subsamples without imposing any additional restrictive assumptions. Also, we insert more levels to the Bayesian Hierarchical Model to introduce subsampling inside the model. We see that the subsampling changes the spatial properties of the dataset that is being used in the model. These properties are also dependent on the subsampling strategy. Specifically, we provide theoretical results which show that if the true spatial model is stationary (or non-stationary) then the spatial data subset model in \eqref{eq:BHM} is also stationary (or non-stationary) under both SRS without replacement and stratified random sampling. We also provide moment results under SRS without replacement and stratified random sampling, and compare and contrast the models. Code to implement the SDSM is made available at \url{https://github.com/sudiptosaha/Spatial-Data-Subset-Model}.

In our simulation study, we have found that the prediction errors decrease as $n$ (subsample size) increases, but the CPU time increases simultaneously. This suggests that one should choose a value of $n$ in a way that it achieves the computational goal (for example, producing inference in 2 minutes) while getting reasonable predictions. This performance is particularly exciting since the CPU time does not scale with $N$ (total number of observations), but rather $n$, which we can choose. The decrease in RMSE have shown an elbow pattern which is why we have plotted the pairwise difference in prediction as the RMSE decreases gradually. This helps us to choose $n$ at which the prediction error starts decreasing at a slower rate. Furthermore, we have analyzed the scatter plots between the true process and the predictions, and saw that the predictions get closer to the 45{\degree} reference line as $n$ increases. We have also done a simulation study where we compared the performance of our model with the full model using a smaller dataset. It turns out the our proposed model performs even better than the full model for large enough $n$, which may be due to the semi-parametric nature of our model. The biggest advantage of our proposed model is scalability. Using simulation, we were able to show that our model is scalable to such large value of $N$ where other methods fail. We also emphasize that our proposed model is fully scalable (for large enough $G$) to a dataset of any size that can be stored and there is no model out there which is fully scalable to any size of the data.

In practice, a model is sometimes chosen simply because it is a traditional choice, and in spatial statistics a Gaussian process with a stationary covariogram is considered a textbook model \citep{banerjee2014hierarchical, cressie2015statistics}. However, this traditional spatial model does not scale naturally to high-dimensions and does not allow for non-stationary covariance functions. Our proposed approach allows spatial statisticians to apply traditional Bayesian spatial statistical models in a way that (i) is fully scalable in the limit and (ii) removes assumptions from the traditional spatial model to allow for the possibly of non-stationary covariance functions. This is achieved through the data subset model framework, which allows one to assume a traditional spatial model for all data with $\delta(\bld{s}_{i}) = 1$ such that $\sum_{i}\delta(\bld{s}_{i}) = n\ll N$. This is exciting solution for those who wish to adopt the more traditional model assumptions because of its ability to easily explain the spatial properties. However, it is important to clarify that while Items (i) and (ii) are attractive features of our model, it does not mean the proposed model will dominate estimation, prediction, and computational metrics. In our application, we demonstrate that the proposed model performs similar to competing methods, but not the best (i.e., our proposed method did not perform the best nor did it perform worst for any of metrics). In particular, we illustrate the use of our model to a daytime Land Surface Temperature (LST) data captured by the MODIS instrument onboard the Terra satellite. We have found that our model outperforms several methods in terms of MAE, RMSE, CRPS, INT, and CVG with $n=2,000$. We emphasize that although our proposed model is not the best, we are able to take a standard spatial model for low-dimensional stationary spatial data and modify it to be competitive for large data and to be able to model even non-stationary data.

Although our proposed model does produce competitive results, it depends on the choice of $n$. As $n$ changes, the model performance also changes. Thus, the choice of $n$ is important and an open problem in our proposed framework. In our empirical results, we see that posterior samples tend to have less variability as the subsample size $n$ grows. This is consistent with the results from \cite{bradley2021approach}. Consequently, the best choice of $n$ is the largest possible value that can be computed. In this article, we focus on applying the data subset model to kriging, which is well-known to be computationally feasible when $n$ is on the order of thousands (e.g., see \citealp{rulliere2018nested}, where $n = 2,000$ appears reasonable). When implementing our model in practice, one should simply choose an $n= 2,000$, or if more computational resources are available than our laptop, $n = 10,000$. In this article we make use of elbow plots simply to demonstrate the sensitivity of the choice of $n$, and we see that the ``elbow'' is often quite small (in the simulations elbow appears at $n = 300$). Of course, in general, when one does not know what $n$ is practical a priori through historical usage, we suggest what is commonly used in the SoD literature \citep{chalupka2013framework}. That is, an ``elbow plot'' can be used, where one selects an $n$ from a sequence that ``passes the elbow''.

Our proposed spatial data subset model is built on the classical Bayesian spatial model. However, one can take the similar approach of applying the subsampling strategy on some other spatial model. One important thing to note that the data subset model approach works on the Bayesian framework. So, one has to choose a Bayesian framework of a spatial model to build a model using this strategy. Therefore, building different spatial models on the data subset model framework can be a potential topic for future research.

\bibliographystyle{apalike}
\bibliography{bibliography.bib}

\appendix
\appendixpage
\section{Technical Results}
In this section, we show the proof of the propositions and the also the proof of covariance between $Z(\bld{s}_i)$ and $Z(\bld{s}_j)$ given $\bld{\theta}$, the variogram and the spatial properties such as sill, nugget and range. In Sections \ref{app:a1} and \ref{app:a2}, we show those proofs when $\text{Pr}(\bld{\delta}|n)$ is defined respectively by SRS without replacement and by stratified random sampling.

\subsection{Proof of Propositions \ref{prop:1} and \ref{prop:2}} \label{app:a1}

\subsubsection*{Proof of Proposition \ref{prop:1}:}

We prove Proposition \ref{prop:1} in two steps. In the first step, we use iterated expectations to derive $\mathbb{E}[Y(\bld{s}_i)|\bld{\nu},\bld{\theta}],Var(Y(\bld{s}_i)|\bld{\nu},\bld{\theta})$ and $Cov(Y(\bld{s}_i),Y(\bld{s}_j)|\bld{\nu},\bld{\theta})$. Then in the second step, we again use iterated expectations to finally derive $\mathbb{E}[Y(\bld{s}_i)|\bld{\theta}],Var(Y(\bld{s}_i)|\bld{\theta})$ and $Cov(Y(\bld{s}_i),Y(\bld{s}_j)|\bld{\theta})$.

From \cite{lohr2010sampling}, pages 52 and 53 we get
\begin{align}
    &\mathbb{E}[\delta(\bld{s}_i)] = \{0\times P(\delta(\bld{s}_i)=0)\} + \{1\times P(\delta(\bld{s}_i)=1)\} = \frac{\binom{N-1}{n-1}}{\binom{N}{n}} = \frac{n}{N}, \label{A1} \\
    &Var(\delta(\bld{s}_i)) = \mathbb{E}\left[\delta(\bld{s}_i)^2\right]-\mathbb{E}[\delta(\bld{s}_i)]^2 = \frac{n}{N}-\left(\frac{n}{N}\right)^2 = \frac{n}{N}\left(1-\frac{n}{N}\right), \label{A2} \\
    &\mathbb{E}[\delta(\bld{s}_i)\delta(\bld{s}_j)] = P(\delta(\bld{s}_j)=1|\delta(\bld{s}_i)=1)P(\delta(\bld{s}_i)=1) = \frac{\binom{N-2}{n-2}}{\binom{N-1}{n-1}}\frac{n}{N} = \frac{n}{N}\frac{n-1}{N-1}, \label{A3} \\
    &Cov(\delta(\bld{s}_i),\delta(\bld{s}_j)) = \mathbb{E}[\delta(\bld{s}_i)\delta(\bld{s}_j)]-\mathbb{E}[\delta(\bld{s}_i)]\mathbb{E}[\delta(\bld{s}_j)] = \frac{n}{N}\left(\frac{n-1}{N-1}-\frac{n}{N}\right). \label{A4}
\end{align}
Using these standard results we can derive the moments of $Y(\bld{s}_i)$ given $\bld{\nu}$ and $\bld{\theta}$. For ease of exposition, we denote $\mathbb{E}_{\bld{\delta}}\left(\mathbb{E}[Y(\bld{s}_i)|\bld{\nu},\bld{\theta},\bld{\delta}]|\bld{\nu},\bld{\theta}\right)=\mathbb{E}_{\bld{\delta}}\left(\mathbb{E}[Y(\bld{s}_i)|\bld{\nu},\bld{\theta},\bld{\delta}]\right)$ for all the expressions with iterated expectation formula. It means that we integrate the inside expression only with respect to the random variable present in the suffix, given all the other variables. Using the formula for iterated expectations we get
\begin{align}
    \mathbb{E}[Y(\bld{s}_i)|\bld{\nu},\bld{\theta}] &= \mathbb{E}_{\bld{\delta}}\left(\mathbb{E}[Y(\bld{s}_i)|\bld{\nu},\bld{\theta},\bld{\delta}]\right) \nonumber \\
    &= \mathbb{E}_{\bld{\delta}}\left[\delta(\bld{s}_i)\{\bld{x}(\bld{s}_i)'\bld{\beta}+\nu(\bld{s}_i)\}+\{1-\delta(\bld{s}_i)\}\tilde{\mu}(\bld{s}_i)\right] \nonumber \\
    &= \mathbb{E}_{\bld{\delta}}[\delta(\bld{s}_i)]\{\bld{x}(\bld{s}_i)'\bld{\beta}+\nu(\bld{s}_i)-\tilde{\mu}(\bld{s}_i)\}+\tilde{\mu}(\bld{s}_i) \nonumber \\
    \implies \mathbb{E}[Y(\bld{s}_i)|\bld{\nu},\bld{\theta}] &= \frac{n}{N}\{\bld{x}(\bld{s}_i)'\bld{\beta}+\nu(\bld{s}_i)-\tilde{\mu}(\bld{s}_i)\}+\tilde{\mu}(\bld{s}_i), \label{A5}
\end{align}
where we have applied \eqref{A1} to get the final expression. Also, recall $\tilde{\mu}(\bld{s}_i)$ is the true mean of $Y(\bld{s}_i)$ for $i=1,\ldots,N$. After integrating out $\bld{y}$ in Equation (2.2) we have the joint distribution of $\bld{\delta},\bld{\nu},\bld{\theta}$ is equal to $\text{Pr}(\bld{\delta}|n)p(\bld{\nu}|\bld{\theta})p(\bld{\theta})$ so that this independence implies $\mathbb{E}_{\bld{\delta}}[\delta(\bld{s}_i)|\bld{\nu},\bld{\theta}]=\mathbb{E}_{\bld{\delta}}[\delta(\bld{s}_i)]$. Similarly, $Var(\delta(\bld{s}_i)|\bld{\nu},\bld{\theta})=Var(\delta(\bld{s}_i))$, and this leads to
\begin{align}
    Var(Y(\bld{s}_i)|\bld{\nu},\bld{\theta}) &= \mathbb{E}_{\bld{\delta}}\left[Var(Y(\bld{s}_i)|\bld{\nu},\bld{\theta},\bld{\delta})\right]+Var_{\bld{\delta}}\left[\mathbb{E}(Y(\bld{s}_i)|\bld{\nu},\bld{\theta},\bld{\delta})\right] \nonumber \\
    &= \mathbb{E}_{\bld{\delta}}\left[\delta(\bld{s}_i)\tau^2+\{1-\delta(\bld{s}_i)\}\tilde{\sigma}^2\right] \nonumber \\
    &\qquad +Var_{\bld{\delta}}\left[\delta(\bld{s}_i)\{\bld{x}(\bld{s}_i)'\bld{\beta}+\nu(\bld{s}_i)\}+\{1-\delta(\bld{s}_i)\}\tilde{\mu}(\bld{s}_i)\right] \nonumber \\
    &= \mathbb{E}_{\bld{\delta}}[\delta(\bld{s}_i)](\tau^2-\tilde{\sigma}^2)+\tilde{\sigma}^2+Var_{\bld{\delta}}[\delta(\bld{s}_i)]\{\bld{x}(\bld{s}_i)'\bld{\beta}+\nu(\bld{s}_i)-\tilde{\mu}(\bld{s}_i)\}^2 \nonumber \\
    Var(Y(\bld{s}_i)|\bld{\nu},\bld{\theta}) &= \frac{n}{N}(\tau^2-\tilde{\sigma}^2)+\tilde{\sigma}^2+\frac{n}{N}\left(1-\frac{n}{N}\right)\{\bld{x}(\bld{s}_i)'\bld{\beta}+\nu(\bld{s}_i)-\tilde{\mu}(\bld{s}_i)\}^2, \label{A6}
\end{align}
where we have applied \eqref{A1} and \eqref{A2} to get the final expression. Again, recall $\tilde{\sigma}^2$ is the true variance of $Y(\bld{s}_i)$ for $i=1,\ldots,N$. The covariance between $Y(\bld{s}_i)$ and $Y(\bld{s}_j)$ given $\bld{\nu}$ and $\bld{\theta}$ is solved in a similar manner as follows:
\begin{align}
    Cov(Y(\bld{s}_i),Y(\bld{s}_j)|\bld{\nu},\bld{\theta}) &=  \mathbb{E}_{\bld{\delta}}\left[Cov(Y(\bld{s}_i),Y(\bld{s}_j)|\bld{\nu},\bld{\theta},\bld{\delta})\right] \nonumber \\
    &\qquad +Cov_{\bld{\delta}}\left[\mathbb{E}(Y(\bld{s}_i)|\bld{\nu},\bld{\theta},\bld{\delta}),\mathbb{E}(Y(\bld{s}_j)|\bld{\nu},\bld{\theta},\bld{\delta})\right]. \label{A7}
\end{align}
The first term in \eqref{A7} is
\begin{align}
    &\mathbb{E}_{\bld{\delta}}\left[Cov(Y(\bld{s}_i),Y(\bld{s}_j)|\bld{\nu},\bld{\theta},\bld{\delta})\right] \nonumber \\
    &= \mathbb{E}_{\bld{\delta}}\biggl[\delta(\bld{s}_i)\delta(\bld{s}_j)\mathbb{E}\left[\{Y(\bld{s}_i)-\bld{x}(\bld{s}_i)'\bld{\beta}-\nu(\bld{s}_i)\}\{Y(\bld{s}_j)-\bld{x}(\bld{s}_j)'\bld{\beta}-\nu(\bld{s}_j)\}|\bld{\nu},\bld{\theta},\bld{\delta}\right] \nonumber \\
    &\qquad \qquad +\delta(\bld{s}_i)[1-\delta(\bld{s}_j)]\mathbb{E}\left[\{Y(\bld{s}_i)-\bld{x}(\bld{s}_i)'\bld{\beta}-\nu(\bld{s}_i)\}\{Y(\bld{s}_j)-\tilde{\mu}(\bld{s}_j)\}|\bld{\nu},\bld{\theta},\bld{\delta}\right] \nonumber \\
    &\qquad \qquad +[1-\delta(\bld{s}_i)]\delta(\bld{s}_j)\mathbb{E}\left[\{Y(\bld{s}_i)-\tilde{\mu}(\bld{s}_i)\}\{Y(\bld{s}_j)-\bld{x}(\bld{s}_j)'\bld{\beta}-\nu(\bld{s}_j)\}|\bld{\nu},\bld{\theta},\bld{\delta}\right] \nonumber \\
    &\qquad \qquad +[1-\delta(\bld{s}_i)][1-\delta(\bld{s}_j)]\mathbb{E}\left[\{Y(\bld{s}_i)-\tilde{\mu}(\bld{s}_i)\}\{Y(\bld{s}_j)-\tilde{\mu}(\bld{s}_j)\}|\bld{\nu},\bld{\theta},\bld{\delta}\right]\biggr]. \label{A8}
\end{align}
The first three terms in \eqref{A8} are zero since given $\bld{\nu},\bld{\theta}$ and $\bld{\delta}$, the parametric model is independent. Only when $\delta(\bld{s}_i)=\delta(\bld{s}_j)$, do we have possible dependence only from the true model $w_{\delta}(\bld{y}_{-\delta})$. Thus,
\begin{align}
    &\mathbb{E}_{\bld{\delta}}\left[Cov(Y(\bld{s}_i),Y(\bld{s}_j)|\bld{\nu},\bld{\theta},\bld{\delta})\right] \nonumber \\
    &= \mathbb{E}_{\bld{\delta}}\bigl[\{1-\delta(\bld{s}_i)\}\{1-\delta(\bld{s}_j)\}\tilde{C}(\bld{s}_i,\bld{s}_j)\bigr] = \mathbb{E}_{\bld{\delta}}[\delta(\bld{s}_i)\delta(\bld{s}_j)-\delta(\bld{s}_i)-\delta(\bld{s}_j)+1]\tilde{C}(\bld{s}_i,\bld{s}_j) \nonumber \\
    &= \Bigl\{\mathbb{E}_{\bld{\delta}}[\delta(\bld{s}_i)\delta(\bld{s}_j)]-\mathbb{E}_{\bld{\delta}}[\delta(\bld{s}_i)]-\mathbb{E}_{\bld{\delta}}[\delta(\bld{s}_j)]+1\Bigr\}\tilde{C}(\bld{s}_i,\bld{s}_j) \nonumber \\
    &\implies \mathbb{E}_{\bld{\delta}}\left[Cov(Y(\bld{s}_i),Y(\bld{s}_j)|\bld{\nu},\bld{\theta},\bld{\delta})\right] = \left(\frac{n}{N}\frac{n-1}{N-1}-2\frac{n}{N}+1\right)\tilde{C}(\bld{s}_i,\bld{s}_j), \label{A9}
\end{align}
where we have applied \eqref{A1} and \eqref{A3} to get Equation \eqref{A9}. We recall $\tilde{C}(\bld{s}_i,\bld{s}_j)$ is the true covariance between $Y(\bld{s}_i)$ and $Y(\bld{s}_j)$.
The second term in \eqref{A7} is given by
\begin{align}
    &Cov_{\bld{\delta}}\left[\mathbb{E}(Y(\bld{s}_i)|\bld{\nu},\bld{\theta},\bld{\delta}),\mathbb{E}(Y(\bld{s}_j)|\bld{\nu},\bld{\theta},\bld{\delta})\right] \nonumber \\
    &= Cov_{\bld{\delta}}[\delta(\bld{s}_i)\{\bld{x}(\bld{s}_i)'\bld{\beta}+\nu(\bld{s}_i)\}+\{1-\delta(\bld{s}_i)\}\tilde{\mu}(\bld{s}_i), \nonumber \\
    & \qquad \qquad \delta(\bld{s}_j)\{\bld{x}(\bld{s}_j)'\bld{\beta}+\nu(\bld{s}_j)\}+\{1-\delta(\bld{s}_j)\}\tilde{\mu}(\bld{s}_j)] \nonumber \\
    &= Cov_{\bld{\delta}}[\delta(\bld{s}_i)\{\bld{x}(\bld{s}_i)'\bld{\beta}+\nu(\bld{s}_i)-\tilde{\mu}(\bld{s}_i)\}+\tilde{\mu}(\bld{s}_i),\delta(\bld{s}_j)\{\bld{x}(\bld{s}_j)'\bld{\beta}+\nu(\bld{s}_j)-\tilde{\mu}(\bld{s}_j)\}+\tilde{\mu}(\bld{s}_j)] \nonumber \\
    &= Cov_{\bld{\delta}}[\delta(\bld{s}_i),\delta(\bld{s}_j)]\{\bld{x}(\bld{s}_i)'\bld{\beta}+\nu(\bld{s}_i)-\tilde{\mu}(\bld{s}_i)\}\{\bld{x}(\bld{s}_j)'\bld{\beta}+\nu(\bld{s}_j)-\tilde{\mu}(\bld{s}_j)\} \nonumber \\
    &= \frac{n}{N}\left(\frac{n-1}{N-1}-\frac{n}{N}\right)\{\bld{x}(\bld{s}_i)'\bld{\beta}+\nu(\bld{s}_i)-\tilde{\mu}(\bld{s}_i)\}\{\bld{x}(\bld{s}_j)'\bld{\beta}+\nu(\bld{s}_j)-\tilde{\mu}(\bld{s}_j)\}, \label{A10}
\end{align}
where we have applied \eqref{A4} to get the final expression. Now, substituting \eqref{A9} and \eqref{A10} into Equation \eqref{A7} yields
\begin{align}
    &Cov(Y(\bld{s}_i),Y(\bld{s}_j)|\bld{\nu},\bld{\theta})= \left(\frac{n}{N}\frac{n-1}{N-1}-2\frac{n}{N}+1\right)\tilde{C}(\bld{s}_i,\bld{s}_j) \nonumber \\
    &\qquad +\frac{n}{N}\left(\frac{n-1}{N-1}-\frac{n}{N}\right)\{\bld{x}(\bld{s}_i)'\bld{\beta}+\nu(\bld{s}_i)-\tilde{\mu}(\bld{s}_i)\}\{\bld{x}(\bld{s}_j)'\bld{\beta}+\nu(\bld{s}_j)-\tilde{\mu}(\bld{s}_j)\}. \label{A11}
\end{align}
We now find the moments of $Y(\bld{s}_i)$ given $\bld{\theta}$ using iterated expectations and using \eqref{A5}, \eqref{A6} and \eqref{A11}. First, we calculate $\mathbb{E}[Y(\bld{s}_i)|\bld{\theta}]$ by substituting \eqref{A5} as follows:
\begin{align}
    \mathbb{E}[Y(\bld{s}_i)|\bld{\theta}] &= \mathbb{E}_{\bld{\nu}}\left[\mathbb{E}(Y(\bld{s}_i)|\bld{\nu},\bld{\theta})\right] \nonumber \\
    &= \mathbb{E}_{\bld{\nu}}\left[p\{\bld{x}(\bld{s}_i)'\bld{\beta}+\nu(\bld{s}_i)-\tilde{\mu}(\bld{s}_i)\}+\tilde{\mu}(\bld{s}_i)\right] \nonumber \\
    &= p\{\bld{x}(\bld{s}_i)'\bld{\beta}-\tilde{\mu}(\bld{s}_i)\}+\tilde{\mu}(\bld{s}_i) \nonumber \\
    &= p\{\bld{x}(\bld{s}_i)'\bld{\beta}\}+(1-p)\tilde{\mu}(\bld{s}_i), \label{A12}
\end{align}
where we recall $p=\frac{n}{N}$. Now, we calculate $Var(Y(\bld{s}_i)|\bld{\theta})$ by substituting \eqref{A5} and \eqref{A6} into the algebra below.
\begin{align}
    Var(Y(\bld{s}_i)|\bld{\theta}) &= \mathbb{E}_{\bld{\nu}}[Var(Y(\bld{s}_i)|\bld{\nu},\bld{\theta})]+Var_{\bld{\nu}}[\mathbb{E}(Y(\bld{s}_i)|\bld{\nu},\bld{\theta})] \nonumber \\
    &=\mathbb{E}_{\bld{\nu}}\left[p(\tau^2-\tilde{\sigma}^2)+\tilde{\sigma}^2+p(1-p)\{\bld{x}(\bld{s}_i)'\bld{\beta}+\nu(\bld{s}_i)-\tilde{\mu}(\bld{s}_i)\}^2\right] \nonumber \\
    &\qquad +Var_{\bld{\nu}}\left[p\{\bld{x}(\bld{s}_i)'\bld{\beta}+\nu(\bld{s}_i)-\tilde{\mu}(\bld{s}_i)\}+\tilde{\mu}(\bld{s}_i)\right] \nonumber \\
    &= p(\tau^2-\tilde{\sigma}^2)+\tilde{\sigma}^2+p(1-p)\mathbb{E}_{\bld{\nu}}\bigl[\nu(\bld{s}_i)^2+2\nu(\bld{s}_i)\{\bld{x}(\bld{s}_i)'\bld{\beta}-\tilde{\mu}(\bld{s}_i)\} \nonumber \\
    &\qquad +\{\bld{x}(\bld{s}_i)'\bld{\beta}-\tilde{\mu}(\bld{s}_i)\}^2\bigr]+p^2Var_{\bld{\nu}}(\nu(\bld{s}_i)) \nonumber \\
    &= p(\tau^2-\tilde{\sigma}^2)+\tilde{\sigma}^2+p(1-p)\left[\sigma^2+\{\bld{x}(\bld{s}_i)'\bld{\beta}-\tilde{\mu}(\bld{s}_i)\}^2\right] + p^2\sigma^2 \nonumber \\
    &= p(\tau^2+\sigma^2-\tilde{\sigma}^2)+\tilde{\sigma}^2+p(1-p)\{\bld{x}(\bld{s}_i)'\bld{\beta}-\tilde{\mu}(\bld{s}_i)\}^2 \nonumber \\
    &= p(\tau^2+\sigma^2)+(1-p)\tilde{\sigma}^2+p(1-p)\{\bld{x}(\bld{s}_i)'\bld{\beta}-\tilde{\mu}(\bld{s}_i)\}^2. \label{A13}
\end{align}
Lastly, we calculate $Cov(Y(\bld{s}_i),Y(\bld{s}_j)|\bld{\theta})$ by substituting \eqref{A5} and \eqref{A11} into the algebra below:
\begin{align}
    &Cov(Y(\bld{s}_i),Y(\bld{s}_j)|\bld{\theta}) \nonumber \\
    &=\mathbb{E}_{\bld{\nu}}[Cov(Y(\bld{s}_i),Y(\bld{s}_j)|\bld{\nu},\bld{\theta})]+Cov_{\bld{\nu}}[\mathbb{E}(Y(\bld{s}_i)|\bld{\nu},\bld{\theta}),\mathbb{E}(Y(\bld{s}_j)|\bld{\nu},\bld{\theta})] \nonumber \\
    &= \mathbb{E}_{\bld{\nu}}[a\tilde{C}(\bld{s}_i,\bld{s}_j)+b\{\bld{x}(\bld{s}_i)'\bld{\beta}+\nu(\bld{s}_i)-\tilde{\mu}(\bld{s}_i)\}\{\bld{x}(\bld{s}_j)'\bld{\beta}+\nu(\bld{s}_j)-\tilde{\mu}(\bld{s}_j)\}] \nonumber \\
    &\quad +Cov_{\bld{\nu}}[p\{\bld{x}(\bld{s}_i)'\bld{\beta}+\nu(\bld{s}_i)-\tilde{\mu}(\bld{s}_i)\}+\tilde{\mu}(\bld{s}_i),p\{\bld{x}(\bld{s}_j)'\bld{\beta}+\nu(\bld{s}_j)-\tilde{\mu}(\bld{s}_j)\}+\tilde{\mu}(\bld{s}_j)] \nonumber \\
    &= a\tilde{C}(\bld{s}_i,\bld{s}_j)+b\bigl[\mathbb{E}_{\bld{\nu}}[\nu(\bld{s}_i)\nu(\bld{s}_j)]+\{\bld{x}(\bld{s}_i)'\bld{\beta}-\tilde{\mu}(\bld{s}_i)\}\{\bld{x}(\bld{s}_j)'\bld{\beta}-\tilde{\mu}(\bld{s}_j)\}\bigr] \nonumber \\
    & \qquad \qquad +p^2Cov_{\bld{\nu}}[\nu(\bld{s}_i),\nu(\bld{s}_j)] \nonumber \\
    &= a\tilde{C}(\bld{s}_i,\bld{s}_j)+b\sigma^2h_{ij}(\phi)+b\{\bld{x}(\bld{s}_i)'\bld{\beta}-\tilde{\mu}(\bld{s}_i)\}\{\bld{x}(\bld{s}_j)'\bld{\beta}-\tilde{\mu}(\bld{s}_j)\}+p^2\sigma^2h_{ij}(\phi) \nonumber \\
    &= a\tilde{C}(\bld{s}_i,\bld{s}_j)+(b+p^2)\sigma^2h_{ij}(\phi)+b\{\bld{x}(\bld{s}_i)'\bld{\beta}-\tilde{\mu}(\bld{s}_i)\}\{\bld{x}(\bld{s}_j)'\bld{\beta}-\tilde{\mu}(\bld{s}_j)\}, \label{A14}
\end{align}
where we recall $a=\frac{n}{N}\frac{n-1}{N-1}-2\frac{n}{N}+1$ and $b=\frac{n}{N}\left(\frac{n-1}{N-1}-\frac{n}{N}\right)$. Equations \eqref{A12}, \eqref{A13} and \eqref{A14} together complete the result.

\subsubsection*{Proof of Proposition \ref{prop:2}:}

To study the spatial characteristics, we calculate the variogram for the de-trended $Y(\bld{s}_i)$ given $\bld{\theta}$ and $\bld{\delta}$. We define the de-trended data $Z(\bld{s}_i)$ given $\bld{\theta}$ and $\bld{\delta}$ as follows:
\begin{equation} \label{A15}
    Z(\bld{s}_i)=\begin{cases}
        Y(\bld{s}_i)-\bld{x}(\bld{s}_i)'\bld{\beta} & \delta(\bld{s}_i)=1\\
        Y(\bld{s}_i)-\tilde{\mu}(\bld{s}_i) & \delta(\bld{s}_i)=0.
    \end{cases}
\end{equation}
We expand Equation \eqref{A15} as
\begin{align}
    Z(\bld{s}_i) &= \delta(\bld{s}_i)\left\{Y(\bld{s}_i)-\bld{x}(\bld{s}_i)'\bld{\beta}\right\}+(1-\delta(\bld{s}_i))\left\{Y(\bld{s}_i)-\tilde{\mu}(\bld{s}_i)\right\} \nonumber \\
    \implies Z(\bld{s}_i) &= Y(\bld{s}_i)-\delta(\bld{s}_i)\{\bld{x}(\bld{s}_i)'\bld{\beta}-\tilde{\mu}(\bld{s}_i)\}-\tilde{\mu}(\bld{s}_i), \label{A16}
\end{align}
given $\bld{\theta}$ and $\bld{\delta}$. Now, if the process is intrinsically stationary, then there exists a $\gamma(\bld{s}_i-\bld{s}_j)$ such that
\begin{align}
    2\gamma(\bld{s}_i-\bld{s}_j) &= \mathbb{E}\left[\{Z(\bld{s}_i)-Z(\bld{s}_j)\}^2|\bld{\theta}\right] \nonumber \\
    &= Var\bigl(Z(\bld{s}_i)-Z(\bld{s}_j)|\bld{\theta}\bigr) \nonumber \\
    &= Var(Z(\bld{s}_i)|\bld{\theta})+Var(Z(\bld{s}_j)|\bld{\theta})-2Cov(Z(\bld{s}_i),Z(\bld{s}_j)|\bld{\theta}), \label{A17}
\end{align}
where $\bld{s}_i-\bld{s}_j$ is called the spatial lag and $2\gamma(\bld{s}_i-\bld{s}_j)$ is called the variogram. So, to find the variogram we need to find $Var(Z(\bld{s}_i)|\bld{\theta})$ and $Cov(Z(\bld{s}_i),Z(\bld{s}_j)|\bld{\theta})$. We calculate $Var(Z(\bld{s}_i)|\bld{\theta})$ as follows:
\begin{equation} \label{A18}
    Var(Z(\bld{s}_i)|\bld{\theta}) = \mathbb{E}_{\bld{\delta}}\left[Var(Z(\bld{s}_i)|\bld{\theta},\bld{\delta})\right]+Var_{\bld{\delta}}\left(\mathbb{E}[Z(\bld{s}_i)|\bld{\theta},\bld{\delta}]\right).
\end{equation}
Notice that, $Z(\bld{s}_i)|\bld{\theta},\bld{\delta}$ is de-trended, which means $\mathbb{E}[Z(\bld{s}_i)|\bld{\theta},\bld{\delta}]=0$. So, the second term of Equation \eqref{A18} is zero. Let's calculate $Var(Z(\bld{s}_i)|\bld{\theta},\bld{\delta})$ from the first term of \eqref{A18} by substituting \eqref{A16} into the below algebra.
\begin{align}
    Var(Z(\bld{s}_i)|\bld{\theta},\bld{\delta}) &= Var(Y(\bld{s}_i)-\delta(\bld{s}_i)\{\bld{x}(\bld{s}_i)'\bld{\beta}-\tilde{\mu}(\bld{s}_i)\}-\tilde{\mu}(\bld{s}_i)|\bld{\theta},\bld{\delta}) \nonumber \\
    &=Var(Y(\bld{s}_i)|\bld{\theta},\bld{\delta}) \nonumber \\
    &=\mathbb{E}_{\bld{\nu}}[Var(Y(\bld{s}_i)|\bld{\nu},\bld{\theta},\bld{\delta})]+Var_{\bld{\nu}}(\mathbb{E}[Y(\bld{s}_i)|\bld{\nu},\bld{\theta},\bld{\delta}]) \nonumber \\
    &= \mathbb{E}_{\bld{\nu}}\left[\delta(\bld{s}_i)\tau^2+\{1-\delta(\bld{s}_i)\}\tilde{\sigma}^2\right] \nonumber \\
    & \qquad +Var_{\bld{\nu}}\left(\delta(\bld{s}_i)\{\bld{x}(\bld{s}_i)'\bld{\beta}+\nu(\bld{s}_i)\}+\{1-\delta(\bld{s}_i)\}\tilde{\mu}(\bld{s}_i)\right) \nonumber \\
    &= \delta(\bld{s}_i)\tau^2+\{1-\delta(\bld{s}_i)\}\tilde{\sigma}^2+Var_{\bld{\nu}}\left(\delta(\bld{s}_i)\nu(\bld{s}_i)\right) \nonumber \\
    &= \delta(\bld{s}_i)\tau^2+\{1-\delta(\bld{s}_i)\}\tilde{\sigma}^2+\delta(\bld{s}_i)^2\sigma^2. \label{A19}
\end{align}
Substituting \eqref{A19} into \eqref{A18} and setting the second term of \eqref{A18} to zero we get
\begin{align}
    Var(Z(\bld{s}_i)|\bld{\theta}) &= \mathbb{E}_{\bld{\delta}}\left[\delta(\bld{s}_i)\tau^2+\{1-\delta(\bld{s}_i)\}\tilde{\sigma}^2+\delta(\bld{s}_i)^2\sigma^2\right] \nonumber \\
    &=p\tau^2+(1-p)\tilde{\sigma}^2+\left\{Var_{\bld{\delta}}(\delta(\bld{s}_i))+\mathbb{E}_{\bld{\delta}}[\delta(\bld{s}_i)]^2\right\}\sigma^2 \nonumber \\
    &= p\tau^2+(1-p)\tilde{\sigma}^2+p\sigma^2 \nonumber \\
    &= p\left(\tau^2+\sigma^2\right)+(1-p)\tilde{\sigma}^2, \label{A20}
\end{align}
where we have applied \eqref{A1} and \eqref{A2}, and we recall $p=\frac{n}{N}$. Now we calculate $Cov(Z(\bld{s}_i),Z(\bld{s}_j)|\bld{\theta})$ using the following expression.
\begin{equation} \label{A21}
    Cov(Z(\bld{s}_i),Z(\bld{s}_j)|\bld{\theta}) = \mathbb{E}_{\bld{\delta}}\left[Cov(Z(\bld{s}_i),Z(\bld{s}_j)|\bld{\theta},\bld{\delta})\right]+Cov_{\bld{\delta}}\left(\mathbb{E}[Z(\bld{s}_i)|\bld{\theta},\bld{\delta}],\mathbb{E}[Z(\bld{s}_j)|\bld{\theta},\bld{\delta}]\right)
\end{equation}
Again, $\mathbb{E}[Z(\bld{s}_i)|\bld{\theta},\bld{\delta}]=0$ and $\mathbb{E}[Z(\bld{s}_j)|\bld{\theta},\bld{\delta}]=0$. So, Equation \eqref{A21} becomes
\begin{align}
    &Cov(Z(\bld{s}_i),Z(\bld{s}_j)|\bld{\theta}) \nonumber \\
    &= \mathbb{E}_{\bld{\delta}}\left[Cov(Z(\bld{s}_i),Z(\bld{s}_j)|\bld{\theta},\bld{\delta})\right] \nonumber \\
    &= \mathbb{E}_{\bld{\delta}}\bigl[Cov(Y(\bld{s}_i)-\delta(\bld{s}_i)\{\bld{x}(\bld{s}_i)'\bld{\beta}-\tilde{\mu}(\bld{s}_i)\}-\tilde{\mu}(\bld{s}_i), \nonumber \\
    & \qquad \qquad \qquad Y(\bld{s}_j)-\delta(\bld{s}_j)\{\bld{x}(\bld{s}_j)'\bld{\beta}-\tilde{\mu}(\bld{s}_j)\}-\tilde{\mu}(\bld{s}_j)|\bld{\theta},\bld{\delta})\bigr] \nonumber \\
    &= \mathbb{E}_{\bld{\delta}}\left[Cov(Y(\bld{s}_i),Y(\bld{s}_j)|\bld{\theta},\bld{\delta})\right]. \label{A22}
\end{align}
First, we calculate $Cov(Y(\bld{s}_i),Y(\bld{s}_j)|\bld{\theta},\bld{\delta})$ in \eqref{A22} using the below expression.
\begin{align}
    Cov(Y(\bld{s}_i),Y(\bld{s}_j)|\bld{\theta},\bld{\delta}) &= \mathbb{E}_{\bld{\nu}}\left[Cov(Y(\bld{s}_i),Y(\bld{s}_j)|\bld{\nu},\bld{\theta},\bld{\delta})\right] \nonumber \\
    &\qquad +Cov_{\bld{\nu}}\left(\mathbb{E}[Y(\bld{s}_i)|\bld{\nu},\bld{\theta},\bld{\delta}],\mathbb{E}[Y(\bld{s}_j)|\bld{\nu},\bld{\theta},\bld{\delta}]\right) \label{A23}
\end{align}
So, \eqref{A23} consists of two terms on the right hand side. We calculate these two terms separately. The first term in \eqref{A23} is
\begin{align}
    &\mathbb{E}_{\bld{\nu}}\left[Cov(Y(\bld{s}_i),Y(\bld{s}_j)|\bld{\nu},\bld{\theta},\bld{\delta})\right] \nonumber \\
    &= \mathbb{E}_{\bld{\nu}}\biggl[\delta(\bld{s}_i)\delta(\bld{s}_j)\mathbb{E}\left[\{Y(\bld{s}_i)-\bld{x}(\bld{s}_i)'\bld{\beta}-\nu(\bld{s}_i)\}\{Y(\bld{s}_j)-\bld{x}(\bld{s}_j)'\bld{\beta}-\nu(\bld{s}_j)\}|\bld{\nu},\bld{\theta},\bld{\delta}\right] \nonumber \\
    &\qquad \qquad +\delta(\bld{s}_i)[1-\delta(\bld{s}_j)]\mathbb{E}\left[\{Y(\bld{s}_i)-\bld{x}(\bld{s}_i)'\bld{\beta}-\nu(\bld{s}_i)\}\{Y(\bld{s}_j)-\tilde{\mu}(\bld{s}_j)\}|\bld{\nu},\bld{\theta},\bld{\delta}\right] \nonumber \\
    &\qquad \qquad +[1-\delta(\bld{s}_i)]\delta(\bld{s}_j)\mathbb{E}\left[\{Y(\bld{s}_i)-\tilde{\mu}(\bld{s}_i)\}\{Y(\bld{s}_j)-\bld{x}(\bld{s}_j)'\bld{\beta}-\nu(\bld{s}_j)\}|\bld{\nu},\bld{\theta},\bld{\delta}\right] \nonumber \\
    &\qquad \qquad +[1-\delta(\bld{s}_i)][1-\delta(\bld{s}_j)]\mathbb{E}\left[\{Y(\bld{s}_i)-\tilde{\mu}(\bld{s}_i)\}\{Y(\bld{s}_j)-\tilde{\mu}(\bld{s}_j)\}|\bld{\nu},\bld{\theta},\bld{\delta}\right]\biggr]. \label{A24}
\end{align}
The first three terms in \eqref{A24} are zero since the parametric model is independent given $\bld{\nu}$, $\bld{\theta}$ and $\bld{\delta}$. Hence,
\begin{align}
    \mathbb{E}_{\bld{\nu}}\left[Cov(Y(\bld{s}_i),Y(\bld{s}_j)|\bld{\nu},\bld{\theta},\bld{\delta})\right] &= \mathbb{E}_{\bld{\nu}}\bigl[\{1-\delta(\bld{s}_i)\}\{1-\delta(\bld{s}_j)\}\tilde{C}(\bld{s}_i,\bld{s}_j)\bigr] \nonumber \\
    &= \{1-\delta(\bld{s}_i)\}\{1-\delta(\bld{s}_j)\}\tilde{C}(\bld{s}_i,\bld{s}_j) \nonumber \\
    &= \left\{\delta(\bld{s}_i)\delta(\bld{s}_j)-\delta(\bld{s}_i)-\delta(\bld{s}_j)+1\right\}\tilde{C}(\bld{s}_i,\bld{s}_j). \label{A25}
\end{align}
The second term in \eqref{A23} is
\begin{align}
    &Cov_{\bld{\nu}}\left(\mathbb{E}[Y(\bld{s}_i)|\bld{\nu},\bld{\theta},\bld{\delta}],\mathbb{E}[Y(\bld{s}_j)|\bld{\nu},\bld{\theta},\bld{\delta}]\right) \nonumber \\
    &= Cov_{\bld{\nu}}\bigl(\delta(\bld{s}_i)\{\bld{x}(\bld{s}_i)'\bld{\beta}+\nu(\bld{s}_i)\}+\{1-\delta(\bld{s}_i)\}\tilde{\mu}(\bld{s}_i), \nonumber \\
    &\qquad \qquad \delta(\bld{s}_j)\{\bld{x}(\bld{s}_j)'\bld{\beta}+\nu(\bld{s}_j)\}+\{1-\delta(\bld{s}_j)\}\tilde{\mu}(\bld{s}_j)\bigr) \nonumber \\
    &= Cov_{\bld{\nu}}\bigl(\delta(\bld{s}_i)\nu(\bld{s}_i),\delta(\bld{s}_j)\nu(\bld{s}_i)\bigr) \nonumber \\
    &= \delta(\bld{s}_i)\delta(\bld{s}_j)Cov_{\bld{\nu}}\bigl(\nu(\bld{s}_i),\nu(\bld{s}_j)\bigr) = \delta(\bld{s}_i)\delta(\bld{s}_j)\sigma^2h_{ij}(\phi). \label{A26}
\end{align}
Substituting \eqref{A25} and \eqref{A26} into \eqref{A23}, and further substituting \eqref{A23} to \eqref{A22} we get
\begin{align}
    &Cov(Z(\bld{s}_i),Z(\bld{s}_j)|\bld{\theta}) \nonumber \\
    &= \mathbb{E}_{\bld{\delta}}\left[Cov(Y(\bld{s}_i),Y(\bld{s}_j)|\bld{\theta},\bld{\delta})\right] \nonumber \\
    &= \mathbb{E}_{\bld{\delta}}\left[\left\{\delta(\bld{s}_i)\delta(\bld{s}_j)-\delta(\bld{s}_i)-\delta(\bld{s}_j)+1\right\}\tilde{C}(\bld{s}_i,\bld{s}_j)+\delta(\bld{s}_i)\delta(\bld{s}_j)\sigma^2h_{ij}(\phi)\right] \nonumber \\
    &= \left(\frac{n}{N}\frac{n-1}{N-1}-2\frac{n}{N}+1\right)\tilde{C}(\bld{s}_i,\bld{s}_j)+\frac{n}{N}\frac{n-1}{N-1}\sigma^2h_{ij}(\phi) \nonumber \\
    &= a\tilde{C}(\bld{s}_i,\bld{s}_j)+(b+p^2)\sigma^2h_{ij}(\phi), \label{A27}
\end{align}
where we have applied \eqref{A1} and \eqref{A3}, and we recall $a=\left(\frac{n}{N}\frac{n-1}{N-1}-2\frac{n}{N}+1\right)$ and $b=\frac{n}{N}\left(\frac{n-1}{N-1}-\frac{n}{N}\right)$. Notice that, $Z(\bld{s}_i)$ for $i=1,\ldots,N$ is non-stationary when the true data model is non-stationary, and $Z(\bld{s}_i)$ for $i=1,\ldots,N$ is weakly and/or intrinsically stationary when the true data model is respectively weakly and/or intrinsically stationary. This concludes the result of Proposition 2.2.
Now, to find the variogram, we substitute \eqref{A20} and \eqref{A27} into \eqref{A17}, and we get
\begin{equation} \label{A28}
    2\gamma(\bld{s}_i-\bld{s}_j) = 2\{p(\tau^2+\sigma^2)+(1-p)\tilde{\sigma}^2\}-2\left\{a\tilde{C}(\bld{s}_i,\bld{s}_j)+(b+p^2)\sigma^2h_{ij}(\phi)\right\}.
\end{equation}
Now, spatial properties like sill, nugget and range, of the de-trended data $Z(\bld{s}_i)$ for $i=1,\ldots,N$ can easily be computed from the variogram. If we assume the true de-trended data process model to be weakly stationary, i.e. $\tilde{C}(\bld{s}_i,\bld{s}_j)=\tilde{C}(\bld{d})$ where $\bld{d}=\bld{s}_i-\bld{s}_j$ is the lag, then the sill is
\begin{align}
    \text{Sill } &= \lim_{\lVert \bld{d} \rVert \to \infty}\gamma(\bld{d}) \nonumber \\
    &=p(\tau^2+\sigma^2)+(1-p)\tilde{\sigma}^2-\left\{a\lim_{\lVert \bld{d} \rVert \to \infty}\left[\tilde{C}(\bld{d})\right]+(b+p^2)\sigma^2\lim_{\lVert \bld{d} \rVert \to \infty}\left[\rho_c(\bld{d};\phi)\right]\right\} \nonumber \\
    &=p(\tau^2+\sigma^2)+(1-p)\tilde{\sigma}^2, \label{A29}
\end{align}
where we substitute $\gamma(\bld{d})$ from Equation \eqref{A28}, and we recall $h_{ij}(\phi)=\rho_c(\bld{d};\phi)$. Also, when $\lVert \bld{d} \rVert \to \infty$, $\tilde{C}(\bld{d})$ and $\rho_c(\bld{d};\phi)$ are 0. The nugget becomes
\begin{align}
    \text{Nugget } &= \lim_{\lVert \bld{d} \rVert \to 0^+}\gamma(\bld{d}) \nonumber \\
    &= p(\tau^2+\sigma^2)+(1-p)\tilde{\sigma}^2-a\lim_{\lVert \bld{d} \rVert \to 0^+}\left[\tilde{C}(\bld{d})\right] \nonumber \\
    &\qquad -(b+p^2)\sigma^2\lim_{\lVert \bld{d} \rVert \to 0^+}\left[\rho_c(\bld{d};\phi)\right] \nonumber \\
    &= p(\tau^2+\sigma^2)+(1-p)\tilde{\sigma}^2-a(\tilde{\sigma}^2-\tilde{\tau}^2)-(b+p^2)\sigma^2 \nonumber \\
    &= \left[p\tau^2+(p-b-p^2)\sigma^2\right]+\left[(1-p-a)\tilde{\sigma}^2+a\tilde{\tau}^2\right], \label{A30}
\end{align}
where $\tilde{\tau}^2$ is the nugget of the true de-trended data process. Notice that, when $\lVert \bld{d} \rVert \to 0^+$, $\tilde{C}(\bld{d})$ is equals to the difference between the true variance and true nugget. To find the range, we calculate the minimum value of $\bld{d}$ for which the variogram reaches the sill. So
\begin{equation} \label{A31}
    \text{Range } = \inf\left\{\bld{d}:\gamma(\bld{d})=\lim_{\bld{d} \to \infty} \gamma(\bld{d})\right\},
\end{equation}
where spatial lag $\bld{d}=\bld{s}_i-\bld{s}_j$. Again, $C(\bld{d})\to 0$ when $\bld{d}\to \infty$, where $C(\bld{d})$ is the covariance of the de-trended data $Z(\bld{s}_i)$. Therefore, range is also defined as
\begin{equation} \label{A32}
    \text{Range } = \inf\{\bld{d}:C(\bld{d})=0\} = \inf\left\{\bld{d}:a\tilde{C}(\bld{d})+(b+p^2)\sigma^2\rho_c(\bld{d};\phi)=0\right\},
\end{equation}
where we use Equation \eqref{A27} to get Equation \eqref{A32} and we recall $h_{ij}(\phi)=\rho_c(\bld{d};\phi)$. Equations \eqref{A29}, \eqref{A30} and \eqref{A32} together show the spatial properties of the de-trended data.

\subsection{Proof of Propositions \ref{prop:3} and \ref{prop:4}} \label{app:a2}

\subsubsection*{Proof of Proposition \ref{prop:3}:}

We prove Proposition \ref{prop:3} also in two steps. In the first step, we derive the expressions for $\mathbb{E}[Y(\bld{s}_i)|\bld{\theta}],Var(Y(\bld{s}_i)|\bld{\theta})$ and $Cov(Y(\bld{s}_i),Y(\bld{s}_j)|\bld{\theta})$ when $\bld{s}_i$ and $\bld{s}_j$ belong to same stratum. In the second step, we derive $Cov(Y(\bld{s}_i),Y(\bld{s}_j)|\bld{\theta})$ when $\bld{s}_i$ and $\bld{s}_j$ belong to different strata. To perform the second step, we first derive $Cov(Y(\bld{s}_i),Y(\bld{s}_j)|\bld{\nu},\bld{\theta})$, and then we derive $Cov(Y(\bld{s}_i),Y(\bld{s}_j)|\bld{\theta})$ using iterated expectations.

Let, $D_1,\ldots,D_R$ represent $R$ disjoint strata such that the spatial domain $D=\cup_{r=1}^R D_r$. The proofs of expectations, variances, and covariances when $\bld{s}_i$ and $\bld{s}_j$ are in same stratum $r$ are same as those were for SRS without replacement, since stratified random sampling applies SRS without replacement within each stratum independently (\citealp{lohr2010sampling}, pg. 77). Hence, mimicking \eqref{A1} through \eqref{A4} for $\bld{s}_i,\bld{s}_j \in D_r$ we get
\begin{align}
    &\mathbb{E}[\delta(\bld{s}_i)]=\frac{n_r}{N_r}, \label{A34} \\
    &Var(\delta(\bld{s}_i))=\frac{n_r}{N_r}\left(1-\frac{n_r}{N_r}\right), \label{A35} \\
    &\mathbb{E}[\delta(\bld{s}_i)\delta(\bld{s}_j)]=\frac{n_r}{N_r}\frac{n_r-1}{N_r-1}, \label{A36} \\
    &Cov(\delta(\bld{s}_i),\delta(\bld{s}_j))=\frac{n_r}{N_r}\left(\frac{n_r-1}{N_r-1}-\frac{n_r}{N_r}\right), \label{A37}
\end{align}
where we recall $n_r$ to be the number of subsamples drawn from $D_r$ with size $N_r$. So, substituting \eqref{A34} through \eqref{A37} into their respective places in \eqref{A12}, \eqref{A13} and \eqref{A14}, we get
\begin{align}
    &\mathbb{E}[Y(\bld{s}_i)|\bld{\theta}] = p_r\{\bld{x}(\bld{s}_i)'\bld{\beta}\}+(1-p_r)\tilde{\mu}(\bld{s}_i), \label{A38} \\
    &Var(Y(\bld{s}_i)|\bld{\theta}) = p_r(\tau^2+\sigma^2)+(1-p_r)\tilde{\sigma}^2+p_r(1-p_r)\{\bld{x}(\bld{s}_i)'\bld{\beta}-\tilde{\mu}(\bld{s}_i)\}^2, \label{A39} \\
    &Cov(Y(\bld{s}_i),Y(\bld{s}_j)|\bld{\theta}) = a_r\tilde{C}(\bld{s}_i,\bld{s}_j)+(b_r+p_r^2)\sigma^2h_{ij}(\phi) \nonumber \\
    &\qquad \qquad \qquad \qquad \qquad +b_r\{\bld{x}(\bld{s}_i)'\bld{\beta}-\tilde{\mu}(\bld{s}_i)\}\{\bld{x}(\bld{s}_j)'\bld{\beta}-\tilde{\mu}(\bld{s}_j)\}, \label{A40}
\end{align}
where we recall $p_r=\frac{n_r}{N_r},a_r=\left(\frac{n_r}{N_r}\frac{n_r-1}{N_r-1}-2\frac{n_r}{N_r}+1\right)$ and $b_r=\frac{n_r}{N_r}\left(\frac{n_r-1}{N_r-1}-\frac{n_r}{N_r}\right)$. Equations \eqref{A38}, \eqref{A39} and \eqref{A40} together conclude the result of the first step.

Now, we move to the second step. When $\bld{s}_i \in D_r$ and $\bld{s}_j\in D_t$ for $r \neq t$, $\mathbb{E}[\delta(\bld{s}_i)\delta(\bld{s}_j)]$ becomes
\begin{equation} \label{A41}
    \mathbb{E}[\delta(\bld{s}_i)\delta(\bld{s}_j)]=\mathbb{E}[\delta(\bld{s}_i)]\mathbb{E}[\delta(\bld{s}_j)]=\frac{n_r}{N_r}\frac{n_t}{N_t},
\end{equation}
since we independently sample from each stratum (\citealp{lohr2010sampling}, pg. 77). We recall $n_r$ to be the number of subsamples drawn from $D_r$ with size $N_r$ and $n_t$ to be the number of subsamples drawn from $D_t$ with size $N_t$. In this case, the covariance between two samples becomes zero because of independence.
\begin{equation} \label{A42}
    Cov(\delta(\bld{s}_i),\delta(\bld{s}_j)) = 0
\end{equation}

So, only the covariance between $Y(\bld{s}_i)$ and $Y(\bld{s}_j)$ changes when $\bld{s}_i$ and $\bld{s}_j$ belong to different strata. Hence, we now derive the the covariance between $Y(\bld{s}_i)$ and $Y(\bld{s}_j)$ given $\bld{\nu}$ and $\bld{\theta}$ only when $\bld{s}_i \in D_r$ and $\bld{s}_j\in D_t$ for $r \neq t$, using the standard results presented in \eqref{A34}, \eqref{A41} and \eqref{A42}. For ease of exposition, we again denote $\mathbb{E}_{\bld{\delta}}\left(\mathbb{E}[Y(\bld{s}_i)|\bld{\nu},\bld{\theta},\bld{\delta}]|\bld{\nu},\bld{\theta}\right)=\mathbb{E}_{\bld{\delta}}\left(\mathbb{E}[Y(\bld{s}_i)|\bld{\nu},\bld{\theta},\bld{\delta}]\right)$ for all the expressions with iterated expectation formula. It means that we integrate the inside expression only with respect to the random variable present in the suffix, given all the other variables. The covariance between $Y(\bld{s}_i)$ and $Y(\bld{s}_j)$ is given by
\begin{align}
    Cov(Y(\bld{s}_i),Y(\bld{s}_j)|\bld{\nu},\bld{\theta}) &=  \mathbb{E}_{\bld{\delta}}\left[Cov(Y(\bld{s}_i),Y(\bld{s}_j)|\bld{\nu},\bld{\theta},\bld{\delta})\right] \nonumber \\
    & \qquad +Cov_{\bld{\delta}}\left[\mathbb{E}(Y(\bld{s}_i)|\bld{\nu},\bld{\theta},\bld{\delta}),\mathbb{E}(Y(\bld{s}_j)|\bld{\nu},\bld{\theta},\bld{\delta})\right]. \label{A43}
\end{align}
To evaluate the first term in \eqref{A43}, we follow \eqref{A8} and \eqref{A9}, and we get
\begin{align}
    &\mathbb{E}_{\bld{\delta}}\left[Cov(Y(\bld{s}_i),Y(\bld{s}_j)|\bld{\nu},\bld{\theta},\bld{\delta})\right] \nonumber \\
    &= \mathbb{E}_{\bld{\delta}}\bigl[\{1-\delta(\bld{s}_i)\}\{1-\delta(\bld{s}_j)\}\tilde{C}(\bld{s}_i,\bld{s}_j)\bigr] \nonumber \\
    &= \mathbb{E}_{\bld{\delta}}[\delta(\bld{s}_i)\delta(\bld{s}_j)-\delta(\bld{s}_i)-\delta(\bld{s}_j)+1]\tilde{C}(\bld{s}_i,\bld{s}_j) \nonumber \\
    &= \Bigl\{\mathbb{E}_{\bld{\delta}}[\delta(\bld{s}_i)\delta(\bld{s}_j)]-\mathbb{E}_{\bld{\delta}}[\delta(\bld{s}_i)]-\mathbb{E}_{\bld{\delta}}[\delta(\bld{s}_j)]+1\Bigr\}\tilde{C}(\bld{s}_i,\bld{s}_j) \nonumber \\
    &\implies \mathbb{E}_{\bld{\delta}}\left[Cov(Y(\bld{s}_i),Y(\bld{s}_j)|\bld{\nu},\bld{\theta},\bld{\delta})\right] = \left(\frac{n_r}{N_r}\frac{n_t}{N_t}-\frac{n_r}{N_r}-\frac{n_t}{N_t}+1\right)\tilde{C}(\bld{s}_i,\bld{s}_j), \label{A44}
\end{align}
where we have applied \eqref{A34} and \eqref{A41} to get \eqref{A44}. Notice that, $\mathbb{E}[\delta(\bld{s}_j)]=\frac{n_t}{N_t}$ since $\bld{s}_j\in D_t$. Now, the second term in \eqref{A43} becomes
\begin{align}
    &Cov_{\bld{\delta}}\left[\mathbb{E}(Y(\bld{s}_i)|\bld{\nu},\bld{\theta},\bld{\delta}),\mathbb{E}(Y(\bld{s}_j)|\bld{\nu},\bld{\theta},\bld{\delta})\right] \nonumber \\
    &= Cov_{\bld{\delta}}\bigl[\delta(\bld{s}_i)\{\bld{x}(\bld{s}_i)'\bld{\beta}+\nu(\bld{s}_i)\}+\{1-\delta(\bld{s}_i)\}\tilde{\mu}(\bld{s}_i), \nonumber \\
    & \qquad \qquad \qquad \delta(\bld{s}_j)\{\bld{x}(\bld{s}_j)'\bld{\beta}+\nu(\bld{s}_j)\}+\{1-\delta(\bld{s}_j)\}\tilde{\mu}(\bld{s}_j)\bigr] \nonumber \\
    &= Cov_{\bld{\delta}}[\delta(\bld{s}_i)\{\bld{x}(\bld{s}_i)'\bld{\beta}+\nu(\bld{s}_i)-\tilde{\mu}(\bld{s}_i)\}+\tilde{\mu}(\bld{s}_i),\delta(\bld{s}_j)\{\bld{x}(\bld{s}_j)'\bld{\beta}+\nu(\bld{s}_j)-\tilde{\mu}(\bld{s}_j)\}+\tilde{\mu}(\bld{s}_j)] \nonumber \\
    &= Cov_{\bld{\delta}}[\delta(\bld{s}_i),\delta(\bld{s}_j)]\{\bld{x}(\bld{s}_i)'\bld{\beta}+\nu(\bld{s}_i)-\tilde{\mu}(\bld{s}_i)\}\{\bld{x}(\bld{s}_j)'\bld{\beta}+\nu(\bld{s}_j)-\tilde{\mu}(\bld{s}_j)\} = 0, \label{A45}
\end{align}
where we have applied \eqref{A42} to get \eqref{A45}. Substituting \eqref{A44} and \eqref{A45} into Equation \eqref{A43} yields
\begin{equation} \label{A46}
    Cov(Y(\bld{s}_i),Y(\bld{s}_j)|\bld{\nu},\bld{\theta})=\left(\frac{n_r}{N_r}\frac{n_t}{N_t}-\frac{n_r}{N_r}-\frac{n_t}{N_t}+1\right)\tilde{C}(\bld{s}_i,\bld{s}_j).
\end{equation}
Now, we find the covariance between $Y(\bld{s}_i)$ and $Y(\bld{s}_j)$ given $\bld{\theta}$ when $\bld{s}_i \in D_r$ and $\bld{s}_j\in D_t$ for $r \neq t$ using iterated expectations. First, we rewrite \eqref{A5} for $\bld{s}_i \in D_r$ and $\bld{s}_j\in D_t$ as follows:
\begin{align}
    &\mathbb{E}[Y(\bld{s}_i)|\bld{\nu},\bld{\theta}] = p_r\{\bld{x}(\bld{s}_i)'\bld{\beta}+\nu(\bld{s}_i)-\tilde{\mu}(\bld{s}_i)\}+\tilde{\mu}(\bld{s}_i), \label{A47} \\
    &\mathbb{E}[Y(\bld{s}_j)|\bld{\nu},\bld{\theta}] = p_t\{\bld{x}(\bld{s}_j)'\bld{\beta}+\nu(\bld{s}_j)-\tilde{\mu}(\bld{s}_j)\}+\tilde{\mu}(\bld{s}_j). \label{A48}
\end{align}
Now, we calculate $Cov(Y(\bld{s}_i),Y(\bld{s}_j)|\bld{\theta})$ by substituting \eqref{A46}, \eqref{A47} and \eqref{A48} into the algebra below:
\begin{align}
    &Cov(Y(\bld{s}_i),Y(\bld{s}_j)|\bld{\theta})=\mathbb{E}_{\bld{\nu}}[Cov(Y(\bld{s}_i),Y(\bld{s}_j)|\bld{\nu},\bld{\theta})]+Cov_{\bld{\nu}}[\mathbb{E}(Y(\bld{s}_i)|\bld{\nu},\bld{\theta}),\mathbb{E}(Y(\bld{s}_j)|\bld{\nu},\bld{\theta})] \nonumber \\
    &=\mathbb{E}_{\bld{\nu}}\left[\left(p_rp_t-p_r-p_t+1\right)\tilde{C}(\bld{s}_i,\bld{s}_j)\right] \nonumber \\
    &\qquad +Cov_{\bld{\nu}}\left[p_r\{\bld{x}(\bld{s}_i)'\bld{\beta}+\nu(\bld{s}_i)-\tilde{\mu}(\bld{s}_i)\}+\tilde{\mu}(\bld{s}_i),p_t\{\bld{x}(\bld{s}_j)'\bld{\beta}+\nu(\bld{s}_j)-\tilde{\mu}(\bld{s}_j)\}+\tilde{\mu}(\bld{s}_j)\right] \nonumber \\
    &=\left(p_rp_t-p_r-p_t+1\right)\tilde{C}(\bld{s}_i,\bld{s}_j)+p_rp_tCov_{\bld{\nu}}[\nu(\bld{s}_i),\nu(\bld{s}_j)] \nonumber \\
    &=\left(p_rp_t-p_r-p_t+1\right)\tilde{C}(\bld{s}_i,\bld{s}_j)+p_rp_t\sigma^2h_{ij}(\phi), \label{A49}
\end{align}
where we recall $p_r=\frac{n_r}{N_r}$ and $p_t=\frac{n_t}{N_t}$. Equations \eqref{A38}, \eqref{A39}, \eqref{A40} and \eqref{A49} together complete the result.

\subsubsection*{Proof of Proposition \ref{prop:4}:}

We start with the de-trended data $Z(\bld{s}_i)$ given $\bld{\theta}$ and $\bld{\delta}$ for $i=1,\ldots,N$ as defined in \eqref{A15} and \eqref{A16}, and the variogram defined in \eqref{A17}. We prove Proposition 2.4 for two scenarios - (a) when $\bld{s}_i$ and $\bld{s}_j$ belong to same stratum and (b) when $\bld{s}_i$ and $\bld{s}_j$ belong to different strata.

When $\bld{s}_i,\bld{s}_j \in D_r$, the $Cov(Z(\bld{s}_i),Z(\bld{s}_i)|\bld{\theta})$ is same as it is for SRS without replacement. So, rewriting \eqref{A27} for $\bld{s}_i,\bld{s}_j \in D_r$ we get
\begin{equation} \label{A50}
    Cov(Z(\bld{s}_i),Z(\bld{s}_j)|\bld{\theta}) = a_r\tilde{C}(\bld{s}_i,\bld{s}_j)+(b_r+p_r^2)\sigma^2h_{ij}(\phi),
\end{equation}
where we recall $p_r=\frac{n_r}{N_r},a_r=\left(\frac{n_r}{N_r}\frac{n_r-1}{N_r-1}-2\frac{n_r}{N_r}+1\right)$ and $b_r=\frac{n_r}{N_r}\left(\frac{n_r-1}{N_r-1}-\frac{n_r}{N_r}\right)$.

Now, we calculate $Cov(Z(\bld{s}_i),Z(\bld{s}_i)|\bld{\theta})$ when $\bld{s}_i \in D_r$ and $\bld{s}_j \in D_t$ for $r \neq t$. We start with rewriting Equation \eqref{A22} as follows:
\begin{equation} \label{A51}
    Cov(Z(\bld{s}_i),Z(\bld{s}_j)|\bld{\theta}) = \mathbb{E}_{\bld{\delta}}\left[Cov(Y(\bld{s}_i),Y(\bld{s}_j)|\bld{\theta},\bld{\delta})\right].
\end{equation}
So, first we calculate $Cov(Y(\bld{s}_i),Y(\bld{s}_j)|\bld{\theta},\bld{\delta})$ when $\bld{s}_i \in D_r$ and $\bld{s}_j \in D_t$ for $r \neq t$ using the following expression:
\begin{align}
    Cov(Y(\bld{s}_i),Y(\bld{s}_j)|\bld{\theta},\bld{\delta}) &= \mathbb{E}_{\bld{\nu}}\left[Cov(Y(\bld{s}_i),Y(\bld{s}_j)|\bld{\nu},\bld{\theta},\bld{\delta})\right] \nonumber \\
    &\qquad \qquad +Cov_{\bld{\nu}}\left(\mathbb{E}[Y(\bld{s}_i)|\bld{\nu},\bld{\theta},\bld{\delta}],\mathbb{E}[Y(\bld{s}_j)|\bld{\nu},\bld{\theta},\bld{\delta}]\right). \label{A52}
\end{align}
The steps for deriving \eqref{A52} will not be different than before since $\bld{\delta}$ is given. So, we substitute \eqref{A25} and \eqref{A26} into \eqref{A52} and we get
\begin{equation} \label{A53}
    Cov(Y(\bld{s}_i),Y(\bld{s}_j)|\bld{\theta},\bld{\delta})=\left\{\delta(\bld{s}_i)\delta(\bld{s}_j)-\delta(\bld{s}_i)-\delta(\bld{s}_j)+1\right\}\tilde{C}(\bld{s}_i,\bld{s}_j)+\delta(\bld{s}_i)\delta(\bld{s}_j)\sigma^2h_{ij}(\phi).
\end{equation}
Finally, substituting \eqref{A53} into Equation \eqref{A51}, we get
\begin{align}
    Cov(Z(\bld{s}_i),Z(\bld{s}_j)|\bld{\theta}) &= \mathbb{E}_{\bld{\delta}}\bigl[\left\{\delta(\bld{s}_i)\delta(\bld{s}_j)-\delta(\bld{s}_i)-\delta(\bld{s}_j)+1\right\}\tilde{C}(\bld{s}_i,\bld{s}_j) \nonumber \\
    &\qquad \qquad +\delta(\bld{s}_i)\delta(\bld{s}_j)\sigma^2h_{ij}(\phi)\bigr] \nonumber \\
    &= (p_rp_t-p_r-p_t+1)\tilde{C}(\bld{s}_i,\bld{s}_j)+p_rp_t\sigma^2h_{ij}(\phi), \label{A54}
\end{align}
where we have applied \eqref{A34} and \eqref{A41} to get \eqref{A54}. We again recall $p_r=\frac{n_r}{N_r}$ and $p_t=\frac{n_t}{N_t}$. Notice that, in both Equations \eqref{A50} and \eqref{A54}, $Cov(Z(\bld{s}_i),Z(\bld{s}_j)|\bld{\theta})$ depends on $\tilde{C}(\bld{s}_i,\bld{s}_j)$, the true covariance between $Z(\bld{s}_i)$ and $Z(\bld{s}_j)$. Hence, $Z(\bld{s}_i)$ for $i=1,\ldots,N$ is non-stationary when the true data model is non-stationary, and $Z(\bld{s}_i)$ for $i=1,\ldots,N$ is weakly and/or intrinsically stationary when the true data model is respectively weakly and/or intrinsically stationary. This concludes the result of Proposition 2.4.

Now, we find the variogram using Equation \eqref{A17}, when (a) $\bld{s}_i,\bld{s}_j \in D_r$, and (b) $\bld{s}_i \in D_r$ and $\bld{s}_j \in D_t$ for $r \neq t$. We start with rewriting \eqref{A20} for $\bld{s}_i \in D_r$ and $\bld{s}_j\in D_t$ as below:
\begin{align}
    &Var(Z(\bld{s}_i)|\bld{\theta}) = p_r\left(\tau^2+\sigma^2\right)+(1-p_r)\tilde{\sigma}^2, \label{A55} \\
    &Var(Z(\bld{s}_j)|\bld{\theta}) = p_t\left(\tau^2+\sigma^2\right)+(1-p_t)\tilde{\sigma}^2. \label{A56}
\end{align}
When $\bld{s}_i,\bld{s}_j \in D_r$, we rewrite \eqref{A28} accordingly and we get
\begin{equation} \label{A57}
    2\gamma(\bld{s}_i-\bld{s}_j) = 2\{p_r(\tau^2+\sigma^2)+(1-p_r)\tilde{\sigma}^2\}-2\left\{a_r\tilde{C}(\bld{s}_i,\bld{s}_j)+(b_r+p_r^2)\sigma^2h_{ij}(\phi)\right\}.
\end{equation}
But when $\bld{s}_i \in D_r$ and $\bld{s}_j \in D_t$ for $r \neq t$, we substitute \eqref{A54}, \eqref{A55} and \eqref{A56} into Equation \eqref{A17} and we get
\begin{align}
    &2\gamma(\bld{s}_i-\bld{s}_j) \nonumber \\
    &= Var(Z(\bld{s}_i)|\bld{\theta})+Var(Z(\bld{s}_j)|\bld{\theta})-2Cov(Z(\bld{s}_i),Z(\bld{s}_j)|\bld{\theta}) \nonumber \\
    &=p_r\left(\tau^2+\sigma^2\right)+(1-p_r)\tilde{\sigma}^2+p_t\left(\tau^2+\sigma^2\right)+(1-p_t)\tilde{\sigma}^2 \nonumber \\
    &\qquad -2\left\{a_{rt}\tilde{C}(\bld{s}_i,\bld{s}_j)+p_rp_t\sigma^2h_{ij}(\phi)\right\} \nonumber \\
    &=(p_r+p_t)(\tau^2+\sigma^2)+(2-p_r-p_t)\tilde{\sigma}^2-2\left\{a_{rt}\tilde{C}(\bld{s}_i,\bld{s}_j)+p_rp_t\sigma^2h_{ij}(\phi)\right\}, \label{A58}
\end{align}
where $a_{rt}=p_rp_t-p_r-p_t+1$. Now, spatial properties like sill, nugget and range, of the de-trended data $Z(\bld{s}_i)$ for $i=1,\ldots,N$ can easily be computed from the variogram given by Equations \eqref{A57} and \eqref{A58}.

When $\bld{s}_i,\bld{s}_j \in D_r$, the sill, nugget and range will be same as it is for SRS without replacement. So, if we assume the true de-trended data process model to be weakly stationary (i.e., $\tilde{C}(\bld{s}_i,\bld{s}_j)=\tilde{C}(\bld{d})$), then rewriting Equations \eqref{A29}, \eqref{A30} and \eqref{A32} for $\bld{s}_i,\bld{s}_j \in D_r$ we get
\begin{align}
    &\text{Sill} = p_r(\tau^2+\sigma^2)+(1-p_r)\tilde{\sigma}^2, \label{A59} \\
    &\text{Nugget} = \left[p_r\tau^2+(p_r-b_r-p_r^2)\sigma^2\right]+\left[(1-p_r-a_r)\tilde{\sigma}^2+a_r\tilde{\tau}^2\right], \label{A60} \\
    &\text{Range } = \inf\left\{\bld{d}:a_r\tilde{C}(\bld{d})+(b_r+p_r^2)\sigma^2\rho_c(\bld{d};\phi)=0\right\}, \label{A61}
\end{align}
where $\bld{d}=\bld{s}_i-\bld{s}_j$ is spatial lag and $\tilde{\tau}^2$ is the nugget of the true de-trended data process.

Now, when $\bld{s}_i \in D_r$ and $\bld{s}_j \in D_t$ for $r \neq t$, we calculate sill, nugget and range from \eqref{A58} assuming the true de-trended data process model to be weakly stationary and we get
\begin{align}
    \text{Sill } &= \lim_{\lVert \bld{d} \rVert \to \infty}\gamma(\bld{d}) \nonumber \\
    &=\frac{p_r+p_t}{2}(\tau^2+\sigma^2)+\left(1-\frac{p_r+p_t}{2}\right)\tilde{\sigma}^2 \nonumber \\
    & \qquad - \left\{a_{rt}\lim_{\lVert \bld{d} \rVert \to \infty}\left[\tilde{C}(\bld{d})\right]+p_rp_t\sigma^2\lim_{\lVert \bld{d} \rVert \to \infty}\left[\rho_c(\bld{d};\phi)\right]\right\} \nonumber \\
    &=\frac{p_r+p_t}{2}(\tau^2+\sigma^2)+\left(1-\frac{p_r+p_t}{2}\right)\tilde{\sigma}^2, \label{A63}
\end{align}
where we substitute $\gamma(\bld{d})$ from Equation \eqref{A58} and we recall $h_{ij}(\phi)=\rho_c(\bld{d};\phi)$. Also, $\tilde{C}(\bld{d})$ and $\rho_c(\bld{d};\phi)$ are zero when $\lVert \bld{d} \rVert \to \infty$. Then nugget becomes
\begin{align}
    \text{Nugget } &= \lim_{\lVert \bld{d} \rVert \to 0^+}\gamma(\bld{d}) \nonumber \\
    &=\frac{p_r+p_t}{2}(\tau^2+\sigma^2)+\left(1-\frac{p_r+p_t}{2}\right)\tilde{\sigma}^2 \nonumber \\
    &\qquad -\left\{a_{rt}\lim_{\lVert \bld{d} \rVert \to 0^+}\left[\tilde{C}(\bld{d})\right]+p_rp_t\sigma^2\lim_{\lVert \bld{d} \rVert \to 0^+}\left[\rho_c(\bld{d};\phi)\right]\right\} \nonumber \\
    &=\frac{p_r+p_t}{2}(\tau^2+\sigma^2)+\left(1-\frac{p_r+p_t}{2}\right)\tilde{\sigma}^2-a_{rt}(\tilde{\sigma}^2-\tilde{\tau}^2)-p_rp_t\sigma^2 \nonumber \\
    &=\left[\frac{p_r+p_t}{2}\tau^2+\frac{p_r+p_t-2p_rp_t}{2}\sigma^2\right]+\left[\frac{p_r+p_t-2p_rp_t}{2}\tilde{\sigma}^2+a_{rt}\tilde{\tau}^2\right], \label{A64}
\end{align}
where $\tilde{\tau}^2$ is the nugget of the true de-trended data process, and when $\lVert \bld{d} \rVert \to 0^+$, $\tilde{C}(\bld{d})$ is equals to the difference between the true variance and true nugget.

Finally, substituting \eqref{A54} in the definition of Range we get
\begin{align}
    \text{Range } &= \inf \left\{\bld{d}:C(\bld{d})=0\right\} \nonumber \\
    &= \inf \left\{\bld{d}:a_{rt}\tilde{C}(\bld{d})+p_rp_t\sigma^2\rho_c(\bld{d};\phi)=0\right\}, \label{A65}
\end{align}
where $\bld{d}$ is the spatial lag, and we recall $a_{rt}=p_rp_t-p_r-p_t+1$ and $h_{ij}(\phi)=\rho_c(\bld{d};\phi)$. Equations \eqref{A59}, \eqref{A60} and \eqref{A61} together show the spatial properties of the de-trended data when $\bld{s}_i,\bld{s}_j \in D_r$, and Equations \eqref{A63}, \eqref{A64} and \eqref{A65} together show the same when $\bld{s}_i \in D_r,\bld{s}_j \in D_t$ for $r \neq t$.

\section{Proof of Full Conditional Distributions} \label{app:b}
In this section, we show the proof of all the full conditional distributions that we use in Algorithm \ref{algo:gibbs}.

We use standard conjugate specifications in our Gibbs Sampler. Namely, we set $\mathcal{IG}(a_{\tau},b_{\tau})$, $\mathcal{IG}(a_{\sigma},b_{\sigma})$ and $\mathcal{IG}(a_{\beta},b_{\beta})$ for the priors of $\tau^2$, $\sigma^2$ and $\sigma_{\beta}^2$ respectively, and a discrete uniform distribution with support $\tilde{\phi}$ for the prior of $\phi$. The details behind the derivation of our full conditional distributions are given below.

\subsection*{Full conditional distribution of $\bld{\nu}_{\delta}$}
The full conditional distribution of $\bld{\nu}_{\delta}=(\nu(\bld{s}_i):\delta(\bld{s}_i)=1)'$ is proportional to the product of the parametric portion of the data subset model and the prior for $\bld{\nu}_{\delta}$. So, we get
\begin{align}
    f(\bld{\nu}_{\delta}|\cdot) &\propto \left\{\prod_{\{i:\delta(\bld{s}_i)=1\}} f\left(Y(\bld{s}_i)|\bld{\nu},\bld{\theta},\bld{\delta}\right)\right\}\pi\left(\bld{\nu}_{\delta}|{\sigma^2,\phi}\right) \nonumber \\
    &\propto \exp\left(-\frac{1}{2\tau^2}(\bld{y}_{\delta}-\bld{X}_{\delta}\bld{\beta}-\bld{\nu}_{\delta})'(\bld{y}_{\delta}-\bld{X}_{\delta}\bld{\beta}-\bld{\nu}_{\delta})\right)\exp\left(-\frac{1}{2\sigma^2}\bld{\nu}_{\delta}'\bld{H}_{\delta}^{-1}(\phi)\bld{\nu}_{\delta}\right) \nonumber \\
    &\propto \exp\left(-\frac{1}{2}\left\{\bld{\nu}_{\delta}'\left(\frac{1}{\tau^2}\bld{I}_n+\frac{1}{\sigma^2}\bld{H}_{\delta}^{-1}(\phi)\right)\bld{\nu}_{\delta}-2\frac{1}{\tau^2}\bld{\nu}_{\delta}'(\bld{y}_{\delta}-\bld{X}_{\delta}\bld{\beta})\right\}\right) \nonumber \\
    \implies \bld{\nu}_{\delta}|\cdot &\sim \mathcal{N}\left\{\left(\bld{I}_n+\frac{\tau^2}{\sigma^2}\bld{H}_{\delta}^{-1}(\phi)\right)^{-1}\left(\bld{y}_{\delta}-\bld{X}_{\delta}\bld{\beta}\right),\left(\frac{1}{\tau^2}\bld{I}_n+\frac{1}{\sigma^2}\bld{H}_{\delta}^{-1}(\phi)\right)^{-1}\right\}, \label{B1}
\end{align}
where $\bld{I}_n$ is the $n \times n$ identity matrix, $\bld{y}_{\delta}=\left(Y(\bld{s}_i):\delta(\bld{s}_i)=1\right)'$, $\bld{X}_{\delta}=\left(\bld{x}(\bld{s}_i):\delta(\bld{s}_i)=1\right)'$ and $\bld{H}_{\delta}(\phi)=\left(h_{jk}\left({\phi}\right):j,k\in \{i:\delta(\bld{s}_i)=1\}\right)$.

\subsection*{Full conditional distribution of $\bld{\nu}_A$}
The full conditional distribution of $\bld{\nu}_A$ has a conditional distribution arising from the joint multivariate distribution of $\bld{\nu}_A$ and $\bld{\nu}_{\delta}$, which is
\begin{equation} \label{B2}
    \begin{pmatrix}
    \bld{\nu}_A\\
    \bld{\nu}_{\delta}
    \end{pmatrix} \sim \mathcal{N}\left\{
    \begin{pmatrix}
    \bld{0}\\
    \bld{0}
    \end{pmatrix},
    \begin{pmatrix}
    \sigma^2\bld{H}_A(\phi) & \sigma^2\bld{H}_m(\phi)\\
    \sigma^2\bld{H}_m(\phi)' & \sigma^2\bld{H}_{\delta}(\phi)
    \end{pmatrix}
    \right\},
\end{equation}
where $\bld{H}_{\delta}(\phi)=(h_{jk}(\phi):j,k\in \{i:\delta(\bld{s}_i)=1\})$, $\bld{H}_A(\phi)=(h_{jk}(\phi):j,k\in A)$ and $\bld{H}_m(\phi)=(h_{jk}(\phi):j\in A,k\in \{i:\delta(\bld{s}_i)=1\})$. Using the formula for conditional distribution we get
\begin{equation} \label{B3}
    \bld{\nu}_A|\bld{\nu}_{\delta},\cdot \sim \mathcal{N}\Bigl\{\bld{H}_m(\phi)\bld{H}_{\delta}^{-1}(\phi)\bld{\nu}_{\delta},\sigma^2\left[\bld{H}_A(\phi)-\bld{H}_m(\phi)\bld{H}_{\delta}^{-1}(\phi)\bld{H}_m(\phi)'\right]\Bigr\}.
\end{equation}

\subsection*{Full conditional distribution of $\bld{\beta}$}
Now, we derive the full conditional distribution of $\bld{\beta}$ in the following way.
\begin{align}
    f(\bld{\beta}|\cdot) &\propto \left\{\prod_{\{i:\delta(\bld{s}_i)=1\}} f\left(Y(\bld{s}_i)|\bld{\nu},\bld{\theta},\bld{\delta}\right)\right\}\pi\left(\bld{\beta}|{\sigma_{\beta}^2}\right) \nonumber \\
    &\propto \exp\left(-\frac{1}{2\tau^2}(\bld{y}_{\delta}-\bld{X}_{\delta}\bld{\beta}-\bld{\nu}_{\delta})'(\bld{y}_{\delta}-\bld{X}_{\delta}\bld{\beta}-\bld{\nu}_{\delta})\right)\exp \left(-\frac{1}{2\sigma_{\beta}^2}\bld{\beta}'\bld{\beta}\right) \nonumber \\
    &\propto \exp \left(-\frac{1}{2}\left\{\bld{\beta}'\left(\frac{1}{\tau^2}\bld{X}_{\delta}'\bld{X}_{\delta}+\frac{1}{\sigma_{\beta}^2}\bld{I}_p\right)\bld{\beta}-2\frac{1}{\tau^2}\bld{\beta}'\bld{X}_{\delta}'\left(\bld{y}_{\delta}-\bld{\nu}_{\delta}\right)\right\}\right) \nonumber \\
    \implies \bld{\beta}|\cdot &\sim \mathcal{N}\left\{\left(\bld{X}_{\delta}'\bld{X}_{\delta}+\frac{\tau^2}{\sigma_{\beta}^2}\bld{I}_p\right)^{-1}\bld{X}_{\delta}'\left(\bld{y}_{\delta}-\bld{\nu}_{\delta}\right),\left(\frac{1}{\tau^2}\bld{X}_{\delta}'\bld{X}_{\delta}+\frac{1}{\sigma_{\beta}^2}\bld{I}_p\right)^{-1}\right\}, \label{B4}
\end{align}
where $\bld{I}_p$ is the $p \times p$ identity matrix.

\subsection*{Full conditional distribution of $\tau^2$}
We derive the full conditional distribution of $\tau^2$ in the similar fashion as follows.
\begin{align}
    f(\tau^2|\cdot) &\propto\left\{\prod_{\{i:\delta(\bld{s}_i)=1\}} f\left(Y(\bld{s}_i)|\bld{\nu},\bld{\theta},\bld{\delta}\right)\right\}\pi(\tau^2) \nonumber \\
    &\propto \left(\frac{1}{\tau^2}\right)^{n/2}\exp\left(-\frac{1}{2\tau^2}(\bld{y}_{\delta}-\bld{X}_{\delta}\bld{\beta}-\bld{\nu}_{\delta})'(\bld{y}_{\delta}-\bld{X}_{\delta}\bld{\beta}-\bld{\nu}_{\delta})\right)\left(\frac{1}{\tau^2}\right)^{a_{\tau}+1} \nonumber \\
    &\qquad \qquad \exp\left(-\frac{b_{\tau}}{\tau^2}\right) \nonumber \\
    &\propto \left(\frac{1}{\tau^2}\right)^{a_{\tau}+\frac{n}{2}+1}\exp\left(-\frac{1}{\tau^2}\left\{b_{\tau}+\frac{1}{2}(\bld{y}_{\delta}-\bld{X}_{\delta}\bld{\beta}-\bld{\nu}_{\delta})'(\bld{y}_{\delta}-\bld{X}_{\delta}\bld{\beta}-\bld{\nu}_{\delta})\right\}\right) \nonumber \\
    \implies \tau^2|\cdot &\sim \mathcal{IG}\left(a_{\tau}+\frac{n}{2},b_{\tau}+\frac{1}{2}(\bld{y}_{\delta}-\bld{X}_{\delta}\bld{\beta}-\bld{\nu}_{\delta})'(\bld{y}_{\delta}-\bld{X}_{\delta}\bld{\beta}-\bld{\nu}_{\delta})\right). \label{B5}
\end{align}

\subsection*{Full conditional distribution of $\sigma^2$}
The full conditional distribution of $\sigma^2$ is proportional to the product of the prior distribution for $\bld{\nu}_{\delta}$ and $\sigma^2$. So
\begin{align}
    f(\sigma^2|\cdot) &\propto f(\bld{\nu}_{\delta}|{\sigma^2,\phi})\pi(\sigma^2) \nonumber \\
    &\propto \left(\frac{1}{\sigma^2}\right)^{n/2}\exp\left(-\frac{1}{2\sigma^2}\bld{\nu}_{\delta}'\bld{H}_{\delta}^{-1}(\phi)\bld{\nu}_{\delta}\right)\left(\frac{1}{\sigma^2}\right)^{a_{\sigma}+1}\exp\left(-\frac{b_{\sigma}}{\sigma^2}\right) \nonumber \\
    &\propto \left(\frac{1}{\sigma^2}\right)^{a_{\sigma}+\frac{n}{2}+1}\exp\left(-\frac{1}{\sigma^2}\left\{b_{\sigma}+\frac{1}{2}\bld{\nu}_{\delta}'\bld{H}_{\delta}^{-1}(\phi)\bld{\nu}_{\delta}\right\}\right) \nonumber \\
    \implies \sigma^2|\cdot &\sim \mathcal{IG}\left(a_{\sigma}+\frac{n}{2},b_{\sigma}+\frac{1}{2}\bld{\nu}_{\delta}'\bld{H}_{\delta}^{-1}(\phi)\bld{\nu}_{\delta}\right). \label{B6}
\end{align}

\subsection*{Full conditional distribution of $\sigma_{\beta}^2$}
Similarly, the full conditional distribution of $\sigma_{\beta}^2$ is proportional to the product of the prior distribution for $\bld{\beta}$ and $\sigma_{\beta}^2$. Thus, we get
\begin{align}
    f(\sigma_{\beta}^2|\cdot) &\propto f\left(\bld{\beta}|\sigma_{\beta}^2\right)\pi(\sigma_{\beta}^2) \nonumber \\
    &\propto \left(\frac{1}{\sigma_{\beta}^2}\right)^{\frac{p}{2}} \exp \left(-\frac{1}{2\sigma_{\beta}^2}\bld{\beta}'\bld{\beta}\right)\left(\frac{1}{\sigma_{\beta}^2}\right)^{a_{\beta}+1}\exp \left(-\frac{b_{\beta}}{\sigma_{\beta}^2}\right) \nonumber \\
    &\propto \left(\frac{1}{\sigma_{\beta}^2}\right)^{a_{\beta}+\frac{p}{2}+1}\exp \left(-\frac{1}{\sigma_{\beta}^2}\left\{b_{\beta}+\frac{1}{2}\bld{\beta}'\bld{\beta}\right\}\right) \nonumber \\
    \implies \sigma_{\beta}^2|\cdot &\sim \mathcal{IG}\left(a_{\beta}+\frac{p}{2},b_{\beta}+\frac{1}{2}\bld{\beta}'\bld{\beta}\right). \label{B7}
\end{align}

\subsection*{Full conditional distribution of $\phi$}
The full conditional distribution for $\phi$ has a different form because of the discrete uniform nature of the prior for $\phi$. We derive the full conditional distribution of $\phi$ as below.
\begin{align}
    f(\phi|\cdot) &\propto f(\bld{\nu}_{\delta}|{\sigma^2,\phi})\pi(\phi) \nonumber \\
    &\propto \det\left(\bld{H}_{\delta}(\phi)\right)^{-1} \exp\left(-\frac{1}{2\sigma^2}\bld{\nu}_{\delta}'\bld{H}_{\delta}^{-1}(\phi)\bld{\nu}_{\delta}\right) \mathbbm{1}\left(\phi \in \tilde{\phi}\right), \label{B8}
\end{align}
where $\mathbbm{1}(\cdot)$ is the indicator function and $\tilde{\phi}$ is the support of the discrete uniform distribution $\pi(\phi)$. Since Equation \eqref{B8} does not have a closed form, we calculate the mass function at $\bld{\nu}_{\delta}$ quantiles for each $\phi \in \tilde{\phi}$ and we sample $\phi$ from $\tilde{\phi}$ with the corresponding probabilities/masses.

\newpage
\section{Empirical investigation of posterior consistency} \label{app:c}
In this section, we show the proof empirically that for large enough $n$, $p(\bld{\nu},\bld{\theta}|\bld{\delta}^*,\bld{y},n)$ reaches $p(\bld{\nu},\bld{\theta}|\bld{\delta},\bld{y},n)$ and it is reasonable to use Algorithm \ref{algo:gibbs}. In particular, we provide empirical evidence for posterior consistency (for large enough $n$) needed for Equation \eqref{eq:heuristic} for the simulated data (defined in Section \ref{subsec:simsetup}) and the LST data (described in Section \ref{subsec:dataset}).

\begin{figure}[b!]
    \centering
    \includegraphics[scale=0.45]{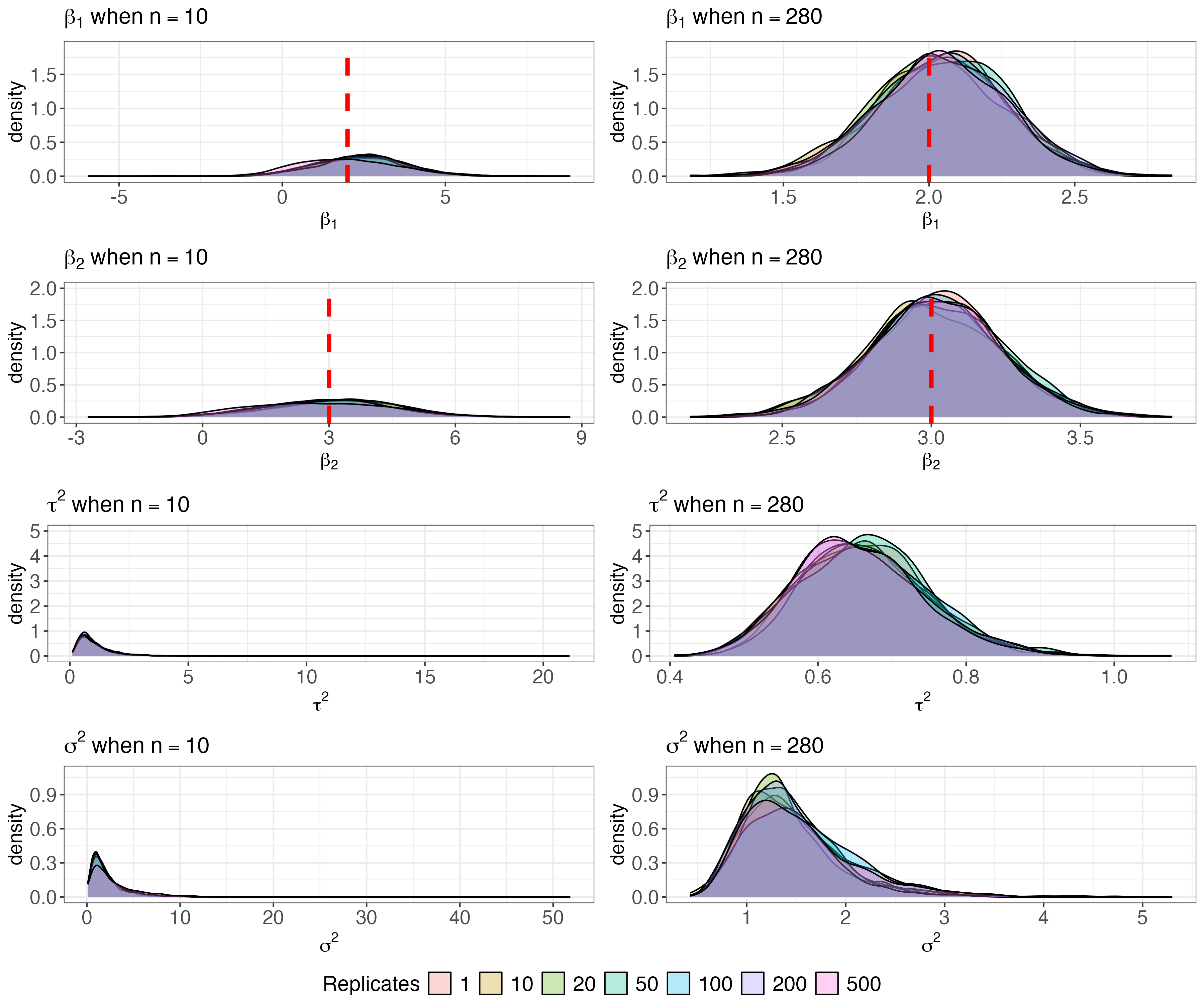}
    \caption{Density plots of MCMC replicates of the parameters from the simulated data when $n=10$ (left column) and $n=280$ (right column) for different number of replicates. The first, second, third and fourth row shows the kernel density comparison of $\beta_1$, $\beta_2$, $\tau^2$ and $\sigma^2$ respectively, for small (left panel) and large (right panel) value of $n$. The red dashed lines indicate the true values $\beta_1=2$ (first row) and $\beta_2=3$ (second row).}
    \label{fig:densityplots}
\end{figure}

\begin{figure}[t]
    \centering
    \includegraphics[scale=0.6]{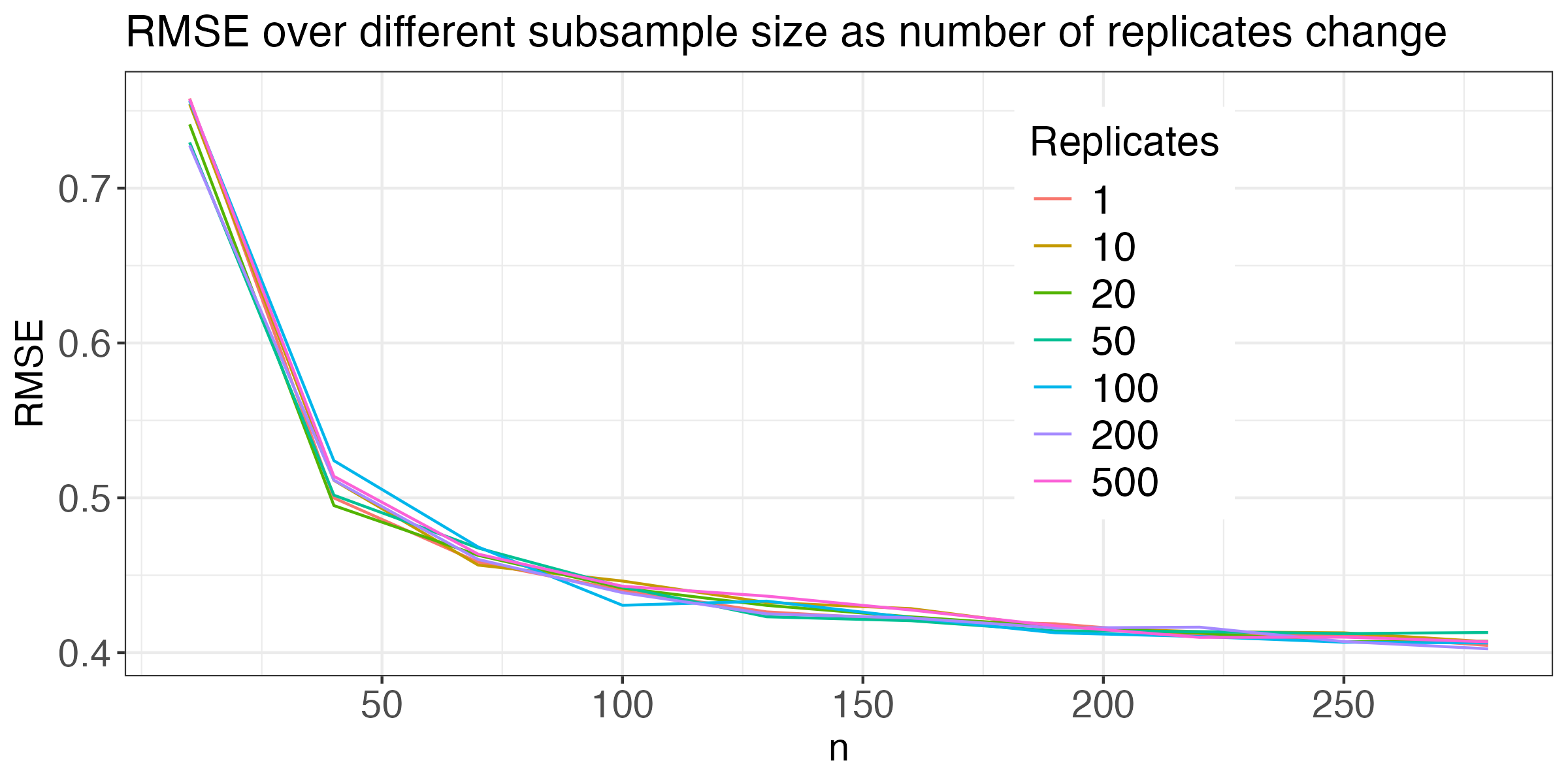}
    \caption{We plot the RMSE over different subsample sizes (on the simulated data) as we change the number of replicates for each choice of subsample size.}
    \label{fig:multiplereplicates}
\end{figure}

\begin{figure}[b!]
    \centering
    \includegraphics[scale=0.45]{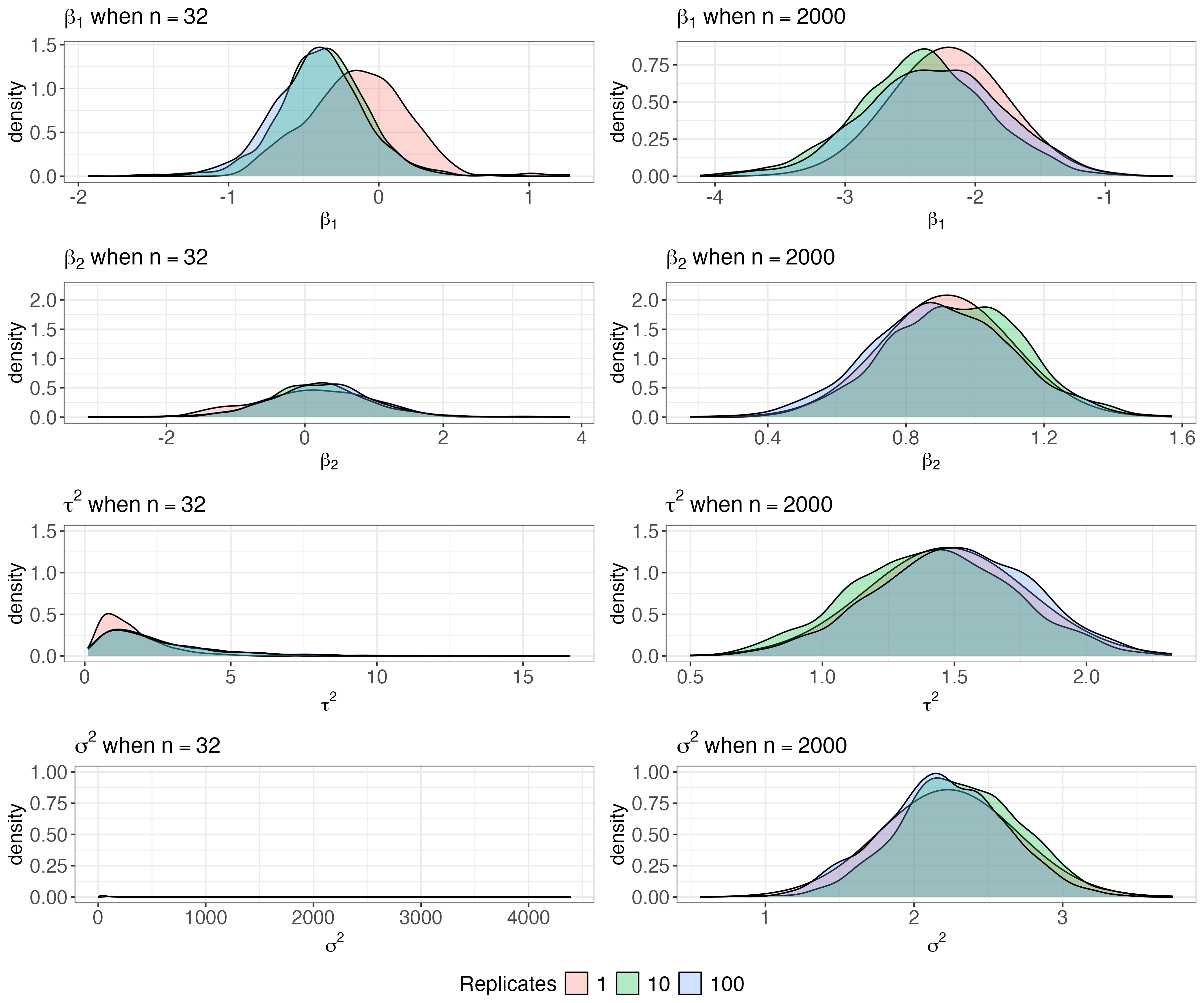}
    \caption{Density plots of MCMC replicates of the parameters from the LST data when $n=32$ (left column) and $n=2,000$ (right column) for different number of replicates. The first, second, third and fourth row shows the kernel density comparison of $\beta_1$, $\beta_2$, $\tau^2$ and $\sigma^2$ respectively, for small (left panel) and large (right panel) values of $n$.}
    \label{fig:densityplots_sat}
\end{figure}

In Figure \ref{fig:densityplots}, we plot the kernel densities of the MCMC replicates of $\beta_1$, $\beta_2$, $\tau^2$ and $\sigma^2$ that we get from the simulated data when the subsample size is 10 and 280 for different number of replicates. The left and right column shows all the kernel density plots when $n=10$ and $n=280$ respectively. The red dashed line in the first (and second) row indicates the true value of $\beta_1$ (and $\beta_2$). We see that both $\beta_1$ and $\beta_2$ contain zero when $n=10$. But when $n=280$, the densities of $\beta_1$ and $\beta_2$ concentrate around their respective true values with lesser variance and contain only non zero values. Similar structure (higher peak and lower variance) can also be seen in case of the kernel densities of $\tau^2$ and $\sigma^2$ when $n=280$ (right panels of third and fourth row). Also, in each plot in the right column (when $n=280$), the densities appear (almost) at the same position irrespective of the number of replicates.

In Figure \ref{fig:multiplereplicates}, we plot the out-of-sample RMSE over different subsample sizes as we change the number of replicates at each choice of $n$. We see that the RMSE decreases as $n$ increases. Moreover, higher number of replicates does not guarantee lower RMSE when $n$ is large enough. Hence, the number of replicates does not seem to affect the RMSE for large enough $n$. All these properties observed from Figure \ref{fig:densityplots} and \ref{fig:multiplereplicates} indicate that the joint distribution $p(\bld{\nu},\bld{\theta}|\bld{\delta}^*,\bld{y},n)$ attains its stationary distribution $p(\bld{\nu},\bld{\theta}|\bld{\delta},\bld{y},n)$ irrespective of the number of replicates for large enough $n$.

In Figure \ref{fig:densityplots_sat}, we plot the kernel densities of the MCMC replicates of $\beta_1$, $\beta_2$, $\tau^2$ and $\sigma^2$ that we get from the LST data when the subsample size is 32 and 2,000 for 1, 10 and 100 replicates. This plot also shows similar behavior as Figure \ref{fig:densityplots} which further indicates that it is reasonable to use Algorithm 1 for the LST data as well.

\end{document}